\documentclass[twocolumn]{aastex63}

\usepackage{lineno}
\linenumbers
\usepackage{graphicx, color}	% Including figure files
\usepackage{amsmath}	% Advanced maths commands
\usepackage{amssymb}	% Extra maths symbols
\usepackage{longtable}
\usepackage{enumitem}
\usepackage{scalerel}

\extrafloats{100}

\renewcommand{\baselinestretch}{1.0}

\mathchardef\mhyphen="2D

\newcommand{\oii}{O\,{\sc ii}}
\newcommand{\oiii}{O\,{\sc iii}}

\newcommand{\mgii}{Mg\,{\sc ii}}

\newcommand{\angstrom}{\text{ \normalfont\AA}}

\mathchardef\mhyphen="2D

\def\lya{Ly$\alpha$}
\def\ly{$\lambda$}
\def\ha{H$\alpha$}
\def\hb{H$\beta$}
\def\hg{H$\gamma$}
\def\hd{H$\delta$}

\def\hi{H\,{\sc i}}
\def\hii{H\,{\sc ii}}

\def\oi{O\,{\sc i}}
\def\oiii{O\,{\sc iii}}

\def\mgii{Mg\,{\sc ii}}

\def\sii{S\,{\sc ii}}

\def\Q0059{Q0059--2735}

\def\S2S3{S2S3}

\definecolor{blk}{rgb}{0.0,0.0,0.0}
\definecolor{red}{rgb}{0.75,0.0,0.0}
\definecolor{yel}{rgb}{0.65,0.65,0.0}
\definecolor{grn}{rgb}{0.0,0.75,0.0}
\definecolor{blu}{rgb}{0.0,0.0,0.75}
\definecolor{gry}{rgb}{0.75,0.75,0.75}

\def\nh{\ifmmode n_\mathrm{\scriptscriptstyle H} \else $n_\mathrm{\scriptscriptstyle H}$\fi}
\def\ne{\ifmmode n_\mathrm{\scriptstyle e} \else $n_\mathrm{\scriptstyle e}$\fi}
\def\Te{\ifmmode T_\mathrm{\scriptstyle e} \else $T_\mathrm{\scriptstyle e}$\fi}
\def\Qh{\ifmmode Q_\mathrm{\scriptstyle H} \else $Q_\mathrm{\scriptstyle H}$\fi}
\def\Uh{\ifmmode U_\mathrm{\scriptstyle H} \else $U_\mathrm{\scriptstyle H}$\fi}
\def\Nh{\ifmmode N_\mathrm{\scriptstyle H} \else $N_\mathrm{\scriptstyle H}$\fi}
\def\Uhhp{\ifmmode U_\mathrm{\scriptstyle H,HP} \else $U_\mathrm{\scriptstyle H,HP}$\fi}
\def\Nhhp{\ifmmode N_\mathrm{\scriptstyle H,HP} \else $N_\mathrm{\scriptstyle H,HP}$\fi}
\def\Uhvhp{\ifmmode U_\mathrm{\scriptstyle H,VHP} \else $U_\mathrm{\scriptstyle H,VHP}$\fi}
\def\Nhvhp{\ifmmode N_\mathrm{\scriptstyle H,VHP} \else $N_\mathrm{\scriptstyle H,VHP}$\fi}
\def\Nion{\ifmmode N_\mathrm{\scriptstyle ion} \else $N_\mathrm{\scriptstyle ion}$\fi}

\def\Zsun{\ifmmode {\rm Z}_{\odot} \else $Z_{\odot}$\fi}
\def\Msun{\ifmmode {\rm M}_{\odot} \else M$_{\odot}$\fi}
\def\kms{\ifmmode {\rm km~s}^{-1} \else km~s$^{-1}$\fi}
\def\Lya{\ifmmode {\rm Ly}\alpha \else Ly$\alpha$\fi}
\def\Lyb{\ifmmode {\rm Ly}\beta \else Ly$\beta$\fi}
\def\Lyg{\ifmmode {\rm Ly}\gamma \else Ly$\gamma$\fi}
\def\Lyd{\ifmmode {\rm Ly}\delta \else Ly$\delta$\fi}
\def\neaod{\ifmmode n_\mathrm{\scriptscriptstyle AOD} \else $n_\mathrm{\scriptscriptstyle AOD}$\fi}
\def\necrit{\ifmmode n_\mathrm{\scriptstyle cr} \else $n_\mathrm{\scriptstyle cr}$\fi}
\def\ncr{\ifmmode n_\mathrm{\scriptstyle cr} \else $n_\mathrm{\scriptstyle cr}$\fi}
\def\nepi{\ifmmode n_\mathrm{\scriptscriptstyle PI} \else $n_\mathrm{\scriptscriptstyle PI}$\fi}
\def\gtorder{\mathrel{\raise.3ex\hbox{$>$}\mkern-14mu\lower0.6ex\hbox{$\sim$}}}
\def\ltorder{\mathrel{\raise.3ex\hbox{$<$}\mkern-14mu\lower0.6ex\hbox{$\sim$}}}

\def\vro{\ifmmode v_\mathrm{\scriptscriptstyle 1, \scriptstyle r} \else $v_\mathrm{\scriptscriptstyle 1, \scriptstyle r}$\fi}
\def\vrc{\ifmmode v_\mathrm{\scriptscriptstyle 2, \scriptstyle r} \else $v_\mathrm{\scriptscriptstyle 2, \scriptstyle r}$\fi}
\def\vzo{\ifmmode v_\mathrm{\scriptscriptstyle 1, \scriptstyle z} \else $v_\mathrm{\scriptscriptstyle 1, \scriptstyle z}$\fi}
\def\vzc{\ifmmode v_\mathrm{\scriptscriptstyle 2, \scriptstyle z} \else $v_\mathrm{\scriptscriptstyle 2, \scriptstyle z}$\fi}

\newcommand{\fescLyC}{\textit{f}$_{\text{esc}}^{\text{ LyC}}$}
\newcommand{\fescLyA}{\textit{f}$_{\text{esc}}^{\text{ Ly}\alpha}$}
\newcommand{\fescMgII}{\textit{f}$_{\text{esc}}^{\text{ MgII}}$}
\newcommand{\fesc}{\textit{f}$_{\text{esc}}$}
\newcommand{\fescLyCPred}{\textit{f}$_{\text{esc,pd}}^{\text{ LyC}}$}

\newcommand{\FLyC}{\textit{F}(LyC)}
\newcommand{\FLyCOBS}{$F_{\scaleto{\text{obs}}{6pt}}(\text{LyC}$)}
\newcommand{\FLyCINT}{$F_{\scaleto{\text{int}}{6pt}}(\text{LyC}$)}

\newcommand{\FMgIIOBS}{$F_{\scaleto{\text{obs}}{6pt}}(\text{MgII}$)}
\newcommand{\FMgIIINT}{$F_{\scaleto{\text{int}}{6pt}}(\text{MgII}$)}
\newcommand{\FLyA}{\textit{F}(\lya)}

\newcommand{\CGeo}{CT$_\text{Geo}$}
\newcommand{\CCont}{CT$_\text{Cont}$}
\newcommand{\vsep}{$V_\text{sep}$}
\newcommand{\CF}{C$_{f}$}
\newcommand{\Tthin}{$\tau_{\text{thin}}$}
\newcommand{\Tthick}{$\tau_{\text{thick}}$}
\newcommand{\VpeakB}{$v^\text{peak}_\text{B}$}
\newcommand{\VpeakR}{$v^\text{peak}_\text{R}$}

\def\ZStar{\ifmmode {\rm Z}_\text{stars} \else $Z_\text{stars}$\fi}
\def\ZGas{\ifmmode {\rm Z}_\text{gas} \else $Z_\text{gas}$\fi}

\shorttitle{}   %update these later
\shortauthors{}
\graphicspath{{./}{figures/}}
\nolinenumbers
\begin{document}

\title{Tracing \lya\ and LyC Escape in Galaxies with \mgii\ Emission}

\author[0000-0002-9217-7051]{Xinfeng Xu}\thanks{E-mail: xinfeng@jhu.edu}
\affiliation{Center for Astrophysical Sciences, Department of Physics \& Astronomy, Johns Hopkins University, Baltimore, MD 21218, USA}

\author[0000-0002-6586-4446]{Alaina Henry}
\affiliation{Center for Astrophysical Sciences, Department of Physics \& Astronomy, Johns Hopkins University, Baltimore, MD 21218, USA}
\affiliation{Space Telescope Science Institute; 3700 San Martin Drive, Baltimore, MD, 21218, USA}

\author[0000-0003-1127-7497]{Timothy Heckman}
\affiliation{Center for Astrophysical Sciences, Department of Physics \& Astronomy, Johns Hopkins University, Baltimore, MD 21218, USA}

%Helped in early stage by alphabetical order
\author[0000-0002-0302-2577]{John Chisholm}
\affiliation{Department of Astronomy, The University of Texas at Austin, 2515 Speedway, Stop C1400, Austin, TX 78712, USA}

\author[0000-0003-0960-3580]{G\'abor Worseck}
\affiliation{Institut f\"ur Physik und Astronomie, Universit\"at Potsdam, Karl-Liebknecht-Str. 24/25, D-14476 Potsdam, Germany}

\author[0000-0003-2491-060X]{Max Gronke}
\affiliation{Max-Planck Institute for Astrophysics, Karl-Schwarzschild-Str. 1, D-85741 Garching, Germany}

\author[0000-0002-6790-5125]{Anne Jaskot}
\affiliation{Department of Astronomy, Williams College, Williamstown, MA 01267, United States}

\author[0000-0003-0503-4667]{Stephan R. McCandliss}
\affiliation{Center for Astrophysical Sciences, Department of Physics \& Astronomy, Johns Hopkins University, Baltimore, MD 21218, USA}

\author[0000-0002-0159-2613]{Sophia R. Flury}
\affiliation{Department of Astronomy, University of Massachusetts Amherst, Amherst, MA 01002, United States}

\author[0000-0002-7831-8751]{Mauro Giavalisco}
\affiliation{Department of Astronomy, University of Massachusetts Amherst, 710 N. Pleasant St., Amherst, MA 01003, USA}

\author[0000-0001-7673-2257]{Zhiyuan Ji}
\affiliation{Department of Astronomy, University of Massachusetts Amherst, 710 N. Pleasant St., Amherst, MA 01003, USA}

%Here are other coauthors by alphabetical order
\author[0000-0001-5758-1000]{Ricardo O. Amorín}
\affiliation{Instituto de Investigacion Multidisciplinar en Ciencia y Tecnologia, Universidad de La Serena, Raul Bitran 1305, La Serena, Chile}

\author[0000-0002-4153-053X]{Danielle A. Berg}
\affiliation{Department of Astronomy, The University of Texas at Austin, 2515 Speedway, Stop C1400, Austin, TX 78712, USA}

\author[0000-0002-2724-8298]{Sanchayeeta Borthakur}
\affiliation{School of Earth \& Space Exploration, Arizona State University, Tempe, AZ 85287, USA}

\author[0000-0003-0068-9920]{Nicolas Bouche}
\affiliation{Univ Lyon, Univ Lyon1, ENS de Lyon, CNRS, Centre de Recherche Astrophysique de Lyon (CRAL) UMR5574, 69230, Saint-Genis-Laval, France}

\author[0000-0003-4166-2855]{Cody Carr}
\affiliation{Minnesota Institute for Astrophysics, School of Physics and Astronomy, University of Minnesota, 316 Church str SE, Minneapolis, MN 55455,USA}

\author[0000-0001-9714-2758]{Dawn K. Erb}
\affiliation{Center for Gravitation, Cosmology and Astrophysics, Department of Physics, University of Wisconsin Milwaukee, 3135 N Maryland Ave., Milwaukee, WI 53211, USA}

\author{Harry Ferguson}
\affiliation{Space Telescope Science Institute; 3700 San Martin Drive, Baltimore, MD, 21218, USA}

\author{Thibault Garel}
\affiliation{Department of Astronomy, University of Geneva, 51 Chemin Pegasi, 1290 Versoix, Switzerland}

\author[0000-0001-8587-218X]{Matthew Hayes}
\affiliation{Department of Astronomy, Oskar Klein Centre; Stockholm University; SE-106 91 Stockholm, Sweden}

\author[0000-0003-3157-1191]{Kirill Makan}
\affiliation{Institut f\"ur Physik und Astronomie, Universit\"at Potsdam, Karl-Liebknecht-Str. 24/25, D-14476 Potsdam, Germany}

\author[0000-0001-8442-1846]{Rui Marques-Chaves}
\affiliation{Department of Astronomy, University of Geneva, 51 Chemin Pegasi, 1290 Versoix, Switzerland}

\author[0000-0001-7016-5220]{Michael Rutkowski}
\affiliation{Department of Physics and Astronomy, Minnesota State University, Mankato, MN, 56001, USA}

\author[0000-0002-3005-1349]{Göran Östlin}
\affiliation{Department of Astronomy, Oskar Klein Centre; Stockholm University; SE-106 91 Stockholm, Sweden}

\author[0000-0002-9946-4731]{Marc Rafelski}
\affiliation{Center for Astrophysical Sciences, Department of Physics \& Astronomy, Johns Hopkins University, Baltimore, MD 21218, USA}
\affiliation{Space Telescope Science Institute; 3700 San Martin Drive, Baltimore, MD, 21218, USA}

\author[0000-0001-8419-3062]{Alberto Saldana-Lopez}
\affiliation{Department of Astronomy, University of Geneva, 51 Chemin Pegasi, 1290 Versoix, Switzerland}
% ASL acknowledge support from Swiss National Science Foundation

\author[0000-0002-9136-8876]{Claudia Scarlata}
\affiliation{Minnesota Institute for Astrophysics, School of Physics and Astronomy, University of Minnesota, 316 Church str SE, Minneapolis, MN 55455, USA}

\author[0000-0001-7144-7182]{Daniel Schaerer}
\affiliation{Department of Astronomy, University of Geneva, 51 Chemin Pegasi, 1290 Versoix, Switzerland}

\author[0000-0002-6849-5375]{Maxime Trebitsch}
\affiliation{Kapteyn Astronomical Institute, University of Groningen, P.O. Box 800, 9700 AV Groningen, The Netherlands}

\author[0000-0003-3097-5178]{Christy Tremonti}
\affiliation{Department of Astronomy, University of Wisconsin–Madison, Madison, WI, 53706, USA}

\author[0000-0002-2201-1865]{Anne Verhamme}
\affiliation{Department of Astronomy, University of Geneva, 51 Chemin Pegasi, 1290 Versoix, Switzerland}

\author[0000-0001-9269-5046]{Bingjie Wang}
\affiliation{Center for Astrophysical Sciences, Department of Physics \& Astronomy, Johns Hopkins University, Baltimore, MD 21218, USA}
%{Industrial and Commercial Bank of China, 725 Fifth Ave., New York, NY 10022, USA}

%\suppressAffiliations



\begin{abstract}
Star-forming galaxies are considered the likeliest source of the \hi\ ionizing Lyman Continuum (LyC) photons that reionized the intergalactic medium at high redshifts.   However, above z $\gtrsim$ 6, the neutral intergalactic medium prevents direct observations of LyC. Therefore, recent years have seen the development of {\it indirect} indicators for LyC that can be calibrated at lower redshifts and applied in the Epoch of Reionization. Emission from \mgii\ \ly\ly 2796, 2803 doublet has been proposed as a promising LyC proxy. In this paper, we present new \textit{Hubble Space Telescope}/\textit{Cosmic Origins Spectrograph} observations for 8 LyC emitter candidates, selected to have strong \mgii\ emission lines. We securely detect LyC emission in 50\% (4/8) galaxies with 2$\sigma$ significance. This high detection rate suggests that strong \mgii\ emitters might be more likely to leak LyC than similar galaxies without strong \mgii. Using photoionization models, we constrain the escape fraction of \mgii\ as $\sim$ 15 -- 60\%. We confirm that the escape fraction of \mgii\ correlates tightly with that of \lya, which we interpret as an indication that the escape fraction of both species is controlled by resonant scattering in the same low column density gas. Furthermore, we show that the combination of the \mgii\ emission and dust attenuation can be used to estimate the escape fraction of LyC statistically. These findings confirm that \mgii\ emission can be adopted to estimate the escape fraction of \lya\ and LyC in local star-forming galaxies and may serve as a useful indirect indicator at the Epoch of Reionization.
%This detection rate is 2.5 times higher than that of the low redshift samples in the literature within the same O32 range: 3 $<$ [\oiii] \ly 5007/[\oii] \ly 3727 $<$ 6.
%Compared to previous studies, our \mgii\ selected galaxies show similar correlations between the escape of LyC and indirect LyC indicators such as O32 and the properties of Lya emission.
%

% galaxies with lower significance detections in the other 2 galaxies

%a similar range of [\oiii] \ly 5007/[\oii] \ly 3727 flux ratios.
%Combined with a larger sample of published LCEs that have high signal-to-noise \mgii\ emission detected, we
%has been proposed to be such an indirect indicator in local LyC emitting galaxies (LCEs)
\end{abstract}



\section{Introduction} 
\label{sec:intro}

%LyC define and challenges
The Epoch of Reionization (EoR) was one of the last major phase transitions of the Universe, marked by the emergence of the first galaxies and the ionization of the neutral hydrogen (\hi) in the intergalactic medium (IGM). Though it has been widely recognized from various observations that the EoR happened at z $\sim$ 6 -- 9 \citep[e.g.,][]{Becker01, Fan06, Banados18, Mason18, Planck20}, the type of sources that were responsible for the majority of ionizing photons remain elusive. 

High luminosity Active Galactic Nuclei (AGN) or quasars are appealing sources of ionizing photons. However, the number of quasars at high$-z$ is too low to reionize the whole universe \citep[e.g.,][]{Hopkins08, Madau15, Matsuoka18, Kulkarni19, Shen20, Trebitsch21}. Although the contribution from unobserved low-luminosity AGN is still debated \citep{Giallongo15, Matsuoka18,Parsa18, Grazian20}, star-forming (SF) galaxies are usually considered the likeliest candidates since they substantially outnumber quasars \citep[e.g.,][]{Madau14,Shen20}. Critically, however, it has not yet been demonstrated that SF galaxies can contribute enough \hi-ioninzing Lyman Continuum (LyC) photons to ionize the intergalactic medium (IGM). Moreover, it is still unclear whether the reionization was dominated by brighter, more massive galaxies, or the more numerous population of faint SF galaxies \citep[e.g.,][]{Finkelstein19, Naidu20}.

The ionizing photon budget is commonly described as the product of three parameters: the UV luminosity density of either AGN or SF galaxies ($\rho_\text{UV}$, ergs s$^{-1}$ Hz$^{-1}$ Mpc$^{-3}$), the ionizing photon production efficiency ($\xi_\text{ion}$, photon ergs$^{-1}$ Hz), and the fraction of ionizing photons that escape from galaxies \citep[\fescLyC, \%, see, e.g.,][]{Madau99, Robertson13, Duncan15}. $\xi_\text{ion}$ and $\rho_\text{UV}$ have been or can be mostly constrained \citep[e.g.,][]{Chevallard18,Tang19, Berg19, Bouwens21}.  Models using current estimations of  $\xi_\text{ion}$ and $\rho_\text{UV}$ suggest that \fescLyC\ needs to be $>$ 5 -- 20\% on average for star-forming galaxies to reionize the universe \citep[e.g.][]{Robertson13, Robertson15,Rosdahl18, Finkelstein19, Naidu20}. However, inferring \fescLyC\ for high-z galaxies is extremely challenging due to the absorption of the LyC photons by 1) its short mean free path to the high incidence of Lyman limit systems for 4 $\lesssim$ z $\lesssim$ 6 \citep[e.g.,][]{Worseck14, Becker21}; and 2) the neutral IGM for z $\gtrsim$ 6 \citep[e.g.,][]{Inoue14}. Thus, observations at lower redshifts must develop indirect indicators for \fescLyC, which can then be applied to EoR galaxies.

During the last decade, considerable efforts have been made towards measuring \fescLyC\ at low redshifts, mainly adopting observations from the \textit{Far Ultraviolet Spectroscopic Explorer} \citep[FUSE, e.g.,][]{Heckman01, Bergvall06, Leitet13, Borthakur14} and \textit{Hubble Space Telescope}/\textit{Cosmic Origins Spectrograph} \citep[HST/COS, e.g.,][]{Leitherer16,Izotov16a, Izotov16b, Puschnig17, Izotov18a,Izotov18b, Wang19, Izotov21, Flury22a, Flury22b}. Various indirect indicators of \fescLyC\ have also been proposed \citep[see a summary in][]{Flury22b}.   For example, the [\oiii] \ly 5007/[\oii] \ly 3727 flux ratio (O32) may be an indicator of the ionization state of the interstellar medium (ISM), with high O32 values suggesting low optical depth in neutral gas \citep[e.g.,][]{Jaskot13,Oey15, Nakajima20}.  Alternatively, star formation rate surface density ($\Sigma_\text{SFR}$) is postulated to relate to galaxy outflows, which can clear holes in the ISM to allow LyC escape \citep[e.g.,][]{Heckman11, Borthakur14, Alex15, Trebitsch17,Saldana-Lopez22}. 

\begin{table*}
	\centering
	\caption{HST Observations and Basic Properties for Galaxies in Our Sample}
	\label{tab:obs}
	\begin{tabular}{lllcllcccc} % four columns, alignment for each
		\hline
		\hline
		ID & RA & Dec 	    & $z^{1}$   &	G140L$^{2}$ &	G160M$^{2}$  & SDSS-u$^{3}$ & NUV$^{4}$ & FUV$^{4}$  &  $E(B-V)_\text{MW}^{5}$  \\
		\hline
		& &         	&           &	(s)         & (s) & (mag) & (mag) & (mag) & \\
		\hline
        J0105+2349  & 01:05:33.74 & +23:49:59.63      &0.3381	    	& 	4500 & 2500	    &21.45  & 21.52 & 21.82 & 0.034\\
        J0152--0431 & 01:52:07.99 & --04:31:17.17     &0.3836		    & 	1700 & 2500	    &21.65  & 21.30 & 21.50 & 0.036\\
        J0208--0401 & 02:08:18.90 & --04:01:36.37     &0.3844		    & 	7200 & 2500	    &21.50  & 21.29 & 21.35 & 0.021\\
        J1103+4834  & 11:03:59.00 & +48:34:55.95      &0.4180		    & 	2000 & 2600	    &21.21  & 21.19 & 21.66 & 0.014\\
        J1105+5947  & 11:05:06.33 & +59:47:41.37      &0.4054		    & 	2000 & 2700	    &21.59  & 21.30 & 22.05 & 0.007 \\
        J1219+4814  & 12:19:47.85 & +48:14:10.59      &0.4203		    & 	1900 & 2600	    &21.26  & 21.23 & 21.48 & 0.011 \\
        J1246+4449$^{6}$  & 12:46:19.49 & +44:49:02.43&0.3222		    & 	1600 & 2000     &20.28  & 20.17 & 20.55 & 0.021 \\
        J1425+5249  & 14:25:35.11 & +52:49:02.18      &0.3870		    & 	2600 & 4900	    &21.90  & 21.58 & 21.92 & 0.007 \\

		\hline
		\hline
	\multicolumn{10}{l}{%
  	\begin{minipage}{16cm}%
	Note. --\\
    	(1)\ \ Redshift of the objects derived from Gaussian fits to the Balmer emission lines.\\
    	(2)\ \ Exposure time in seconds for HST COS/G140L and G160M gratings (HST-GO: 15865, PI: Henry).\\
    	(3)\ \ The u-band magnitudes from SDSS photometry.\\
    	(4)\ \ Near-UV and far-UV band magnitude from GALEX photometry, respectively.\\
    	(5)\ \ Milky way dust extinction obtained from Galactic Dust Reddening and Extinction Map \citep{Schlafly11} at NASA/IPAC Infrared Science Archive.\\
    	(6)\ \ The COS/G140L observation of J1246+4449 is from the archival data (LzLCS survey, GO: 15626, PI: Jaskot).\\
    	
  	\end{minipage}%
	}\\
	\end{tabular}
	\\ [0mm]
	
\end{table*}

One of the leading indirect indicators is \lya\ emission \citep[e.g.,][]{Verhamme15, Henry15, Dijkstra16, Verhamme17, Jaskot19, Gazagnes20, Kakiichi21}. \lya\ photons resonantly scatter so that their emergence from galaxies is strongly influenced by the reservoir of neutral hydrogen in/around the galaxy. The scattering alters the intrinsic \lya\ emission line profile and, at the same time, imprints valuable information about \hi\ onto the modified line profile.  In particular, the presence of double-peaked \lya\ profiles, and the separation between the blue and red peaks have been demonstrated to correlate with \fescLyC\ \citep[e.g.,][]{Izotov18b,Gazagnes20}. However, for high-z galaxies (z $\gtrsim$ 4), the neutral IGM can absorb a large amount of the \lya\ photons, in particular their blue peaks (usually the weaker peak). Therefore, the interpretation of \lya\ profiles at high-z can be challenging \citep[e.g.,][]{Stark11,Schenker14, Gronke21, Hayes21}.

In this paper, we further study \mgii\ as an indirect indicator for tracing the escape of \lya\ and LyC. This idea has been proposed in \cite{Henry18} and also studied in \cite{Chisholm20}. \mgii\ has a doublet transition at 2796.35, 2803.53\angstrom, and is similar to \lya\ as an indirect indicator, because 1) \mgii\ is a low-ionization transition, and has an ionization potential (to destroy \mgii) as 15.03 eV, which is close to that of destroying \hi\ ($\simeq$ 13.6 eV). 2) \mgii\ is also a resonant line, whose line profile could contain information about the neutral gas of the galaxy. Therefore, \mgii\ can be used to trace the neutral hydrogen in/around the galaxy. Furthermore, compared to \lya, \mgii\ has three advantages: 1) Given that the IGM is mostly unpolluted by metals in the EoR \citep[][]{Rafelski14}, \mgii\ is much less attenuated by the neutral IGM than \lya, 2) The doublet line ratio of \mgii\ \ly 2796/\mgii\ \ly 2803 is related to the \mgii\ optical depth \citep{Chisholm20} and is insensitive to the dust extinction given the close wavelengths of the two lines. 3) At high-$z$, the \mgii\ \ly\ly 2796, 2803 doublet is redshifted to wavelengths observable by {\it James Webb Space Telescope} (JWST). These advantages, combined with \mgii's moderate brightness \citep[$\sim$ 10-- 60\% of \hb,][]{Guseva13}, could allow us to detect \mgii\ emission in the distant universe.
%\mgii\ is much less attenuated by the neutral IGM than \lya

The structure of the paper is as follows. In Section \ref{sec:obs}, we introduce the sample selection of \mgii, as well as the observations and data reductions. We also show various basic measurements from the spectra in Section \ref{sec:obs}. In Section \ref{sec:analysis}, we discuss the two main geometries that allow \mgii, \lya, and LyC photons to escape, and then present how to measure their escape fractions, separately. In Section \ref{sec:results}, we present possible correlations between \mgii\ and \lya\ and LyC, and we also discuss how we can predict \fescLyC\ from the observed \mgii\ emission lines. We then discuss the high detection rates of LyC in our sample and [\sii] deficiency in Section \ref{sec:discuss}. In Section \ref{sec:conclude}, we conclude the paper and discuss possible future work. 

%We adopt a cosmology with H$_{0}$ = 69.6 km s$^{-1}$ Mpc$^{-1}$, $\Omega_m$ = 0.286, and $\Omega_{\Lambda}$ = 0.714, and we use Ned Wright's Javascript Cosmology Calculator website \citep{Wright06}. 

%searching low-z analogs from indirect indicators, LzLCS, Izotov, Wang et al.

%The possible indirect indicators they find, O32, LyA define etc, advantage and challenges

%MgII advantage and previous work.
%MgII absorbed does not affect the escape fraction what calculated

%detection rate of LCEs from strong MgII emitters are higher than other samples.

%define HST/COS

%define O32

%1. For comparing the mean flux and the RMS flux: When the difference between the blue and red is small in all objects, we're saying that they all have about the same uncertainty due to background subtraction.  Realize-- the threshold you're talking about is pretty arbitrary.  It is saying that the background variations are around 1/3 of the measured flux, but one could always set a different threshold. 

\begin{table}
	\centering
	\caption{Derived Properties for Galaxies in Our Sample}
	\label{tab:params}
	\begin{tabular}{lcccccc} % four columns, alignment for each
		\hline
		\hline
		ID  	    & CFWHM$^{a}$  &  r$_{50} ^{b}$ & SFR$^{c}$ & $\Sigma_{SFR}^{d}$ & M$_\text{UV}^{e}$  \\
		\hline
		& (\arcsec)& (kpc) & (M$\odot$/yr) & ($^{d}$) & (mag)\\
		\hline
        J0105+2349   & 1.08 & 0.85 & 16.6 & 1.8 & -19.94\\
        J0152--0431  & 0.18 & 0.66 & 13.6 & 2.5 & -20.27\\
        J0208--0401  & 0.54 & 0.79 & 13.2 & 1.7 & -20.32\\
        J1103+4834   & 0.72 & 0.70 & 24.3 & 4.0 & -20.64\\
        J1105+5947   & 0.36 & 0.68 & 6.9  & 1.2 & -20.24\\
        J1219+4814   & 0.90 & 1.25 & 18.0 & 0.9 & -20.37\\
        J1246+4449   & 1.44 & 0.82 & 25.2 & 3.0 & -20.85\\
        J1425+5249   & 0.72 & 1.19 & 9.2  & 0.5 & -19.83\\

		\hline
		\hline
	\multicolumn{6}{l}{%
  	\begin{minipage}{8cm}%
	Note. --\\
    	(a)\ \ FWHM measured in the cross-dispersion direction around the galaxy's \lya\ emission lines.\\
    	(b)\ \ Half-light radius of the galaxy measured from the HST/COS NUV acquisition images (see Figure \ref{fig:acq}).\\
    	(c)\ \ Star formation rate of the galaxy derived from \hb\ emission lines \citep{Kennicutt12}.\\
    	(d)\ \ Star formation rate surface density in units of M$\odot$/yr/kpc$^2$.\\
    	(e)\ \ Absolute AB magnitude measured around 1500\angstrom\ rest-frame and are corrected for Milky Way extinction.\\
    	%, which is calculated by SFR/($\pi$r$_{50}^{2}$)
  	\end{minipage}%
	}\\
	\end{tabular}
	\\ [0mm]
	
\end{table}
%Kennicutt+12 linear SFR =       16.6425      13.6356      13.1748      24.2566      6.86981      18.0104      25.1536      9.17924
%R50_kpc =      0.852372     0.659870     0.792943     0.695001     0.682765      1.25506     0.824547      1.19425
%Y50_kpc =       1.46121     0.263948     0.792943      1.11200     0.546212      1.39451      1.88468      1.06155
%Kennicutt+12 Sigma =       1.82290      2.49207      1.66749      3.99634      1.17275     0.909914      2.94423     0.512177
%mag_ABS =      -19.9370     -20.2690     -20.3247     -20.6394     -20.2402     -20.3737     -20.8455     -19.8311

\begin{figure*}
\center
	\includegraphics[angle=0,trim={0.7cm 0.7cm 0.7cm 0.7cm},clip=true,width=1\linewidth,keepaspectratio]{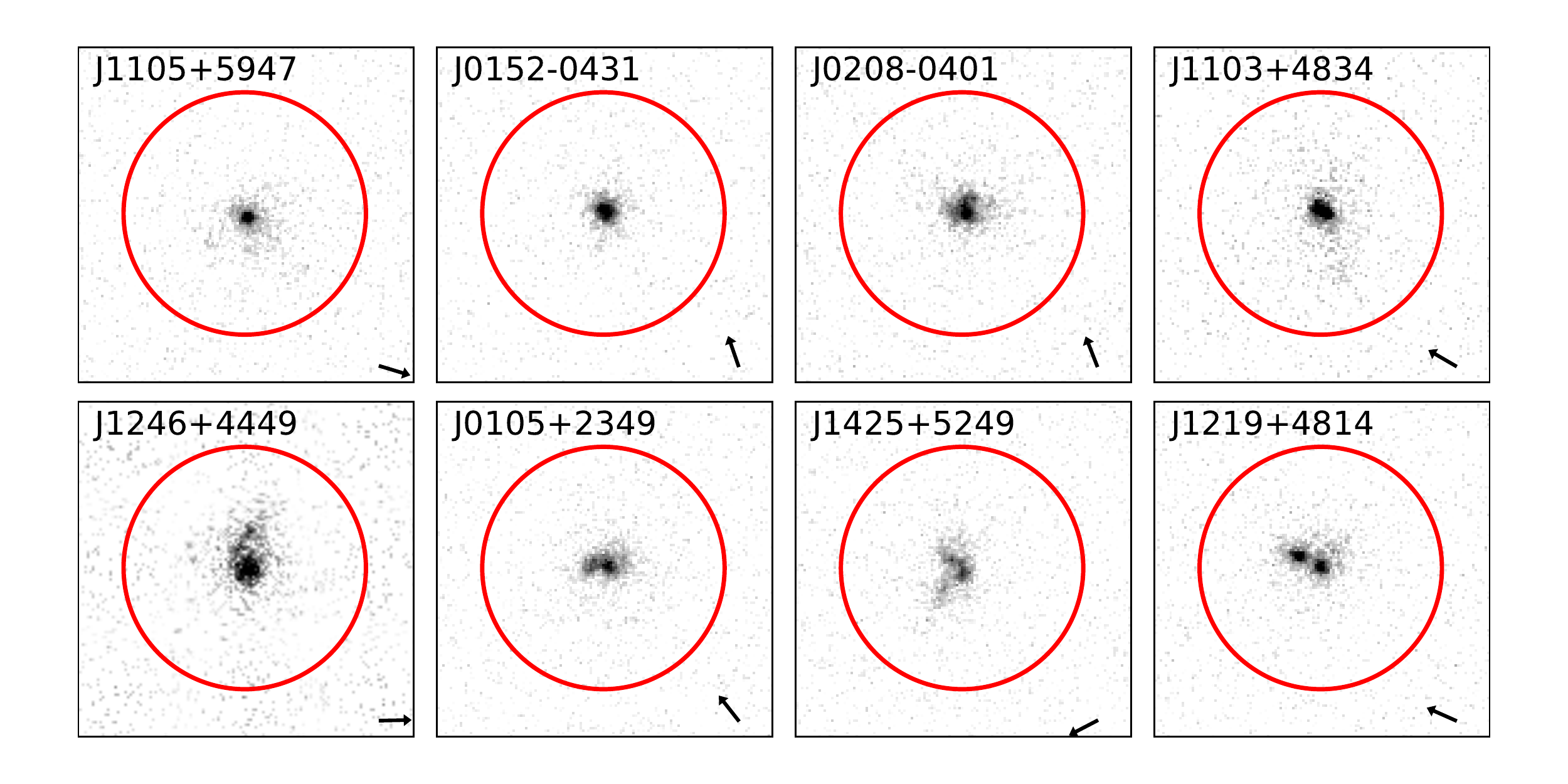}% trim: left lower right upper

\caption{\normalfont{The HST/COS NUV acquisition images for galaxies in our sample. For each panel, we overlay the COS aperture with a diameter of 2\farcs5 as the red circle, and denote the spectral dispersion axis by the black arrow. Objects are ordered by the measured escape fraction of Lyman continuum (\fescLyC, see Section \ref{sec:SED}), while the last two objects have \fescLyC\ as upper limits.} }
\label{fig:acq}
\end{figure*}
%The full scale of the figure is 3.5\arcsec\ $\times$ 3.5\arcsec.
%North is also aligned to the top.

%Consider merging this to the bottom section?
% and can probe new parameter space for possible strong LyC leakers (see Figure \ref{fig:EWMgII-O32} and discussion below)

%(hereafter, EW(\mgii) stands for the summed EW of the doublet)

%GP? Need to show the NUV image to check.

%Add a figure for O32-MgII-EW plots and show the differences between ours and Izotovs (+LzLCS MgII objects and Henry+18 objects in caption)

%Add a figure for O32-R23 to show the differences

%Talk about how we correct the data for MW and internal extinction (how this is calculated from Balmer lines, add a little bit about unusual Balmer decrements? So we exlucde Ha in the calculations if Ha/Hb is inconsistent with Case B).

%3.3: State how we calculate optical lines parameters and add a table for all measured line flux (correct for extinction), and (another) table for Te and ne of the galaxies.

\begin{figure*}
\center
	\includegraphics[angle=0,trim={0.5cm 0.2cm 0.2cm 0.0cm},clip=true,width=1\linewidth,keepaspectratio]{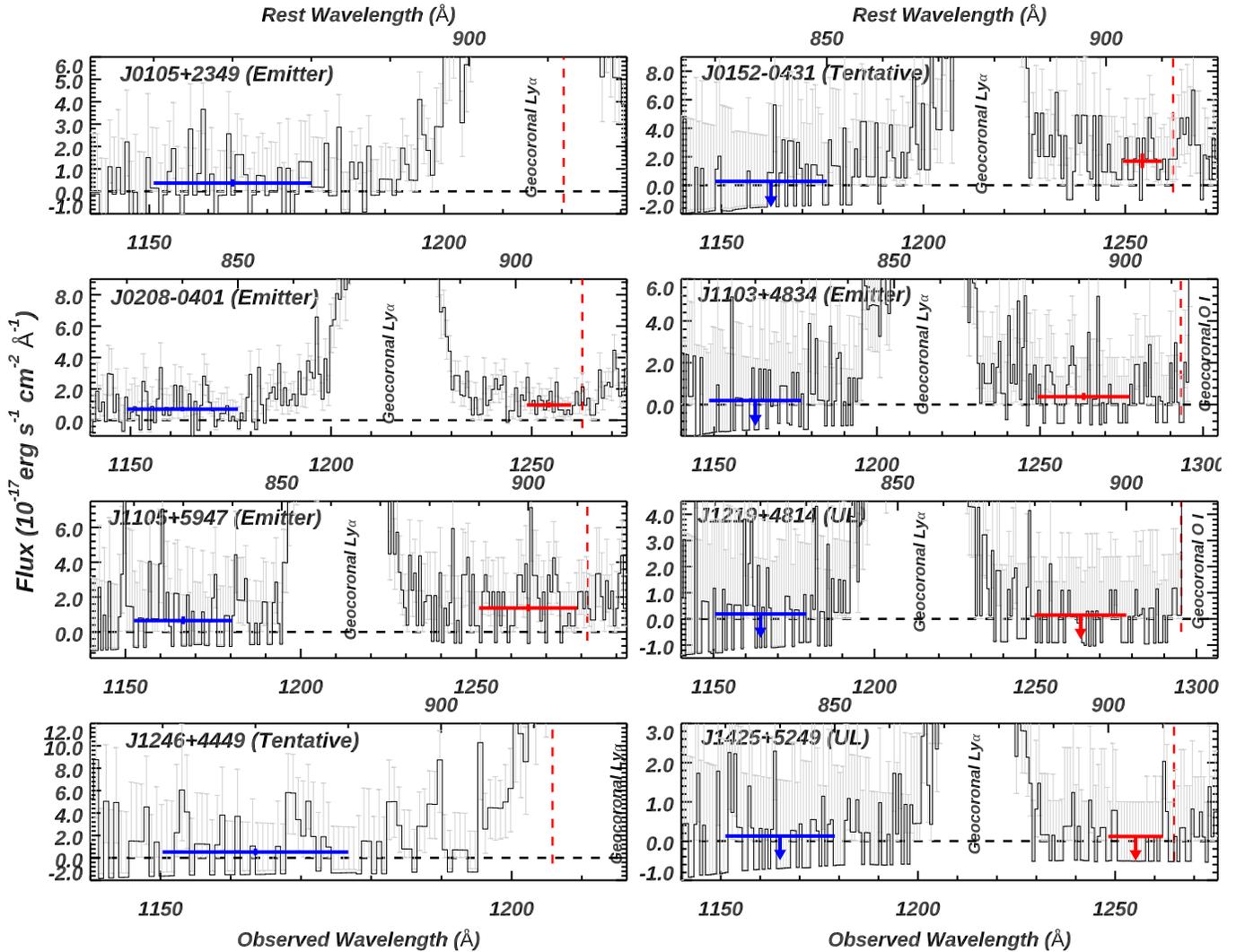}% trim: left lower right upper

\caption{\normalfont{The reduced spectra for the LyC regions (corrected for Milky Way extinction). The HST/COS spectra are in black, while their errors are shown in gray. The location of the Lyman limit is indicated by the vertical red dashed lines. Strong geocoronal lines (\lya\ and \oi\ \ly 1302) are labeled.  To minimize the contamination from geocoronal \lya, we measure the flux of LyC from the blue and red windows that are away from the geocoronal \lya. For each galaxy, the mean LyC flux is calculated by averaging the flux within a window that is denoted as the horizontal blue or red lines. The corresponding uncertainties are indicated by the vertical solid line, or as arrows if no detections. Based on the measured LyC flux, we classify and label galaxies into three categories, i.e., LyC emitters, tentative LyC emitters, and upper limits (UL) on LyC. See details in Section \ref{sec:meaLyC}.} }  
\label{fig:Spec1}
\end{figure*}
%Objects in this Figure and Figure \ref{fig:Spec2} are sorted by decreasing \fescLyC\ (see Section ?). \textbf{Top:} The LyC spectral regions, where the measured mean LyC flux are calculated by averaging the flux within a window that is equal to the size of horizontal red lines. The corresponding errors are indicated by the vertical red solid lines. The location of Lyman limit is indicated by the red dashed lines. Strong geocoronal lines (\lya\ and \oi\ \ly 1302) are labeled and our fitted models for them are shown as the orange lines (see Appendix \ref{app:geo}). \textbf{Middle:} The \lya\ spectral regions, where zero velocity is indicated by blue dashed lines. \textbf{Bottom:} The \mgii\ spectral regions, where we label the \mgii\ \ly\ly 2796 and 2803 by blue and red dashed lines, respectively.

\iffalse
\begin{figure*}
\center
	\includegraphics[angle=0,trim={0.5cm 0.2cm 0.2cm 0.0cm},clip=true,width=1\linewidth,keepaspectratio]{./Figures/SEDFitsLyC.pdf}% trim: left lower right upper

\caption{\normalfont{The reduced spectra for the LyC regions. Captions and labels are the same as Figure \ref{fig:Spec1}. } }
\label{fig:Spec2}
\end{figure*}
\fi

\section{Observations, Data Reduction, and Basic Measurements}
\label{sec:obs}

\subsection{Sample Selection}
\label{sec:sample}

To study \mgii\ along with the observed \lya\ and LyC features, we need to detect three different spectral regions for each galaxy. Therefore, we have selected our sample of \mgii\ emitters from SDSS-IV/eBOSS, and obtained follow-up observations of their \lya\ and LyC spectral regions with HST/COS. The main selection criteria are: 
1) we require GALEX NUV detections, with NUV $<$ 22 AB;
2) we choose galaxies with \hb\ equivalent width $>$ 5\angstrom\ to remove passive galaxies;
3) the redshifts of the galaxies were chosen to be in the range of 0.32 $<$ z $<$ 0.45, which allows us to detect the \mgii\ \ly\ly 2796, 2803 doublet from the SDSS/eBOSS spectra, the \lya\ lines from the COS/G160M spectra, and LyC features from the COS/G140L spectra; 4) Then, we require \mgii\ emission at $\gtrsim$ 7$\sigma$ significance, in order to ensure that we are not proposing follow-up of spurious detections;
5) we require \mgii\ equivalent width (EW) $\gtrsim$ 10\angstrom\ (for the summed doublet lines). This selection returns the most extreme \mgii\ emitters, reaching relatively unexplored parameter space; and 
6) galaxies with AGN activity were excluded using the line ratios and the BPT diagrams \citep{Baldwin81}. 
These criteria lead to a parent sample of 22 objects from eBOSS. Then we select 8 objects with NUV magnitudes brighter than 21.6, which can be observed in relatively short time by HST (1 -- 4 orbits). The final selected objects are listed in table \ref{tab:obs}.

\subsection{COS Observations and Reduction}
\label{sec:reduction}
The eight galaxies in our sample were observed during HST Cycle 27 through project HST-GO-15865 (PI: Henry). This program obtained COS G140L and G160M observations that probed the LyC and \lya\ regions, respectively. One of the eight galaxies (J1246+4449) overlaps with the Low-redshift Lyman Continuum Survey (LzLCS, HST GO: 15626, PI: Jaskot, \cite{Flury22a}), so we use the archival LyC observations for this object (but we still acquired new G160M observations). The details of observations are listed in Table \ref{tab:obs}. 
%Consider showing the ACQ image to characterize them as GP galaxies.

In Figure \ref{fig:acq}, we show the HST/COS acquisition images for galaxies in our sample. We overlay the COS aperture size (2\farcs5 in diameter) as the red circles. The compact NUV light profiles of our galaxies are contained within the COS aperture. These compact features are similar to the Green Pea (GP) galaxies \citep[e.g.,][]{Henry15, Henry18, Jaskot19}. We present several basic galaxy properties in Table \ref{tab:params}.

To get robust estimates of the LyC flux from the HST/COS G140L spectra, we follow the reduction methods in previous publications \citep[][]{Wang19,Flury22a}. We use the standard CALCOS pipeline (v3.3.9), and further estimate the dark current and scattered geocoronal \lya\ background adopting the custom software FaintCOS \citep{Makan21}. For HST/COS G160M data, we directly download the reduced data from the HST/MAST archive. For each galaxy, we have also checked the 2D G160M spectra that \lya\ is completely in the COS extraction aperture. 

We then correct both G140L and G160M spectra for Milky Way extinction using the Galactic Dust Reddening and Extinction Map \citep{Schlafly11} at NASA/IPAC Infrared Science Archive assuming the extinction law from \cite{Cardelli89}. The final reduced spectral regions of LyC (from G140L data) are shown in Figure \ref{fig:Spec1}, while the \lya\ regions (from G160M data) are shown in Figure \ref{fig:SEDFitsLyA1}.

%Schlafly1 refers Schlegel98,
%binned to 0.18angstrom for LyA region, 0.72 angstrom for LyC region (since only G140L data)
% Multiple visits for the same object were added to improve signal-to-noise ratios, and the spectra were binned to about 0.6\angstrom. 

%Therefore, we do not conduct additional aperture corrections between the UV spectra from COS and the optical spectra from SDSS.

%\textbf{Talk about the size of galaxies and why there is no need to do aperture correction for SDSS and COS. But I did scaled the SDSS data of J1246 by 1.29 while comparing with the SDSS photometry.}

%\textbf{(check pulse height in Bingjie paper and or talk about the time variations)}
%Citing Worseck+2016 methods for the additional reduction methods (pulse height, scattered light, dark current, etc.). Add the figure about COS acq image and overlay the aperture size. Discuss consistency between COS spectra and GALEX photometry. Also show LyA and LyC spectra.

\subsection{Basic Measurements from COS and SDSS Spectra}

In the following subsections, we discuss how we conduct basic measurements of the HST/COS and SDSS data. 

%These measurements are listed in Table \ref{tab:SDSS}.

\begin{table*}
	\centering
	\caption{Measurements about Lyman Continuum}
	\label{tab:escape}
	\begin{tabular}{lcccccccc} % four columns, alignment for each
		\hline
		\hline
		Object 	    & \FLyC$^{1}_\text{B}$   & \FLyC$^{1}_\text{R}$       &	\textit{F}(1100)$^{2}$    &	\textit{F}(1300)$^{2}$    & \FLyC/\textit{F}(1100)$^{3}$  & \FLyC/\textit{F}(1300)$^{3}$ & \fescLyC(\hb)$^{4}$    &   \fescLyC (SED)$^{5}$  \\
		\hline
		         	&  (E--17)            &	(E--17)               &	(E--17)               & (E--17)                   & (\%)  & (\%)       & (\%)              &   (\%)         \\
		\hline
        J0105+2349      &\textbf{0.37$^{+0.17}_{-0.15}$}		&NA		                    & 	12.00$^{+0.43}_{-0.32}$ & 	9.9$^{+0.62}_{-0.33}$	    &3.1$^{+1.4}_{-1.3}$    &3.7$^{+1.7}_{-1.5}$    & 1.2$^{+0.8}_{-0.7}$   & 0.7$^{+0.3}_{-0.3}$\\
        J0152--0431     &$<$ 0.55	                &\textbf{1.69$^{+0.54}_{-0.48}$}	    & 	8.71$^{+1.91}_{-0.65}$	& 	10.37$^{+0.97}_{-0.53}$     &19.4$^{+6.4}_{-7.0}$   &16.3$^{+5.4}_{-4.7}$   & 8.6$^{+4.7}_{-4.5}$   & 9.8$^{+3.1}_{-2.7}$\\
        J0208--0401     &0.71$^{+0.15}_{-0.13}$		&\textbf{0.97$^{+0.18}_{-0.17}$}        & 	11.71$^{+0.36}_{-0.28}$	& 	10.03$^{+0.53}_{-0.31}$     &8.3$^{+1.5}_{-1.5}$    &9.7$^{+1.8}_{-1.7}$    & 5.8$^{+2.3}_{-2.2}$   & 5.3$^{+1.0}_{-0.9}$\\        
        J1103+4834      &$<$ 0.41           		&\textbf{0.38$^{+0.18}_{-0.16}$}        & 	15.43$^{+1.1}_{-0.62}$	& 	11.64$^{+2.36}_{-0.76}$     &2.5$^{+1.2}_{-1.1}$    &3.3$^{+1.7}_{-1.4}$    & 1.3$^{+0.8}_{-0.8}$   & 2.5$^{+1.2}_{-1.0}$\\   
        J1105+5947      &0.65$^{+0.24}_{-0.20}$		&\textbf{1.37$^{+0.21}_{-0.19}$}        & 	10.07$^{+0.78}_{-0.45}$	& 	7.28$^{+1.68}_{-0.51}$      &13.6$^{+2.2}_{-2.2}$   &18.8$^{+5.2}_{-2.9}$   & 13.0$^{+6.1}_{-6.0}$  & 13.6$^{+2.1}_{-1.9}$\\ 
        J1219+4814      &$<$ 0.41           		&\textbf{$<$ 0.30}                              & 	10.62$^{+0.93}_{-0.51}$	& 	8.92$^{+2.74}_{-0.76}$      &$<$ 2.8                &$<$ 3.4        & $<$1.4                & $<$1.3\\ 
        J1246+4449$^{(*)}$      &\textbf{0.52$^{+0.30}_{-0.29}$}		&NA                         & 	32.86$^{+1.14}_{-0.84}$	& 	25.42$^{+1.5}_{-0.8}$       &1.6$^{+0.9}_{-0.9}$    &2.0$^{+1.2}_{-1.1}$    & 1.1$^{+0.8}_{-0.8}$   & 0.8$^{+0.5}_{-0.5}$\\ 
        J1425+5249      &$<$ 0.28	                &\textbf{$<$ 0.28}       & 	7.66$^{+0.57}_{-0.34}$  & 	6.21$^{+1.05}_{-0.39}$	    &$<$3.7                &$<$4.5                                  & $<$2.4                     & $<$2.3\\         
		\hline
		\hline

	\multicolumn{9}{l}{%
  	\begin{minipage}{17.5cm}%
	Note. --\\
    	(1) \ \ The measured LyC flux density in units of 10$^{-17}$ ergs s$^{-1}$ cm$^{-2}$ \AA$^{-1}$. \FLyC$_\text{B}$ and  \FLyC$_\text{R}$ represent LyC flux measured at the blue and red side of geocoronal \lya, respectively (see Figure \ref{fig:Spec1} and Section \ref{sec:meaLyC}). The final adopted \FLyC\ are shown in boldface font. For J0105+2349 and J1246+4449, their Lyman limit is $<$ 1250\angstrom\ (in observed frame). Therefore, they do not have \FLyC$_\text{R}$ reported. For \FLyC\ that are consistent with zero, a 2$\sigma$ upper limit is reported.\\
    	(2)\ \ The continuum flux measured around 1100\angstrom\ and 1300\angstrom, which are corrected by Milky Way dust extinction, but not by the internal extinction of the galaxy.\\
    	(3)\ \ The flux ratio of LyC to the continuum at around 1100\angstrom\ and 1300\angstrom.\\    	
    	%(4)\ \ \textbf{The stellar extinction derived from the SED fittings (see Section \ref{sec:SED}).}\\
    	(4)\ \ The absolute escape fraction of LyC from \hb\ method (see Section \ref{sec:Hb}).\\
    	(5)\ \ The absolute escape fraction of LyC from SED fitting method (see Section \ref{sec:SED}).\\
    	$^{(*)}$\ \ J1246+4449 is also analyzed in LzLCS \citep{Flury22a}. They reported \fescLyC(\hb) = 0.9$^{+0.5}_{-0.4}$ and \fescLyC(SED) = 0.5$^{+1.0}_{-0.2}$, which are consistent with our values within errorbars.
    	% (from --0.33 to --0.90)
 \\
    	
  	\end{minipage}%
	}\\
	\end{tabular}
	\\ [0mm]
	
\end{table*}

\subsubsection{The Observed LyC Flux}
\label{sec:meaLyC}
In the literature, flux in the Lyman continuum (\FLyC) is usually measured as the average flux over a window ($\sim$ 20\angstrom\ in the rest-frame) close to and shortward of the Lyman limit \citep[e.g.,][]{Izotov21,Flury22a,Flury22b}. Due to the redshift range of our galaxies, a few of the LyC regions are close to geocoronal \lya\ emission lines. To minimize the contamination from geocoronal \lya, we measure \FLyC\ from two windows that are both away from the geocoronal lines. The blue window is set to be 20\angstrom\ wide in the rest-frame, and $<$ 1180\angstrom\ in the observed frame. The red window is between 1250\angstrom\ and Lyman limit [912\angstrom~$\times$~(1+z)] in the observed frame. These two sets of windows are shown as the blue and red horizontal lines in Figure \ref{fig:Spec1}. Then we measure the average flux within the blue and red window and report them as \FLyC$_\text{B}$ and \FLyC$_\text{R}$, respectively, in Table \ref{tab:escape}. The 1$\sigma$ error bars are computed by sampling from the Poisson distributions of the background and science spectra following the methodology in \cite{Feldman98}. For galaxies with the Lyman limit close to 1250\angstrom, the widths of the red windows are narrower (in J0152--0431, J0208--0401, and J1425+5249). For galaxies that have Lyman limits $<$ 1250\angstrom, no \FLyC$_\text{R}$ is reported (J0105+2349 and J1246+4449) due to the heavy contamination from geocoronal \lya.

%The exception is J0152--0431, where its measured \FLyC$_\text{R}$ is higher than \FLyC$_\text{B}$ by 2$\sigma$. This could be because 1) the lower SNR in the blue window (G140L's throughput drops by 40\% from 1250\angstrom\ to 1150\angstrom), 2) the Lyman 'bump' that have been reported in \citep{Inoue10}, where they explain the peak of LyC flux at 900\angstrom\ from the escaped nebular LyC.

%They suggest that if the escaped nebular LyC could the bound–free nebular emission has strong peaks
In Table \ref{tab:escape}, for the 6 galaxies that have \FLyC\ measured in both windows, 5 out of 6 galaxies have \FLyC$_\text{B}$ and \FLyC$_\text{R}$ are all consistent within 2$\sigma$ (except J0152--0431). Given this general consistency of measured LyC flux from two windows, we adopt the \FLyC$_\text{R}$ when there exists a red window. This is also the common choice in literature, i.e., measuring LyC flux as close as possible to the Lyman limit. For the two galaxies (J0105+2349 and J1246+4449) that do not have \FLyC$_\text{R}$, we adopt their measured \FLyC$_\text{B}$. These adopted \FLyC\ values are shown as boldface font in Table \ref{tab:escape}.

%For J0152--0431, we increase the lower error bar of \FLyC$_\text{R}$ such that it is consistent with the measured \FLyC$_\text{B}$ within 1$\sigma$. 
%in the subsequent analyses
%This leads to \FLyC$_\text{R}$ = 1.71$^{+0.82}_{-0.90}$ $\times$ 10$^{-17}$ ergs s$^{-1}$ cm$^{-2}$ \AA$^{-1}$.

%But we note its red window is only 10\angstrom\ wide (see Figure \ref{fig:Spec1}).

%To minimize the fluctuations when calculating the LyC flux over a narrow window, we choose to adopt \FLyC\ from the window that has $\sim$ 30\angstrom\ width and is also closest to Lyman limit. Therefore, for J1103+4834, J1105+5947, and J1219+4814, we adopt their measured \FLyC$_\text{R}$, while for the other galaxies, we adopt their measured \FLyC$_\text{B}$ in the subsequent analyses.

%This not only reduces the possible fluctuations of flux measured in a narrow window. 

We take the same criteria to define a LyC detection as in LzLCS \citep{Flury22a}, i.e., galaxies with LyC flux detected $>$ 2$\sigma$ level. In this case, 4 out of our 8 galaxies are classified as detections. For J0152--0431, its \FLyC$_\text{R}$ and \FLyC$_\text{B}$ are only consistent within 3$\sigma$, and \FLyC$_\text{B}$ is consistent with 0 (which could be due to the absorption of Lyman series from an IGM cloud at z $\sim$ 0.26). Therefore, we treat it along with J1246+4449 (who has $\sim$1.8$\sigma$ on \FLyC) as the two tentative cases. For the other two objects (J1219+4814 and J1425+5249), \FLyC\ is consistent with zero, so we report their 2$\sigma$ upper limits in Table \ref{tab:escape}. Overall, for our sample, the LyC detection rate ($>$ 2$\sigma$) is at least 4/8 = 50\%.

%See IGM for J0152 at: 

%measuring LyC flux at shorter wavelengths could be okay, but LyC flux may be wavelength dependent, see McCandliss17.
%LyC flux, window, and 0208 caution

%We also choose the widths and positions of the windows to minimize the contamination from geo-coronal \lya\ and \oi\ emission (see Section \ref{sec:reduction} and Appendix \ref{app:geo}).

%For observations taken by COS/G140L grating, these lines primarily stay the same shape and only vary in amplitudes. Therefore, to remove the possible contamination, we took the geo-coronal line models from HST/COS website\footnote{https://www.stsci.edu/hst/instrumentation/cos/calibration/\ airglow}. We then vary amplitude and wavelength center of the model to fit the the geocoronal lines in each of our galaxy. These fitted geocoronal line models are shown as the orange lines in the top panel in Figures \ref{fig:Spec1} and \ref{fig:Spec2}. Finally, we subtract the fitted geocoronal line from the spectra when calculating the LyC flux. See Appendix \ref{app:geo} for more details.

\begin{figure*}
\center
	\includegraphics[angle=0,trim={0.2cm 0.1cm 0.0cm 8.5cm},clip=true,width=1\linewidth,keepaspectratio]{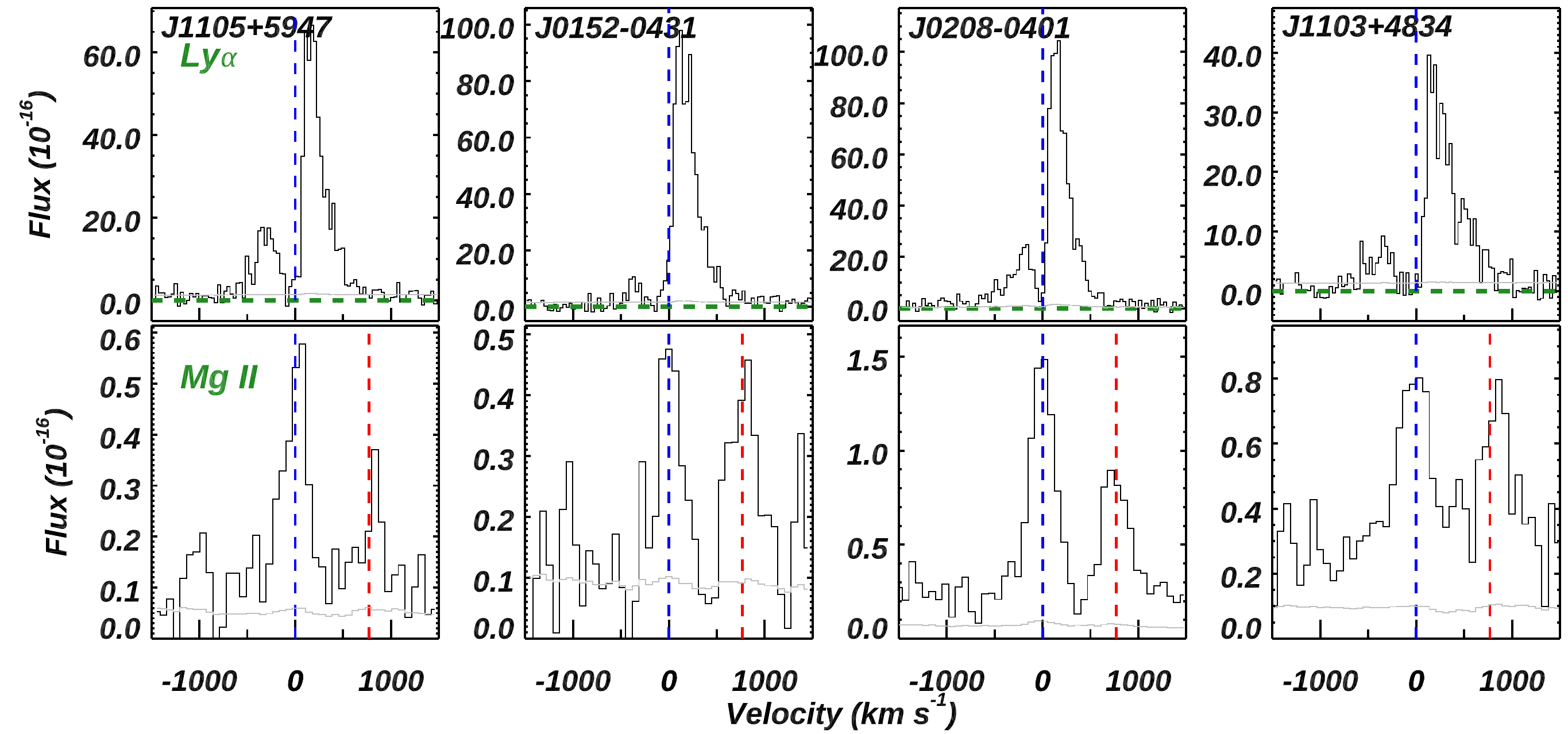}% trim: left lower right upper
	
	\includegraphics[angle=0,trim={0.0cm 0.1cm 0.0cm 8.5cm},clip=true,width=1\linewidth,keepaspectratio]{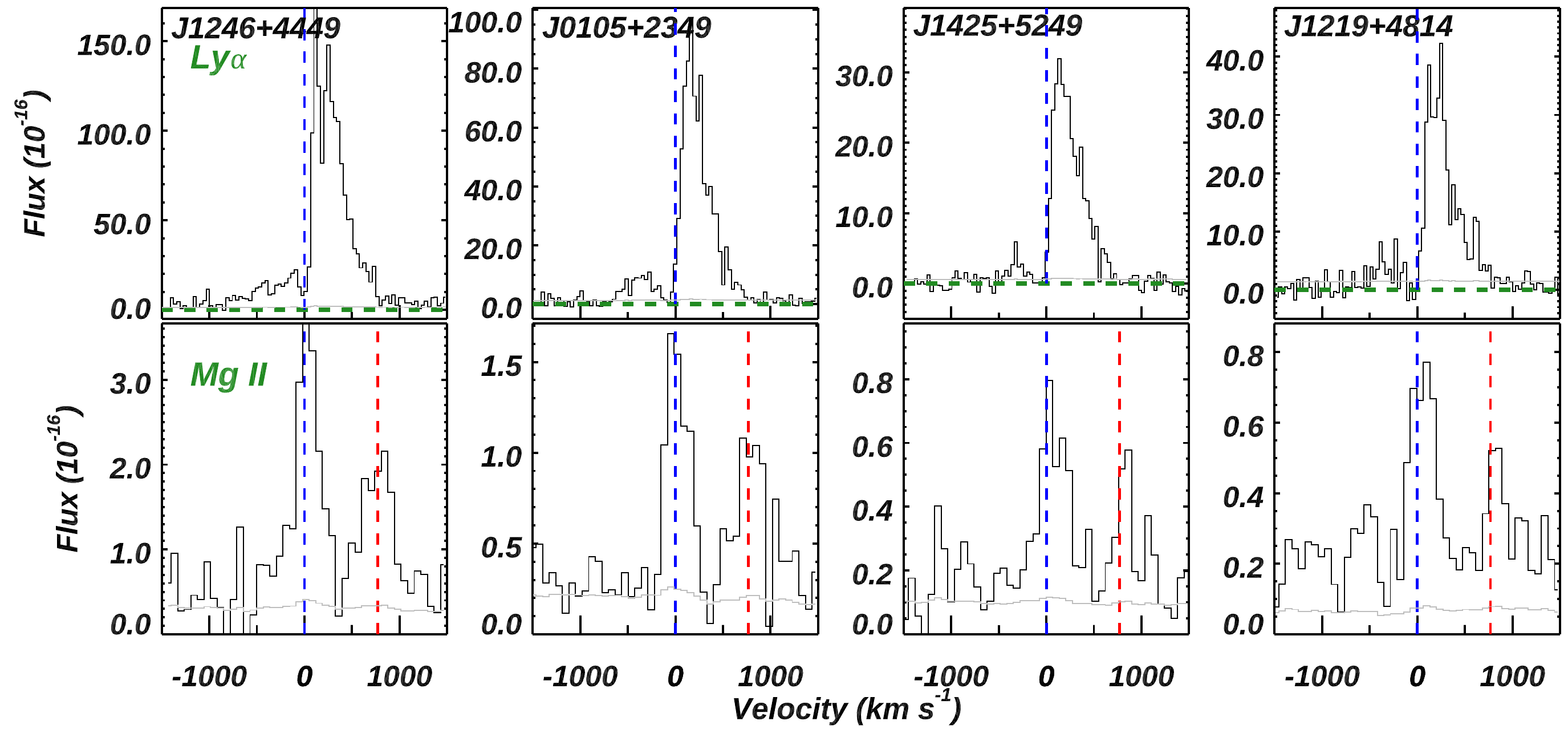}% trim: left lower right upper

\caption{\normalfont{Comparisons of the observed \lya\ and \mgii\ profiles in velocity space, with data taken from HST/COS and SDSS, respectively. The y-axes are in units of 10$^{-16}$ ergs s$^{-1}$ cm$^{-2}$ \AA$^{-1}$. The data and corresponding errors are shown in black and gray. Objects are ordered by measured \fescLyC\ in Section \ref{sec:SED}, while the last two objects have \fescLyC\ as upper limits. The blue lines represent the $v$ = 0 km s$^{-1}$ for \lya\ or \mgii\ \ly 2796, while the red lines represent $v$ = 0 km s$^{-1}$ for \mgii\ \ly 2803. The green dashed lines represent the positions of flux = 0 for \lya\ panels.  } }
\label{fig:SEDFitsLyA1}
\end{figure*}

\subsection{Measurements of the \lya\ features}
\label{sec:meaLyA}
We measure the continuum flux\footnote{In HST/COS G160M spectra for our galaxies, the continuum flux for \lya\ regions usually approaches 0.} for the \lya\ spectral region by adopting a linear fit to the continuum $\sim$ $\pm$ 2000 km~s$^{-1}$ around the systemic velocity. Then we measure the flux, equivalent widths, and peak velocities from the \lya\ profiles. For each measured parameter, the corresponding errors are estimated through a Monte Carlo (MC) simulation where we perturb the spectrum 10$^{4}$ times according to the observed 1$\sigma$ uncertainties. These measurements are reported in Table \ref{tab:LyA}. The \lya\ profiles for our galaxies are shown in Figure \ref{fig:SEDFitsLyA1}.

We have also tested possible contaminations from stellar features (e.g., stellar \lya\ absorption) by subtracting the \lya\ spectral regions by our best-fitted SED models discussed in Section \ref{sec:SED} and remeasure the line parameters for \lya. We find that EW(\lya) only has small changes, while flux and peak velocities stay almost unchanged. This is because our \lya\ emission lines are strong compared to the stellar absorption. Therefore, we conclude the contaminations from stellar absorption are minimal, and we do not correct \lya\ by stellar features. This is also to be consistent with previous studies that we will compare in Section \ref{sec:results}.

For all objects in our sample, we detect double-peaked features in \lya. The velocity separation of the peaks has been found to correlate with LyC photon escape \citep[e.g.,][and see Section \ref{sec:CorrLyC}]{Verhamme15, Gronke16, Orlitova18, Jaskot19, Gazagnes20}. We measure the velocities of both peaks compared to the systemic velocity and also report them in Table \ref{tab:LyA}. Note that some of the galaxies' \lya\ profiles may exhibit more than two peaks, and we return to this feature in Section \ref{sec:CorrLyC}.

%We measure the continuum flux\footnote{In HST/COS G160M spectra for our galaxies, the continuum flux for \lya\ regions usually approaches 0.} for the \lya\ spectral region by adopting a linear fit to the continuum $\sim$ $\pm$ 2000 km~s$^{-1}$ around the systemic velocity.

\subsection{Measurements of the SDSS Spectra}
\label{sec:MeaSDSS}
We retrieved the SDSS spectra for each of the galaxies in our sample and conducted a variety of measurements from the optical emission lines as follows. The \mgii\ regions for these galaxies are shown in Figure \ref{fig:SEDFitsLyA1}.
%Discuss slit loss somewhere
%explain that we do not need the slit loss corrections for all objects except for J1246 (29\% difference).

1). We derive the internal dust extinction of the observed galaxies from the Balmer series.  We follow the methodology discussed in \cite{Henry21} to fit the higher S/N Balmer line flux ratios (\ha/\hb, \hg/\hb, \hd/\hb) with two parameters, i.e., the internal $E(B-V)$ and stellar absorption. The latter is a nuisance parameter, so its inclusion contributes to an accurate determination of the uncertainties on $E(B-V)$.  We adopt the extinction law from \cite{Cardelli89}, and assume an electron temperature (\Te) of 10,000 K.\footnote{The derived $E(B-V)$ is only slightly dependent on $\Te$. For example, adopting $\Te$ = 15,000 K will increase the derived $E(B-V)$ by only $\sim$ 0.02.} Two out of our eight galaxies (J0152--0431 and J0208--0401) have unphysical Balmer decrements in \ha/\hb, which cannot be fitted simultaneously with the \hg/\hb\ and \hd/\hb\ ratios. Their \ha\ features from the SDSS spectra also look asymmetric or clipped, which has previously been found in the literature \citep[e.g., J1248+4259 in][]{Izotov18b}. We attribute this to systematic errors in the SDSS data processing pipelines. Therefore, for these two galaxies, we exclude \ha\ in the above fitting process. The best-fitted internal $E(B-V)$ values are shown in Table \ref{tab:SDSS}.

%For the former, we adopt extinction law from \cite{Cardelli89}. For the latter, \cite{Henry21} has shown that in various Starburst99 \citep{Leitherer99} models, young stellar populations exhibit \hb\ stellar absorption equivalent widths (EW$_{ste}^{H\beta}$) in a range of $\sim$ 0 -- 6\angstrom, while other Balmer lines track \hb. Therefore, we similarly adopt EW$_{ste}^{H\gamma}$ and EW$_{ste}^{H\delta}$ that are equal to EW$_{ste}^{H\beta}$, and EW$_{ste}^{H\alpha}$ = 2/3 $\times$ EW$_{ste}^{H\beta}$. The observed flux ratios are corrected by both internal $E(B-V)$ and stellar absorption while fitting to the theoretical values. 
%Overall, this two-parameter fitting provides better estimates of the dust extinction, marginalizing over the EW$_{ste}^{H\beta}$ as an unconstrained nuisance parameter.

%The relative intensities of the Balmer lines are nearly independent of both density and temperature, so they can be adopted to solve for the internal dust extinction.
%(FYI, for Te=5,000 K and Te=20,000 K (high and low metallicity, respectively), the Ha/Hb ratio = 3.04 and  2.75, respectively.) for Te = 10000, ratio = 2.86.
% In my 2021 paper, we found that the stellar absorption was never really constrained, but rather it was a nuisance parameter

%This is mainly because they have unusual Balmer decrements in \ha, i.e., \ha/\hb\ flux ratio is much less than 2.86, which cannot be explained by dust extinctions \textbf{(mention Ha seems clipped in our case, and mention the Izotov object clipped)}. 

2). Next, when measuring optical emission lines, we correct the SDSS spectra for both the Milky Way and internal extinction, as well as the stellar absorption. After that, for lines of interest, we derive the continuum, measure the lines (i.e., flux, EW, and peak velocities), and calculate their errors adopting the same method discussed in Section \ref{sec:meaLyA}. For the \mgii\ doublet, we divide the spectrum into two regions corresponding to each line, using the midpoint between the two lines, i.e., 2799.1\angstrom. Additionally, the Mg II line fluxes are not corrected for internal dust extinction, since a robust correction is difficult to discern when \mgii\ photons could be resonantly scattered like \lya\ \citep[e.g.,][]{Henry18, Chisholm20}. We return to this question in Section \ref{sec:dust}.

3). For each galaxy, we compare the magnitudes from SDSS photometry in different bands (u,g,r,i,z) to the ones derived from SDSS spectra. Since our galaxies are compact (see Figure \ref{fig:acq}), these two sets of magnitudes are consistent for most of our galaxies. One exception is J1246+4449, where we scale the flux from its SDSS spectra by a factor of 1.29 to match its SDSS r-band magnitude. For other galaxies, we do not apply aperture corrections to their SDSS spectra.

%within the errors ($<$ 15\%) 
%Since our analyses later utilize the information from both HST/COS (LyC and \lya) and SDSS (\mgii) spectra. We check 
%One exception is J1246+4449, where we scale the flux from its SDSS spectra by a factor of 1.29 to match its SDSS r-band photometry.

4). We then derive electron density and temperature (\ne\ and \Te), and metallicity from optical emission lines using \textit{PyNeb} \citep{Luridiana15}. For \ne, since [\oii] \ly\ly 3727, 3729 are not resolved in SDSS spectra and [\sii] \ly\ly 6716, 6731 doublet usually has low signal-to-noise ratio (SNR), we assume \ne\ = 100 cm$^{-3}$ as a characteristic of GP galaxies that show strong \mgii\ emission \citep{Henry18}. For \Te, we use [\oiii] \ly 4363 along with [\oiii] \ly\ly 4959, 5007 lines to determine T[\oiii], while T[\oii] is scaled from T[\oiii] as discussed in \cite{Andrews13}. Given \ne\ and \Te\ for each galaxy, we adopt the direct-method using \textit{PyNeb} to calculate O$^{++}$/H$^{+}$ ionic abundance from intensity ratio of [\oiii] (4959+5007)/\hb, and O$^{+}$/H$^{+}$ ionic abundance from [\oii] 3727/\hb. We then add them to obtain the total oxygen abundance, which are listed in Table \ref{tab:SDSS}. 

%We use \textit{PyNeb} \citep{Luridiana15} with the optical emission lines to determine the electron density and temperature (\ne\ and \Te). 
%Typical uncertainties on these values are $\sim$ 0.1 dex, accounting for inhomogeneous temperature and possible depletion onto dust grains.

%potential depletion onto dust grains (larger at higher metallicities) and biases due to temperature inhomogeneities (larger at lower metallicities). 

%Add a table 1 for all main observation information (coordinates, z, O32, SDSS-mag-obs, GALEX-mag-obs, SDSS-mag-mea, GALEX-mag-mea, and errors).

%e.g., Izotov18a, page 4: The data were reduced with the CALCOS pipeline vXXX adapted to handle spectra obtained at COS Lifetime Position X, and custom software to improve the background subtraction and co-addition for faint targets (Worseck et al. 2016).

%Note: GALEX photometry aperture diameter = 10'', and SDSS photometry diameter = 3''. For spectra, SDSS is 3'' fiber while BOSS is 2'' fiber. Need to be careful here.

%GALEX photometry aperture: http://www.galex.caltech.edu/researcher/faq.html
%SDSS photometry aperture https://www.sdss.org/dr12/algorithms/spectrophotometry/

\begin{figure*}
\center
	\includegraphics[angle=0,trim={3cm 0.3cm 0.3cm 0.3cm},clip=true,width=1\linewidth,keepaspectratio]{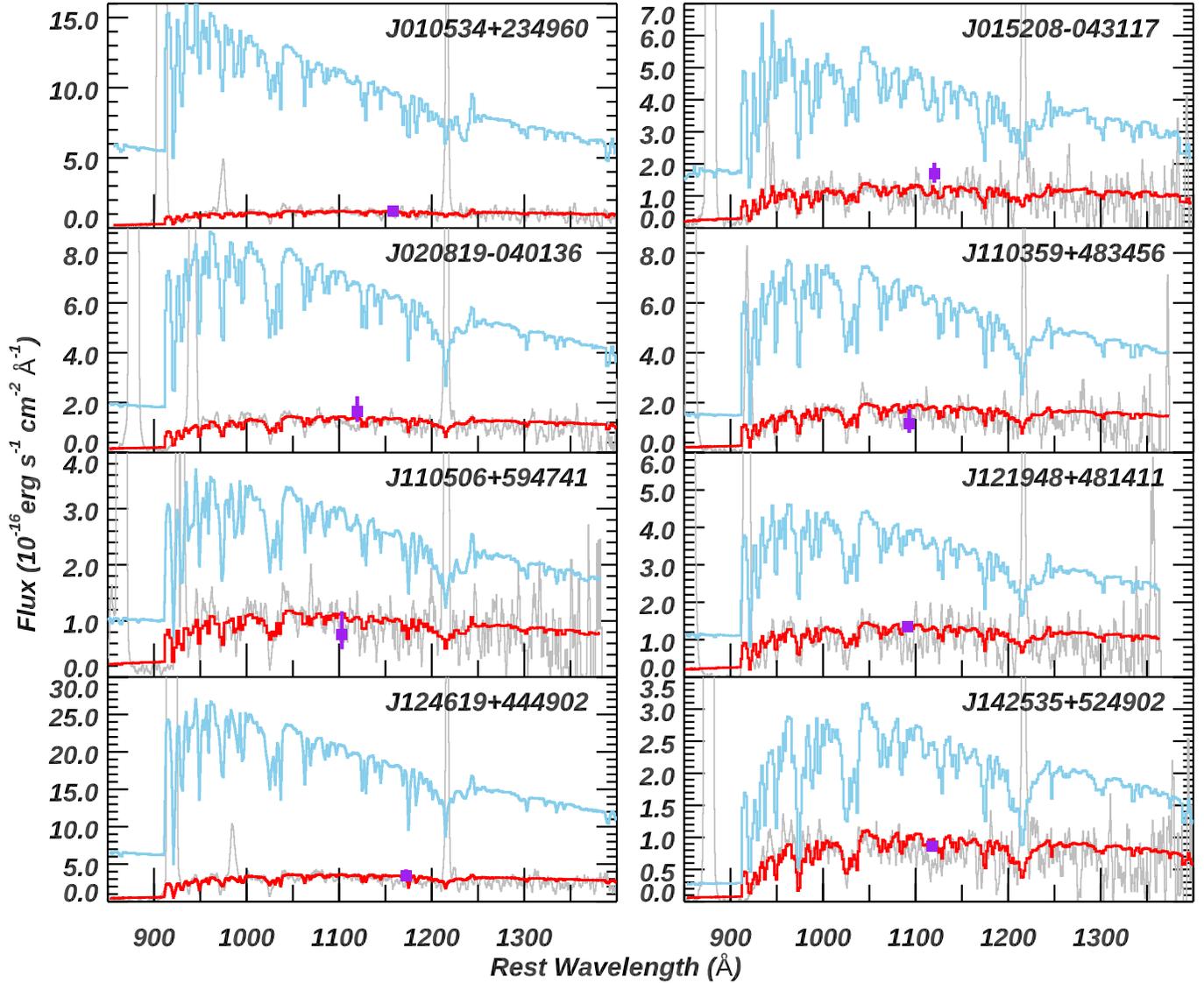}% trim: left lower right upper

\caption{\normalfont{A comparison of the observed data and our best fitting BPASS models. The COS G140L spectra are shown in grey lines, and photometric data from GALEX FUV are shown in purple squares. All data are dereddened by Milky Way extinction (see Section \ref{sec:reduction}). The red curves represent the BPASS models and are dereddened by only Milky Way extinction. The blue curves are the models dereddened for both Milky Way and internal extinction, which represents the intrinsic flux (see Section \ref{sec:SED}). We have resampled the data to bins of $\sim$ 2.5\angstrom\ to match the model's resolution. } }
\label{fig:SEDFitsAll}
\end{figure*}
%The vertical and horizontal dashed lines represent the Lyman limit at 912\angstrom\ and the zero flux's positions.

\section{Analysis}
\label{sec:analysis}
In this section, we focus on measuring three different absolute escape fractions (\fesc) for galaxies in our sample. We first estimate the LyC escape fractions (\fescLyC) from two different methods in Section \ref{Mea:LyC}. Then we determine the \mgii\ escape fractions (\fescMgII) from CLOUDY models in Section \ref{Sec:MgII}. Finally, we derive the \lya\ escape fractions (\fescLyA) in Section \ref{Mea:LyA}. We study possible correlations between these escape fractions in Section \ref{sec:results}.

% and for other comparison samples (if their \fesc\ values were not reported in the literature)

\begin{table*}
	\caption{Measurements about \lya\ Emission Lines}
	\label{tab:LyA}
	\begin{tabular}{lcccccccc} % four columns, alignment for each
		\hline
		\hline
		Object 	    & \FLyA$^{(1)}_\text{B}$   & \FLyA$^{(1)}_\text{R}$       &	EW(\lya)$_\text{B}$    &	EW(\lya)$_\text{R}$    & \VpeakB$^{(2)}$  & \VpeakR$^{(2)}$  & \fescLyA$^{(3)}$     \\
		\hline
		         	&  (E17)            &	(E17)               &	(\AA)               & (\AA)                   & (km s$^{-1}$)   & (km s$^{-1}$)                &                       \\
		\hline

        J0105+2349      &142$\pm$9  &1270$\pm$12		&16$\pm$2   &75$\pm$15	   & 	--273$\pm$17 & 	166$\pm$15     & 0.22$^{+0.02}_{-0.02}$  \\
        J0152--0431     &80$\pm$11  &1326$\pm$15		&18$\pm$4   &131$\pm$13	   & 	--367$\pm$23 & 	124$\pm$11     & 0.36$^{+0.04}_{-0.04}$  \\
        J0208--0401     &343$\pm$6  &1283$\pm$10 		&25$\pm$3   &69$\pm$10	   & 	--195$\pm$10 & 	169$\pm$15     & 0.43$^{+0.04}_{-0.04}$  \\
        J1103+4834      &77$\pm$12  &647$\pm$16		    &3$\pm$0.4   &22$\pm$5     & 	--347$\pm$20 & 	133$\pm$12     & 0.13$^{+0.01}_{-0.01}$  \\
        J1105+5947      &221$\pm$11  &805$\pm$12		&24$\pm$2   &55$\pm$8      & 	--341$\pm$19 & 	175$\pm$16     & 0.59$^{+0.06}_{-0.06}$  \\ 
        J1219+4814      &74$\pm$10  &667$\pm$15		    &24$\pm$2   &95$\pm$13	   & 	--226$\pm$14 & 	252$\pm$23     & 0.18$^{+0.02}_{-0.02}$  \\
        J1246+4449      &280$\pm$11  &2119$\pm$13		&7$\pm$1   &39$\pm$6       & 	--90$\pm$10  & 	116$\pm$11     & 0.22$^{+0.02}_{-0.02}$  \\ 
        J1425+5249      &29$\pm$5  &572$\pm$6	    	&5$\pm$0.8   &82$\pm$12	   & 	--322$\pm$20 & 	136$\pm$12     & 0.23$^{+0.02}_{-0.02}$  \\

		\hline
		\hline
	\multicolumn{8}{l}{%
  	\begin{minipage}{14cm}%
	Note. --\\
    	(1) \ \ The integrated \lya\ flux in units of 10$^{-17}$ ergs s$^{-1}$ cm$^{-2}$. The values have been corrected by Miky Way dust extinction but not by the internal extinction of the galaxy. \FLyA$_\text{B}$ and  \FLyA$_\text{R}$ represent \lya\ flux of the blue and red peaks, respectively (see Figure \ref{fig:SEDFitsLyA1} and Section \ref{sec:meaLyA}).\\
    	(2)\ \ \VpeakB\ and \VpeakR\ represents the velocity of the blue and red peak of the observed \lya\ profiles, respectively.\\
    	(3)\ \ The escape fraction of \lya\ (see Section \ref{Mea:LyA})\\
% A typical errorbar is $\sim$ 30 km s$^{-1}$ (for one velocity pixel).
 \\
    	
  	\end{minipage}%
	}\\
	\end{tabular}
	\\ [0mm]
	
\end{table*}

\subsection{Determining LyC Escape Fractions}
\label{Mea:LyC}
% from Spectra Energy Distribution Fitting

In the literature, two main methods have been adopted to infer the intrinsic LyC fluxes, that when compared to observations,  give \fescLyC: 1) estimating the intrinsic LyC flux from from Spectral Energy Distribution (SED) fitting \citep[e.g.,][]{Izotov21}, and 2) estimating it from \hb\ emission lines \citep[e.g.,][]{Izotov16a}. We adopt these two methods and discuss them in detail below. In Section \ref{sec:ratio}, we compare the derived \fescLyC\ values with the flux ratio between the non-ionizing continuum and LyC \citep[e.g.,][]{Flury22b}, which is sometimes referred as another proxy for \fescLyC.

%and 3) estimating \fescLyC\ from the flux ratio between the LyC region and the redward region $\sim$ 1100\angstrom\ in the rest-frame, i.e., \FLyC/F$_{1100}$. We adopt all three methods and discuss them in details below.

\subsubsection{Determining \fescLyC\ from SED fitting}
\label{sec:SED}
To fit the SEDs, we adopt the Binary Population and Spectral Synthesis (BPASS) models \citep[version 2.2.1,][]{Stanway18}, which include binary star populations. Comparing to other standard models such as Starburst99 \citep{Leitherer99}, BPASS has been suggested to better represent both the local and high$-z$ galaxies \citep[e.g.,][]{Steidel16}. We use the Prospector spectra fitting code which can take spectroscopic data from the UV to Far-IR rigorously using a flexible spectroscopic calibration model \citep{Johnson21}. Currently, the only BPASS model available within Prospector is the ``-bin-imf135all\_100" model, i.e., assuming a Salpeter IMF\footnote{While a Salpeter IMF is only valid to $\sim$0.4\Msun\ and extrapolation will result in overestimated stellar masses relative to newer IMFs in, e.g., \cite{Chabrier03}, this does not impact our UV-only fits later, which are dominated by the high mass stars.} \citep{Salpeter55} with lower and upper mass cutoffs of 0.1M$_\odot$ and 100 M$_\odot$, respectively, and the slope $\alpha$ = --2.35. In the remainder of the paper, a BPASS SED specifically stands for this model. 

%The IMFs determined for the disk and the spheroid yield the mass-to-light ratios of a factor of 1.8–1.4 smaller than for a Salpeter IMF

%See BPASS manual at /Volumes/HHD1/Astronomy/LyCProject/S10_BPASS/BPASSv2.2.1_README_Manual/BPASSv2.2.1_Manual.pdf

We fit the extinction corrected G140L data from HST/COS for each galaxy. We adopt the non-parametric star formation history (SFH) model from Prospector, which assumes SFR is constant within each of a user defined set of temporal age bins. There are four parameters, i.e., total stellar mass over all ages, stellar metallicity, stellar dust attenuation, and the age bins. For each galaxy, the total stellar mass varies between 10$^8$ -- 10$^{10.5}$ M$_\odot$ with a step size of 0.05 dex, metallicity varies $\pm$ 0.3 dex around the measured abundance from optical emission lines in Section \ref{sec:MeaSDSS}, with a step size of 0.02 dex, and stellar dust optical depth varies between 0.1 to 1.0 \citep[at 5500\angstrom, see][]{Conroy09} with a step size of 0.05 dex. We adopt age bins of [0.0, 6.5], [6.5, 7.0], [7.0, 7.5], [7.5, 8.0], [8.0, 9.0], [9.0, 10.0], where each pair represents the lower and upper lookback time in units of log(years). We assume the dust extinction law of \cite{Calzetti00}. We have also tested other dust extinction laws such as \cite{Cardelli89}, and only found minor differences on the fitted parameters.
%See age bins at: https://github.com/bd-j/prospector/issues/178
%Also at: https://prospect.readthedocs.io/en/latest/sfhs.html

Before the SED fitting, we mask out spectra regions $\pm$ 550 km~s$^{-1}$ around ISM absorption lines, strong emission lines of the galaxy, Milky Way (MW) absorption lines, and Geocoronal emission lines. A few exceptions include: 1) For \lya\ of the galaxy, we mask out $\pm$ 1500 km~s$^{-1}$ to remove the contributions from wide \lya\ emission, 2) For \lya\ and \oi\ \ly 1302 of the MW, we mask out $\pm$ 1500 km~s$^{-1}$ due to broad absorption and strong Geocoronal lines, and 3) We also mask regions with $\lambda$ $<$ 1100\angstrom\ or $\lambda$ $>$ 1900\angstrom\ in the observed frame due to the decreasing throughput of the gratings.

%We then go through the spectra and by-eye mask out any of the other features that could be contaminating the spectra. 

We compare the final fitted SEDs with the observed spectra in Figure \ref{fig:SEDFitsAll}. The COS spectra are shown as gray lines. The internal dust reddened and unreddened models are shown as red and blue lines, respectively. The unreddened models are then adopted to measure the intrinsic LyC flux [i.e., \FLyCINT] given the same windows as we measure the observed LyC flux (\FLyCOBS, see Section \ref{sec:meaLyC}). Overall, our SED models fit the COS spectra well. We then calculate the absolute \fescLyC\ =\FLyCOBS/\FLyCINT\, and the results are shown in the last column of Table \ref{tab:escape}. 

For each of our galaxies, we also report the best fitting stellar dust attenuation in Table \ref{tab:SDSS}. This SED-derived quantity is $\sim$ a factor of two compared with the nebular attenuation (derived in Section \ref{sec:MeaSDSS}). This is consistent with what is suggested in \cite{Shivaei20}.
% The GALEX FUV and NUV flux for each galaxy are shown as purple squares with vertical error bars.

\subsubsection{Determining \fescLyC\ from \hb\ Emission}
\label{sec:Hb}
The \hb\ emission line can be used to estimate the intrinsic LyC emission, as it is a direct measure of the number of ionizing photons absorbed by \hi.  However, while the \hb\ flux is proportional to the total number of ionizing photons, our LyC measurement is in a small wavelength window just shortward of 912\angstrom. Hence, the relationship between the \hb\ flux and the intrinsic LyC flux at our wavelengths of interest depends on the slope of the hydrogen-ionizing spectrum, and thereby the age of the stellar population.  Following \cite{Izotov16b,Flury22a}, we account for this age dependence using EW(\hb),  which shows a strong relationship with \textit{F}(\hb)/\FLyC\ in stellar population models. Note \FLyC\ is the modelled flux at 900\angstrom\ in the rest-frame \citep{Izotov16b}, which is close to our LyC measurement windows (see Figure \ref{fig:Spec1}).

\begin{figure}
\center
	\includegraphics[angle=0,trim={0.1cm 0.5cm 0.3cm 0.0cm},clip=true,width=1\linewidth,keepaspectratio]{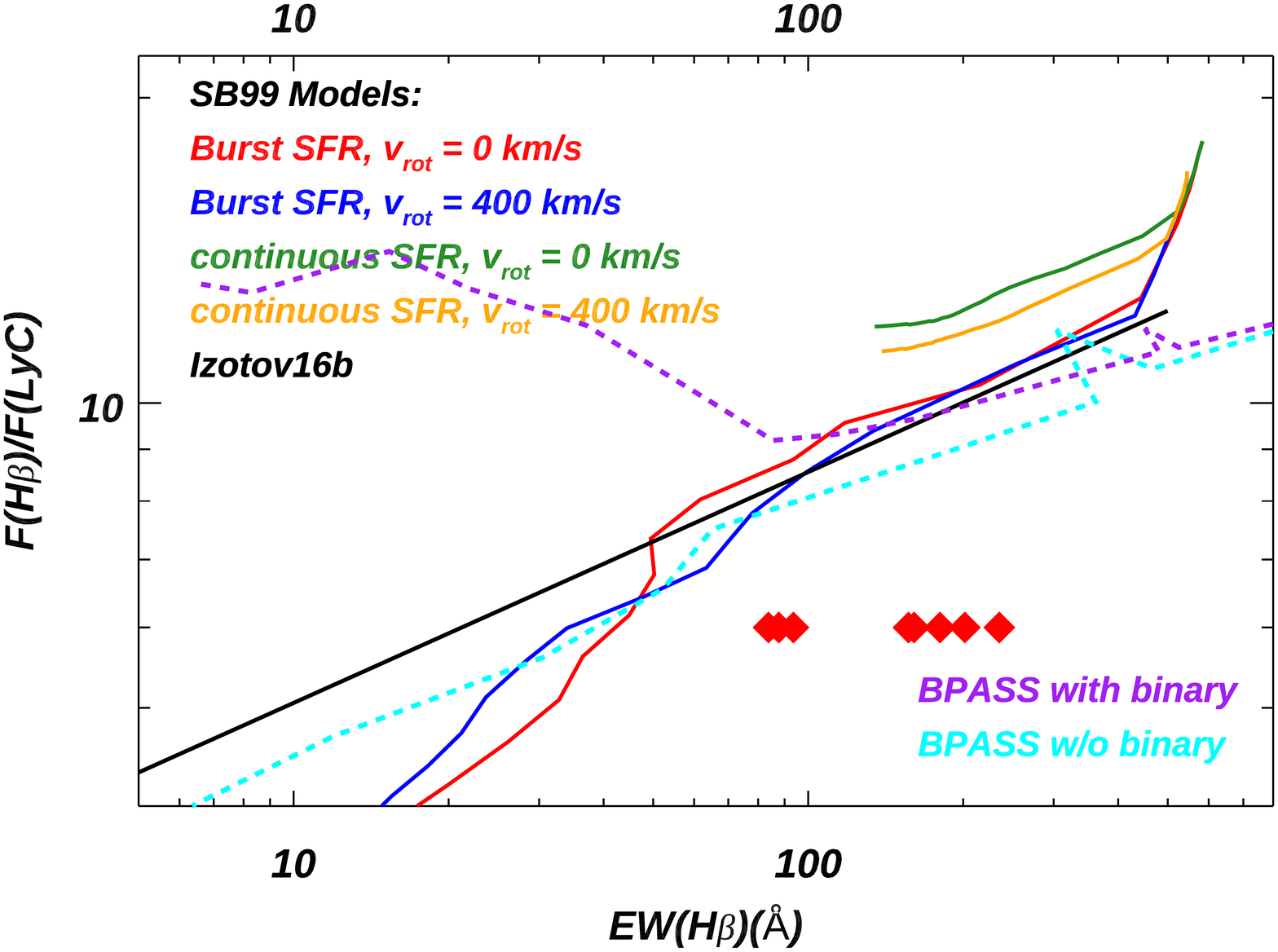}% trim: left lower right upper

\caption{\normalfont{Comparisons of \hb\ -- LyC relationships between different stellar population models. \FLyC\ is the model flux at 900\angstrom\ (restframe). Models from \textit{Starburst99} \citep{Leitherer99} are listed in the top-left corner, where we choose either burst or continuous SFR with different rotation speeds. We show the fitted curve reported in \cite{Izotov16b} as the black line. For BPASS models with burst SFRs, we show the cases with and without binary stars as purple and light blue lines, respectively. Most of the models are consistent when EW(\hb) $\gtrsim$ 80\angstrom, but the trend from BPASS models with binary stars deviates from other models at EW(\hb) $<$ 80\angstrom. Since all galaxies in our sample have EW(\hb) $>$ 80\angstrom\ (shown as the red diamonds at arbitrary y-positions), our derived intrinsic flux of LyC is only weakly dependent on the choice of models. See Section \ref{sec:Hb}.} }
\label{fig:Hb}
\end{figure}

We reproduce this relationship under different models as shown in Figure \ref{fig:Hb}. Four different models from Starburst99 \citep{Leitherer99} are shown, given burst or continuous SFR and different rotation velocities of stars. The relationship adopted in \cite{Izotov16b} is shown as the black line. We also present the relationships from two BPASS models with burst SFR, i.e., one with and another without binary stars. Most of the models are consistent ($\sim$ 10\% difference) when EW(\hb) $\gtrsim$ 80\angstrom, but the trend given by BPASS models with binary stars deviates from other models at EW(\hb) $<$ 80\angstrom. This is because BPASS models with binary stars have harder ionizing spectra at older ages due to the co-evolution of binary stars \citep[see Figure 38 in][]{Eldridge17}. All SB99 models also deviate from BPASS ones at EW(\hb) $\gtrsim$ 450\angstrom, which suggest that SB99 models produce harder ionizing spectra for very young galaxies. Since all galaxies in our sample have 80\angstrom\ $<$ EW(\hb) $<$ 300\angstrom\ (see red diamonds in Figure \ref{fig:Hb}), our derived intrinsic flux of LyC [i.e., \FLyCINT] is only weakly dependent on the choice of models. Therefore, we take the relationship in \cite{Izotov16b} (black line) and get:
%To be consistent with the literature, we choose to use the same trend as \cite{Izotov16b} to derive \FLyCINT\ from \hb,
%The physical reason is that the older stellar populations have a significant contribution from binary effects (extended main sequence, accretion onto compact objects, etc.) 
%which finally lead to a higher \textit{F}(\hb)/\FLyC\ ratio

\begin{equation}\label{eq:Hb}
    f_{esc}^{LyC} = \frac{F_\text{obs}(LyC)}{F_\text{obs}(LyC) + F_\text{int}(LyC)}
    %\fescLyC\ = \frac{\FLyCOBS}{(\FLyCOBS + \FLyCINT)}
\end{equation}
where \FLyCINT\ is the intrinsic (i.e., absorbed) flux of LyC derived from the \hb\ emission line, and \FLyCOBS\ is the observed (i.e., leaked) LyC flux. 

Since the LyC photons that escape from the galaxy do not contribute to the \hb\ emission line, the $F_\text{int}(LyC)$ used in Equation (\ref{eq:Hb}) is only an initial estimate \citep[e.g.,][]{Izotov16b,Flury22a}.  The observed \hb\ EW that we compare to Figure \ref{fig:Hb} must be corrected to account for the ionizing photons that escape.    Therefore, we solve \fescLyC\ iteratively, i.e., using the initial estimate of \fescLyC\ to correct EW(\hb) and recompute \fescLyC\ until the solution converges.  The corresponding errors are calculated from MC simulations while we perturb the measured \hb\ flux, EW and observed LyC flux by their 1$\sigma$ uncertainties, and recalculate \fescLyC. The resulting \fescLyC\ values and errors are shown in the second to last column of Table \ref{tab:escape}.

%Given the relationship between F(\hb)/\FLyC\ and EW(\hb), we can estimate the intrinsic flux of LyC under the assumption that \fescLyC\ is low \citep[e.g.,][]{Izotov16b}.

%Note from Sophia
%Anne and I attributed this difference to the role of stripped stars and accretion-augmented stars, which only become significant at later times in a burst evolution (i.e., lower Hbeta EW). This result was a matter of some debate in the LzLCS team as many collaborators felt the increase in F(Hbeta)/F_lamb(FLyC) at low Hbeta EW for the BPASS models was physically unrealistic. 
%Or could be due to the extinction is not well constrained for old stars at long wavelengths in BPASS.

%\newcommand{\fescLyC}{\textit{f}$_{\text{esc}}^{\text{ LyC}}$}

%F(Hb)/F(LyC) is a proxy for the hardness of the ionizing spectrum

\begin{figure*}
\center
	\includegraphics[angle=0,trim={0.3cm 1.2cm 0.0cm 1.0cm},clip=true,width=0.5\linewidth,keepaspectratio]{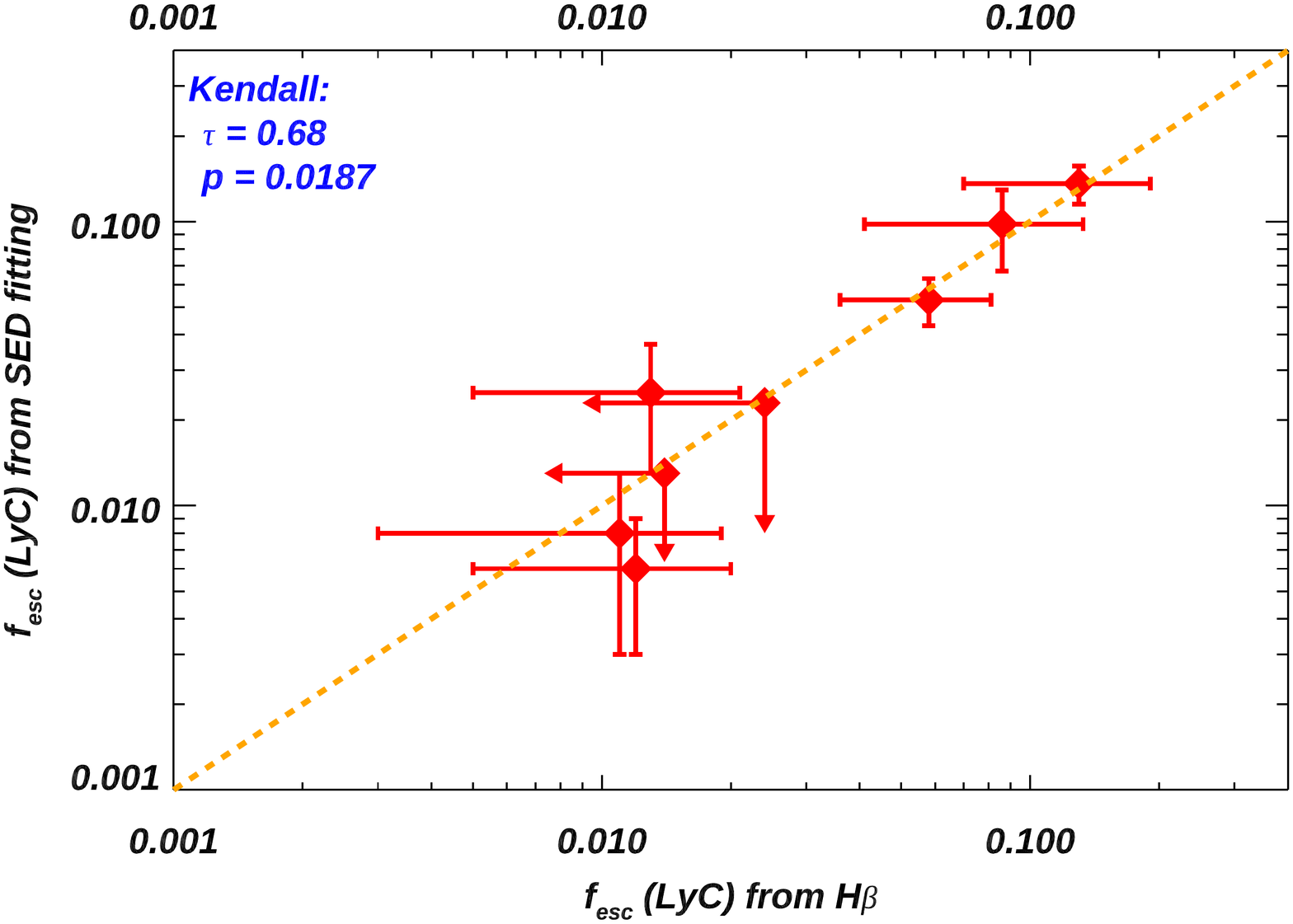}% trim: left lower right upper
	\includegraphics[angle=0,trim={0.3cm 1.2cm 0.2cm 1.0cm},clip=true,width=0.5\linewidth,keepaspectratio]{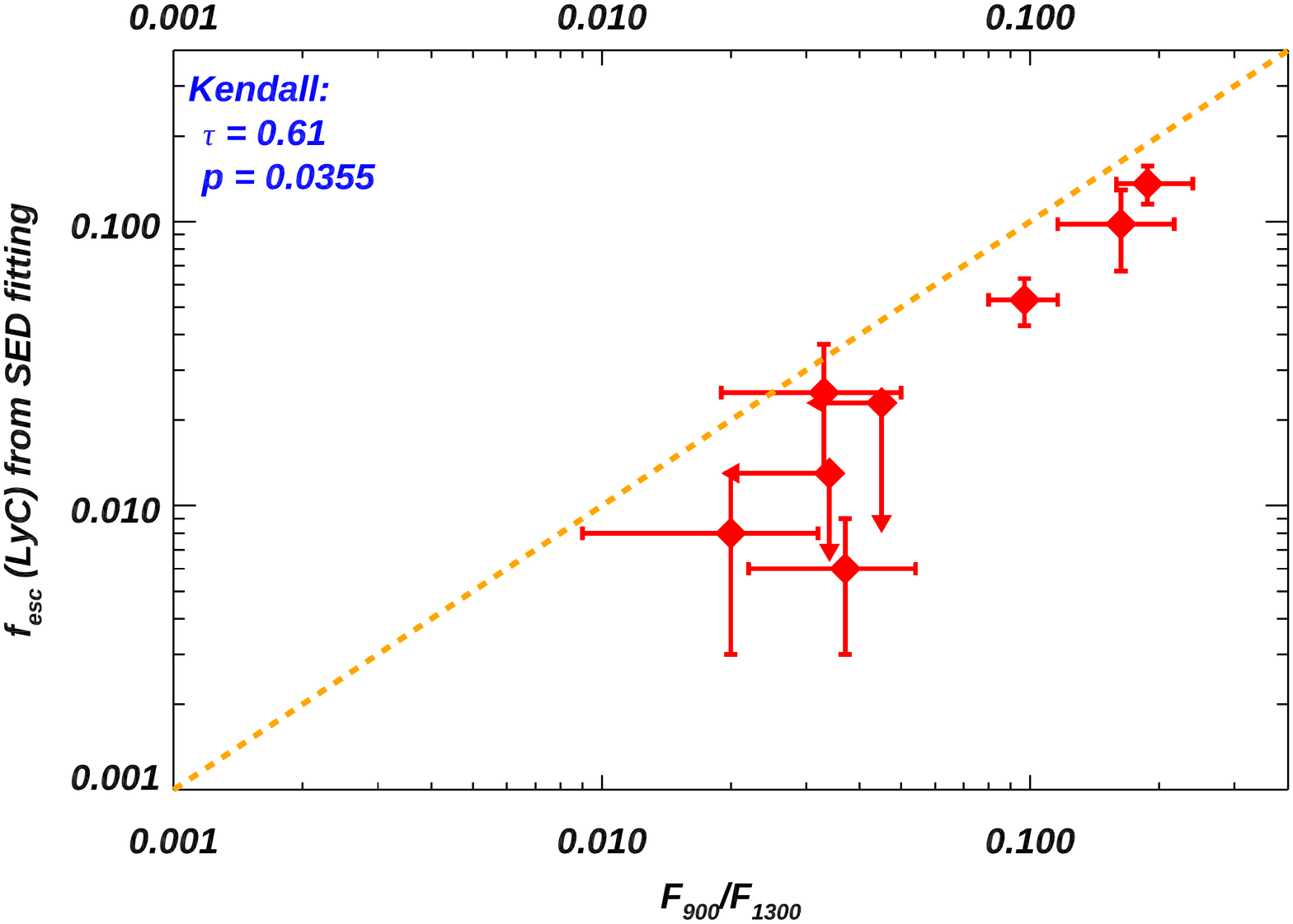}% trim: left lower right upper

\caption{\normalfont{\fescLyC\ measured from two different methods, and comparisons with UV continuum ratios measured at 900 and 1300\angstrom\ respectively. \textbf{Left: }We compare the \fescLyC\ values  derived from SED fitting (Section \ref{sec:SED}) and the \hb\ method (Section \ref{sec:Hb}). \textbf{Right:} We compare the \fescLyC\ values derived from SED fitting and UV continuum flux ratios (\FLyCOBS/\textit{F}(1300)). For each figure, the Kendall's $\tau$ coefficient between the x and y values and the probability of the null hypothesis ($p$) are shown at the top-left corner. The objects with upper limits are denoted by the arrows, where we have taken account of these upper limits in the Kendall $\tau$ test following \cite{Akritas96}.} }
\label{fig:compfesc}
\end{figure*}

\subsubsection{Comparison of Different \fescLyC\ Inferences}
\label{sec:ratio}

In the left panel of Figure \ref{fig:compfesc}, we compare the \fescLyC\ results from the two methods in Sections \ref{sec:SED} and \ref{sec:Hb}.  The \fescLyC\ values with corresponding errors are shown as diamonds (or arrows for upper limits). We also calculate the Kendall's $\tau$ coefficient and the probability of the null hypothesis ($p$), where we take account of the upper limits following \cite{Akritas96}. These are shown at the top-left corner of each panel. The derived \fescLyC\ values from the SED fitting and \hb\ methods are consistent within 1$\sigma$, which enhances our confidence in the \fescLyC\ measurements. This consistency also suggests that the dust destruction of LyC photons within HII regions is not substantial in our galaxies. Otherwise, LyC photons destroyed by dust would not contribute to the \hb\ emission line, and would cause the \hb-derived \fescLyC\ to be systematically higher than the SED-derived value.

Our galaxies' derived \fescLyC\ values are $\sim $1 -- 14\% in the cases where LyC is detected.  In two of the galaxies, \fescLyC\ $\gtrsim$ 10\%, which is significant compared to the values needed for galaxies in the EoR \citep[5 -- 20\%, e.g.,][]{Robertson13, Rosdahl18, Finkelstein19, Naidu20}.
%The characteristic 1$\sigma$ uncertainties in $\tau$ estimated by bootstrapping are $\sim$ 0.1.

In the right panel of Figure \ref{fig:compfesc}, we also compare the derived \fescLyC\ from SED fits to the ratio of ionizing to non-ionizing  UV continuum flux, i.e., \FLyCOBS/\textit{F}(1300). The latter is a purely observational (model-independent) ratio. \textit{F}(1300) is the measured extinction-corrected flux at 1300\angstrom\ in the rest-frame. We find that \fescLyC\ positively correlates with \FLyCOBS/\textit{F}(1300) for our galaxies. Note that it is not necessary to have \fescLyC\ $\simeq$ \FLyCOBS/\textit{F}(1300), since the latter is not a direct measurement of the LyC escape fraction \citep[e.g.,][]{Flury22a}. We show \FLyCOBS/\textit{F}(1300) as well as \FLyCOBS/\textit{F}(1100) in Table \ref{tab:escape}.

%The results are shown in the 4th -- 7th column of Table \ref{tab:escape}.

%x-axis BPASS, left and right y-axis Hb and F900/F1100

%\textbf{4.1: Introduce our SED fittings using Prospector: which data (UV+optical), which database (BPASS instead of SB99), IMF, SFH models, etc. }

%\textbf{Notes: 1) Izotov fit to only optical data and only use SB99 in his 2016, 2018, 2021 papers.
%2) We do not fit the nebular emission lines, since Prospector assumes f$_{esc}$ = 0 when turning emission lines on.  }

%\textbf{Add figures about our fitting results (UV+optical).}

%\section{Characterizing Escape Fraction From LyC and LyA}

%\textbf{5.1: Estimate f$_{esc,LyC}$ from three methods (?): f900/f1100,  Hb methods, and SED fittings.}

\begin{figure}
\center
	\includegraphics[angle=0,trim={0.0cm 3.3cm 0.0cm 0.0cm},clip=true,width=1\linewidth,keepaspectratio]{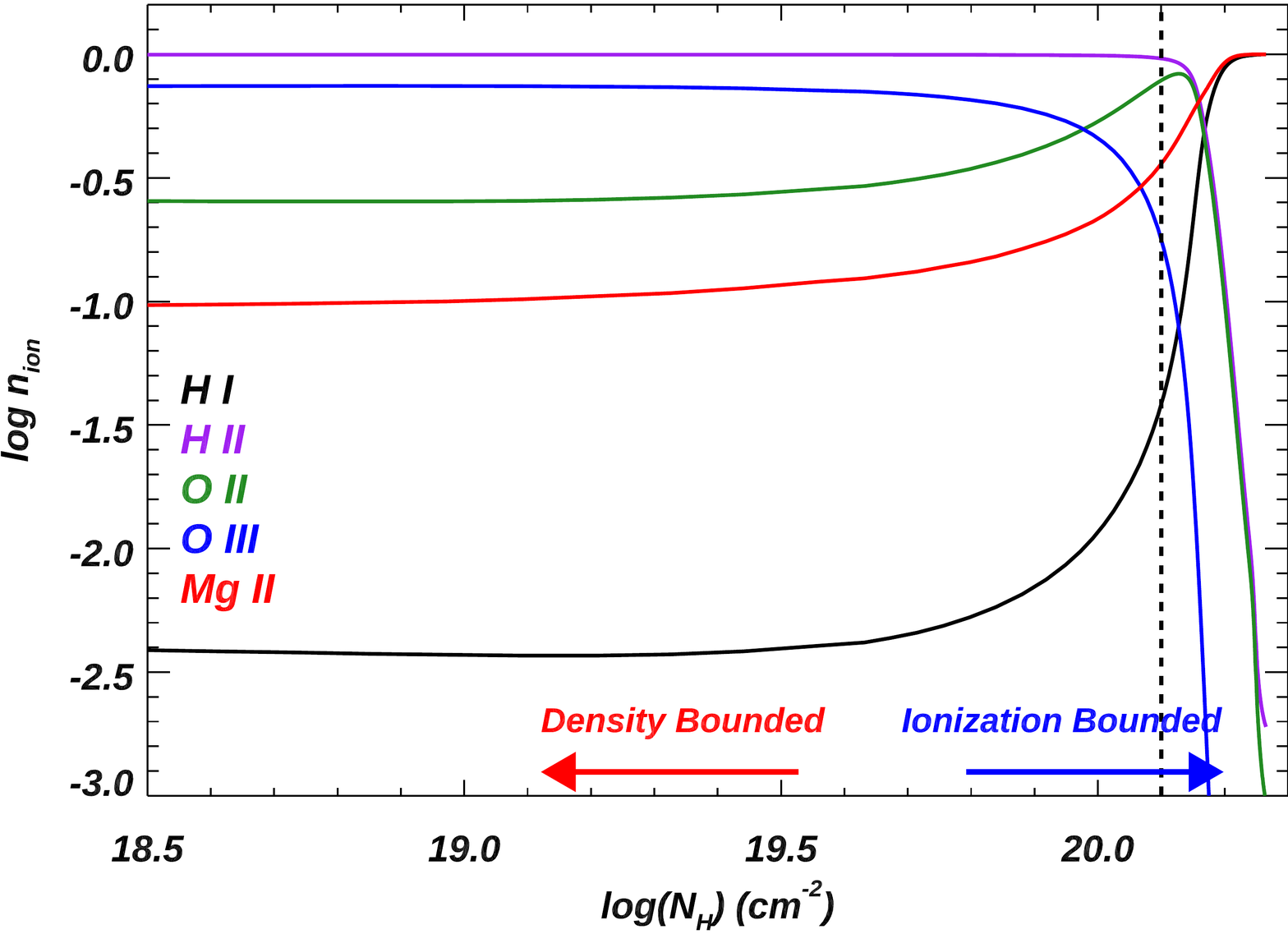}% trim: left lower right upper
	
	\includegraphics[angle=0,trim={0.0cm 0.8cm 0.0cm 0.6cm},clip=true,width=1\linewidth,keepaspectratio]{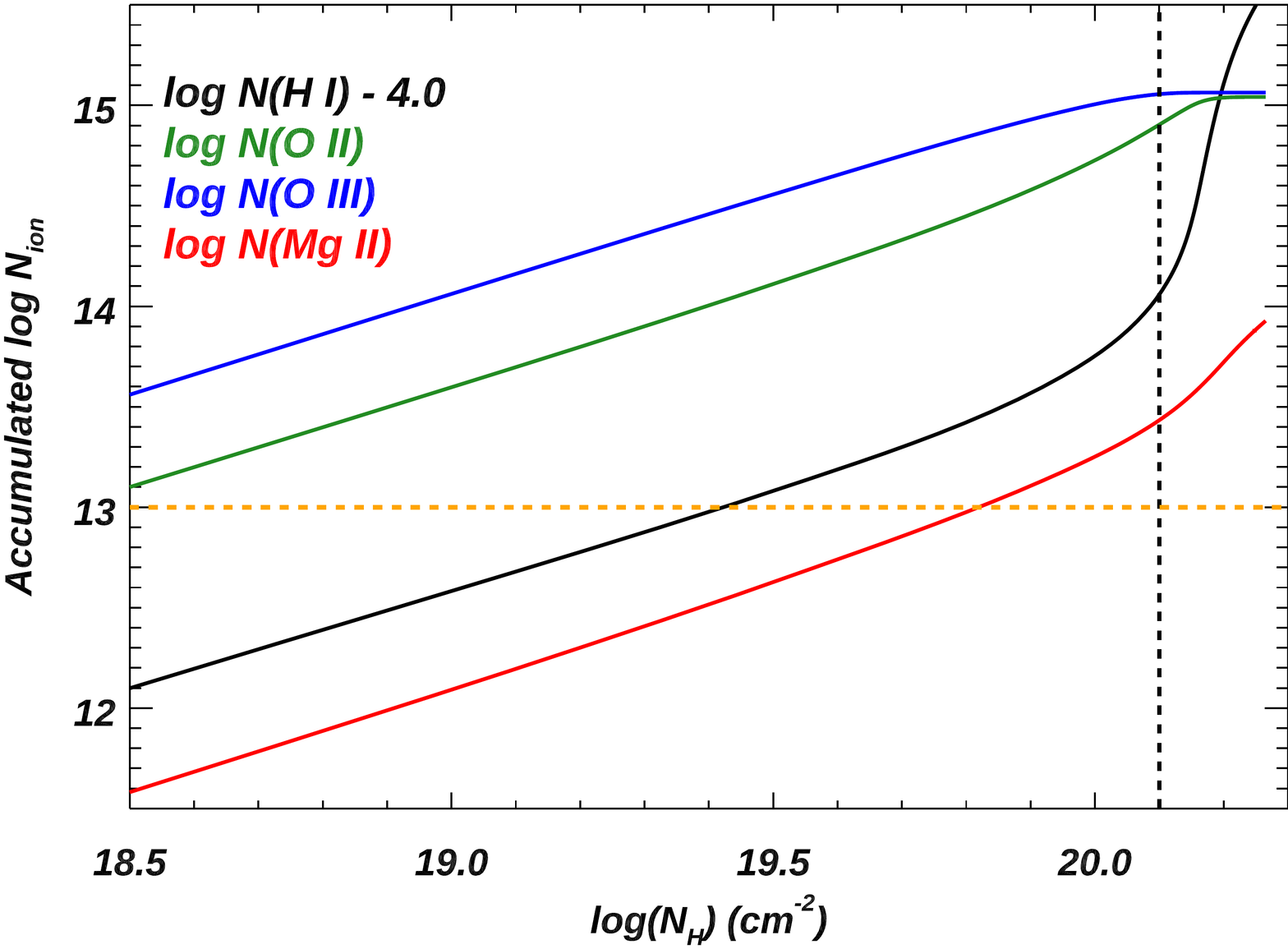}% trim: left lower right upper

\caption{\normalfont{Physical structure of a photoionized cloud at the transition between \hi\ and \hii\ zones. Simulations are done using CLOUDY, and the adopted parameters are provided in Section \ref{sec:MgIITwoCases}. The position of the hydrogen Str$\ddot{\text{o}}$mgren radius is indicated by the vertical black dashed line. The x-axis is the total hydrogen column density (\Nh). \textbf{Top:} Comparison between \Nh\ and the ion population for certain element in the y-axis, e.g., n(\oiii) = the number of \oiii\ ions / total oxygen ions in all states. The approximated regions for the two limiting cases of density and ionization bounds are indicated by the two arrows separately. \textbf{Bottom:} Accumulated \Nion\ for \hi, \oii, \oiii, and \mgii. For the curve of log(N(\hi)), we scale it down by 4.0 dex to include it into the figure. The horizontal orange line represents N(\hi) = 17.0 cm$^{-2}$, where the cloud becomes optically thick to ionizing photons (See discussion in Section \ref{sec:MgIITwoCases}).} }
\label{fig:zoneplot}
\end{figure}

\begin{figure}
\center
	\includegraphics[angle=0,trim={0.0cm 0.8cm 0.0cm 0.0cm},clip=true,width=1\linewidth,keepaspectratio]{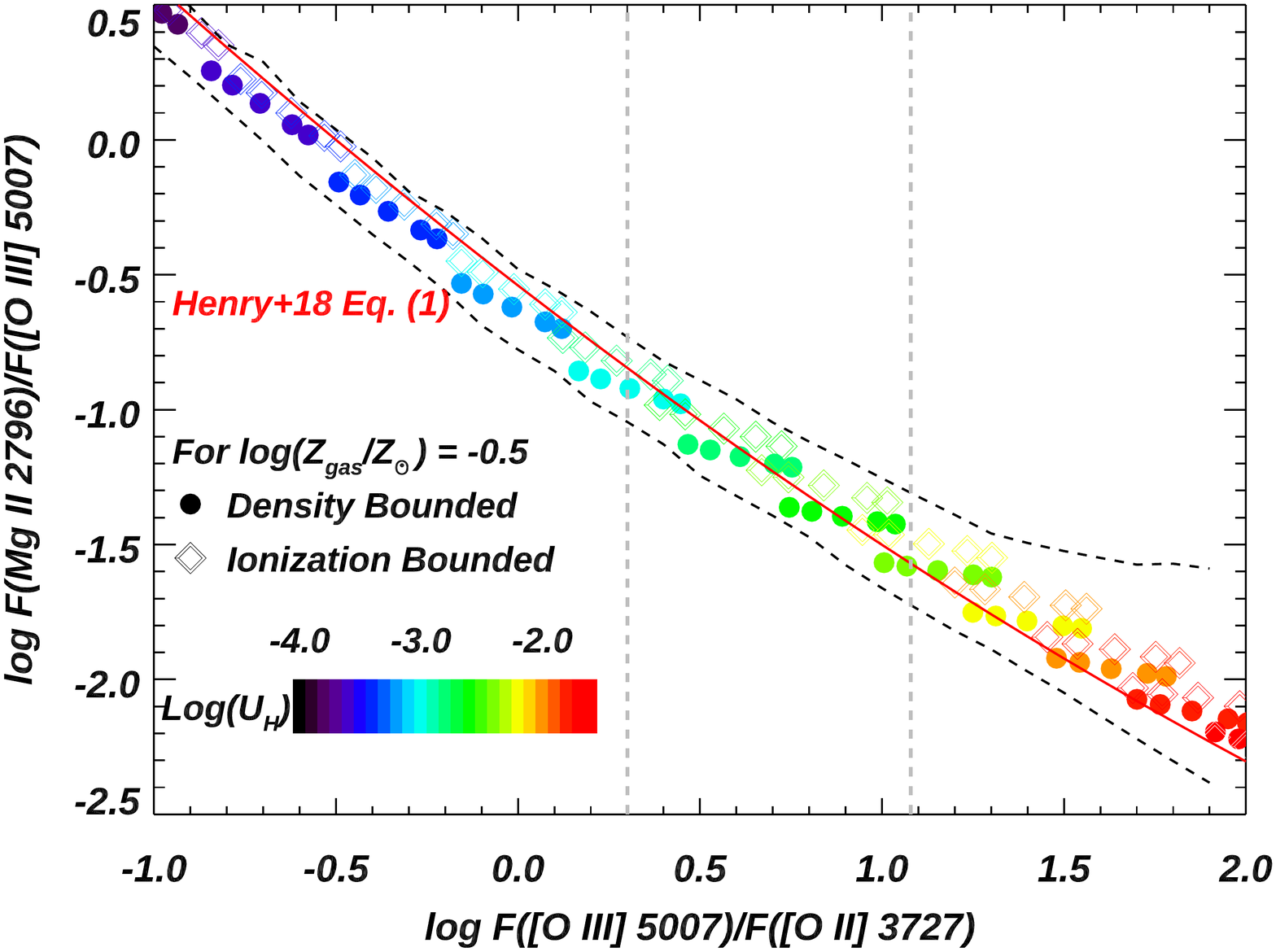}% trim: left lower right upper

\caption{\normalfont{The tight correlation between the flux ratios of \mgii \ly 2796/[\oiii] and O32, which is based on the grid models from CLOUDY simulations. The input parameters are discussed in Section \ref{sec:MgIITwoCases}. Different colors are for different ionization parameter [log(\Uh)]. The circles and diamond symbols are for the density- and ionization-bounded cases, respectively. Each grid point [i.e., at certain log(\Uh) and gas-phase metallicity] is plotted for five stellar metallicities (\ZStar\ = 0.001, 0.002, 0.004, 0.006, and 0.008). In these models, higher log(\Uh) at lower stellar metallicities moves the line ratios downwards along the sequence. The two black dashed lines represent the envelopes for all grid models varying log(\Uh) and \ZStar. The solid red line indicates the polynomial fits from \cite{Henry18}, which matches our correlations well. The galaxies discussed in this paper have O32 falling between the two gray vertical lines. This correlation for \mgii\ \ly 2803 has the same trend, but differs by a factor of about 2. See details in Section \ref{sec:MgIITwoCases}. } }
\label{fig:MgII-O32}
\end{figure}

\subsection{Calculating \mgii\ Escape Fractions from CLOUDY models}
\label{Sec:MgII}

Previous studies have shown that photoionization models can be used to infer the escape fraction of \mgii\ (\fescMgII; e.g., \citealt{Guseva13, Henry18, Chisholm20}).  %\citep[e.g., ][]{Guseva13, Henry18, Chisholm20}. 
By using multiple strong optical emission lines that are set by the ionization parameter, metallicity, density, and ionizing spectra, one can predict the intrinsic flux of \mgii\ [\FMgIIINT], which, combined with the observed flux of \mgii\ [\FMgIIOBS], gives the corresponding \fescMgII. However, previously published photoionization models for \mgii\ are only based on ionization-bounded (IB) cases, while models for density-bounded (DB) and mixed cases have not been studied in the literature. We briefly discuss these two cases below.

%State the two bounded cases here.
The mechanisms for \mgii, \lya, and LyC photons (hereafter, photons specifically stands for these three types) to escape from the galaxy are not fully understood. As discussed in literature \citep[e.g.,][]{Jaskot13, Zackrisson13, Reddy16b,  Chisholm20,Kakiichi21}, there are possibly two major scenarios: 1) photons escape from DB cloud/nebulae. In this case, the cloud is smaller than its hydrogen Str$\ddot{\text{o}}$mgren radius, and the intense radiation from massive stars ionizes all the gas in the cloud. Therefore, in this case, photons can escape through optically thin gas \citep[e.g.,][]{Chisholm20}, and, alternatively 2) photons escape through a clumpy geometry. In this case, most of the cloud remains neutral (i.e., IB) and is optically thick to escaping photons. Thus, photons escaping from these neutral paths suffer from intense scattering and are finally destroyed by dust. But there are low column density channels \citep[possibly holes in ISM, e.g.,][]{Heckman11, Saldana-Lopez22} that allow photons to escape. 
%The combinations of these two scenarios are also possible for each galaxy.
%Shorten this

Recent observations of high detection rates of LyC escape from galaxies with high O32 values \citep[e.g.,][]{Gazagnes18, Izotov18b, Izotov21, Ramambason20,Flury22a,Saldana-Lopez22} suggest that some clouds around those galaxies should be density-bounded (i.e., covering fraction of DB clouds \protect{$>$} 0). Therefore, studies of photoionization models in DB cases are necessary. In Section \ref{sec:MgIITwoCases}, we first discuss the \mgii\ photon production in CLOUDY models. Then in Section \ref{sec:MgIIFit}, we show the updated photoionization models to constrain \fescMgII\ based on both DB and IB case, separately.

%that allow \mgii\ photons (and similarly ionizing photons) to escape
%We follow similar approaches and constrain the \fescMgII\ values for our galaxies as follows.

\subsubsection{\mgii\ Photon Productions in CLOUDY Models}
\label{sec:MgIITwoCases}
%

%These are also the two escape types of MgII photons, see Kakiichi and Max's 2019 paper and Zackrisson13. Discuss more here.

In Figure \ref{fig:zoneplot}, we illustrate these two limiting cases of DB and IB by showing the physical structure of an ionized cloud that is at the transition between \hi\ and \hii\ zones. The vertical black line indicates position of the hydrogen Str$\ddot{\text{o}}$mgren radius, i.e. log(\Nh) = log(\Uh)+23.1 [see Equation (3) in \cite{deKool02}]. We adopt the photoionization code CLOUDY [version c17.01, \citep{Ferland17}] with input parameters that are typical for galaxies in our sample. These include a BPASS SED with constant SFR and an age of 5 Myr, a stellar metallicity of log(\ZStar/\Zsun) = -1.5 (assuming \Zsun\ = 0.02), a gas-phase metallicity of log(\ZGas/\Zsun) = -1.5, electron number density log(\ne) = 2.4 (cm$^{-3}$), and ionization parameter log(\Uh) = -3.0. The vertical axis represents the ion population for a certain element, e.g., n(\oiii) = the number of \oiii\ ions / total oxygen ions in all states. The horizontal axis is the total hydrogen column density [i.e., \Nh\ = N(\hi) + N(\hii)], which is a proxy for the depth into the cloud. The ionizing spectrum illuminates the cloud from the left. 
%a Salpeter IMF with an upper mass cutoff of 100 M$_\odot$
%Orion nebular dust distribution, to account for depletion
%of gas on to grains (Baldwin et al. 1991). 

If we truncate the CLOUDY model at radii far interior to the hydrogen Str$\ddot{\text{o}}$mgren radius, the cloud is highly ionized and is analogous to a density-bounded nebula. This limiting case is indicated by the red arrow in Figure \ref{fig:zoneplot}. In this region, hydrogen is fully ionized, while the dominant form for oxygen is \oiii\ and n(\mgii) is $\sim$ 10\% of the total magnesium (red line in Figure \ref{fig:zoneplot}). When we truncate the model at radii closer to (or after) the hydrogen Str$\ddot{\text{o}}$mgren radius, the cloud transitions to be ionization-bounded (see the blue arrow in Figure \ref{fig:zoneplot}). The ions in this region change drastically from higher to lower or neutral ionization states, e.g., \hii\ to \hi\ and \oiii\ to \oii. Additionally, n(\mgii) increases quickly near the IB region, and becomes the dominant ion for magnesium when  \Nh\  is $\sim$ 0.1 dex larger than at the  
 Str$\ddot{\text{o}}$mgren radius. 

%Note there is no definite boundary between DB and IB cases.
%Overall, these two cases are consistent with recent high detection rates of LyC escape from galaxies with high O32 values \citep[e.g.,][\Fluryab]{Izotov16a, Izotov16b, Izotov18a, Izotov18b, Izotov21}. These observations suggest that the clouds around those galaxies could be at least partially density-bounded. 

In the bottom panel of Figure \ref{fig:zoneplot}, we show the accumulated log column density (\Nion) of \hi, \mgii\, \oii, and \oiii\ from the CLOUDY model described above. For N(\hi), we scale the curve down by 4.0 dex to include it in the figure. The horizontal orange line represents log N(\hi) = 13.0 + 4.0 = 17.0 (cm$^{-2}$), where the cloud becomes optically thick to ionizing photons. From log N(\hi) $\sim$ 16.0 -- 17.5, the curves of N(\mgii) and N(\hi) are parallel (in red and black, respectively). This means that N(\mgii) \textit{can} be adopted to trace N(\hi) in a wide range of conditions from DB to IB regions (before the cloud is a lot deeper than the hydrogen Str$\ddot{\text{o}}$mgren radius). 
This parallel in a wide range is because \mgii\ and \hi\ are optically thin at similar \Nh\ for our galaxies \citep[results of both metallicity, ionization potential and oscillator strengths, see][]{Chisholm20}. 

We also find the variations in metallicity or log(\Uh) in the models only change the relative positions between the two curves or the positions of the hydrogen Str$\ddot{\text{o}}$mgren radius. Therefore, our main conclusion in this sub-section remains unchanged, i.e., the curves of \mgii\ and \hi\ stay in parallel before hydrogen becomes mostly neutral.

%Note this parallel property also holds for the parameter spaces discussed next in Section \ref{sec:MgIIFit}.

%This is consistent with the two theoretical models, i.e., \mgii\ (and \lya, and LyC) escape from either DB nebulae or the clumpy geometry.
%HI, HII vs DB, IB vs optically thin.

%This region has been known to be highly ionized, optically thin, and allow escape for at least a fraction of ionizing photons (cite?).

%One caveat is that the possible different correlations between density- and ionization-bounded cases have not been studied. Therefore, we reproduce their results below and present the correlations in both bounded cases.

%We have also tested the correlations without or with dust (by adding "grain orion" in CLOUDY). We found the resulting N(\mgii) only have minor differences ($<$ 0.03 dex).

\subsubsection{Updated \mgii--O32 Correlations and \fescMgII}
\label{sec:MgIIFit}
%Add a sentence why MgII/OIII works. O32 ~ O3/MII, so OII-MgII?

\cite{Henry18} has determined a tight relationship to derive \FMgIIINT\ from the extinction-corrected flux of [\oiii] \ly 5007 and [\oii] \ly 3727 (see their Figure 4). Their models only considered the IB case in a slab geometry. However, various publications \citep[e.g.,][]{Izotov16a, Izotov16b, Izotov18a, Izotov18b, Izotov21,Flury22a,Flury22b} found higher detection rates of LyC escape from galaxies with increasing O32 values. This suggests that DB scenarios may be more prominent in LyC emitters (LCEs). Therefore, it is important to re-calibrate the relationship considering both DB and IB cases. 

We run grid models in CLOUDY, adopting the same SED, IMF and log(\ne) as in Section \ref{Sec:MgII} and Figure \ref{fig:zoneplot}. We vary the stellar metallicity \ZStar\ = 0.001, 0.002, 0.004, 0.006, 0.008, the gas-phase metallicity log(\ZGas/\Zsun) = -1.5, -1.0, -0.5, and the ionization parameter log(\Uh) from -4.0 to -1.0 in steps of 0.1 dex. For DB, we stop the model at log(\Nh) of 1.0 dex shallower than the hydrogen Str$\ddot{\text{o}}$mgren layer, and for IB, we stop the model at log(\Nh) of 0.1 dex deeper than the layer. Note that for these models with different metallicities, the abundance ratio of log(Mg/O) stays the same as the solar ratio in GASS10 \citep[\protect$\sim$ --1.1, see][]{Grevesse10}. This is as expected since both magnesium and oxygen are alpha elements (created in core-collapse supernovae)\footnote{If magnesium is more heavily depleted onto dust than oxygen, the constant Mg/O ratio might not be a good assumption. But \cite{Henry18} showed that this was not a big concern.}.

\begin{table}
	\centering
	\caption{Fitted Parameters for the \mgii--O32 Correlations$^{1}$}
	\label{tab:MgII}
	\begin{tabular}{cccccccc} % four columns, alignment for each
		\hline
		\hline
		\ly(\mgii)  	    & Metal$^{2}$     &	A$_{0}$  & A$_{1}$  & A$_{2}$    &   B$_{0}$  &   B$_{1}$     &  B$_{2}$ \\
        \hline
        2796      &-1.5		& -0.52	    & -0.93    & 0.09   & -0.45 & -0.97 & 0.076\\
        2796      & -1.0	& -0.55	    & -0.94    & 0.10   & -0.46 & -0.97 & 0.074\\
        2796      & -0.5	& -0.61	    & -0.94    & 0.12   & -0.49 & -0.96 & 0.074\\
        \hline
        2803      & -1.5	& -0.81	    & -0.93    & 0.09   & -0.75 & -0.96 & 0.082\\
        2803      & -1.0	& -0.84	    & -0.94    & 0.11   & -0.76 & -0.97 & 0.077\\
        2803      & -0.5	& -0.91	    & -0.93    & 0.13   & -0.79 & -0.96 & 0.077\\

		\hline
		\hline
	\multicolumn{8}{l}{%
  	\begin{minipage}{8cm}%
	Note. --\\
	    (1)\ \ See definitions of parameters in Equations \ref{eq:MgII-O32_part1} and \ref{eq:MgII-O32_part2}. \\ 
    	(2)\ \ The logarithm of gas-phase metallicity relative to solar metallicity, i.e., log(\ZGas/\Zsun).\\
    	
  	\end{minipage}%
	}\\
	\end{tabular}
	\\ [0mm]
	
\end{table}

An example of the correlation between \mgii\ and O32 is shown in Figure \ref{fig:MgII-O32}. Different colors are for models with different log(\Uh) values. The circle and diamond symbols are for the DB and IB cases, respectively. Each color has a set of 5 symbols which stands for models with 5 different \ZStar\ values. The two black dashed lines represent the envelopes for all models for the full range of log(\Uh) and \ZStar. The solid red line is the fitted correlation in Equation (1) of \cite{Henry18}, which matches our dispersion well. Overall, different models from DB and IB only move the \mgii--O32 correlations along the line. This leads to one important advantage of the correlation, i.e., it is robust without any knowledge about the actual limiting scenarios (either DB or IB or mixed) of the observed galaxy.

Figure \ref{fig:MgII-O32} demonstrates that, given the observed extinction corrected flux of [\oiii] \ly 5007 and [\oii] 3727, we can predict the intrinsic flux ratio of \mgii/[\oiii] \ly 5007. We conduct polynomial fits to our full set of BPASS grid models under each log(\ZGas/\Zsun) value similar to the Equation (1) -- (3) of \cite{Henry18} as follows:

\begin{equation}\label{eq:MgII-O32_part1}
\begin{aligned} 
    & R_{2796} \text{(DB)} & = A_{2} \times O32^2 + A_{1} \times O32 +A_{0} \\
    & R_{2796} \text{(IB)} & = B_{2} \times O32^2 + B_{1} \times O32 +B_{0}
\end{aligned}    
\end{equation}
where 

\begin{equation}\label{eq:MgII-O32_part2}
\begin{aligned} 
    & R_{2796}  &= log(F(\text{Mg II } \lambda2796)/F(\text{[O III] } \lambda5007))\\
    & O32       &= log(F(\text{[O III ]} \lambda5007)/F(\text{[O II] } \lambda3727))
\end{aligned}  
\end{equation}
Similar formulae exist for \mgii\ \ly 2803. The fitted parameters for different metallicities are shown in Table \ref{tab:MgII}. Note that these correlations work best for O32 $\lesssim$ 20 (Figure \ref{fig:MgII-O32}). 

\begin{figure*}
\center
	\includegraphics[angle=0,trim={0.0cm 0.8cm 0.1cm 0.0cm},clip=true,width=0.5\linewidth,keepaspectratio]{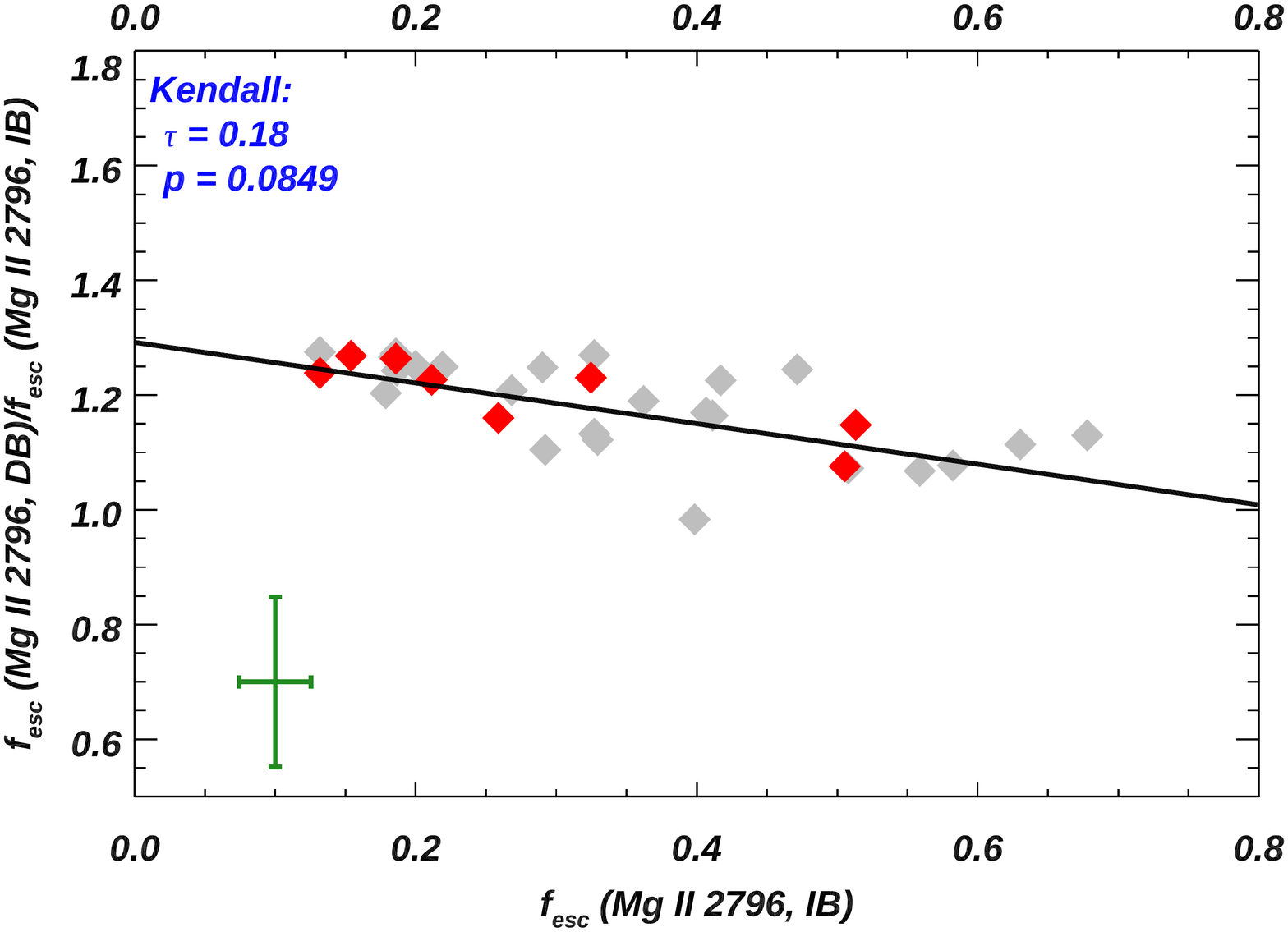}% trim: left lower right upper
	\includegraphics[angle=0,trim={0.0cm 0.8cm 0.1cm 0.0cm},clip=true,width=0.5\linewidth,keepaspectratio]{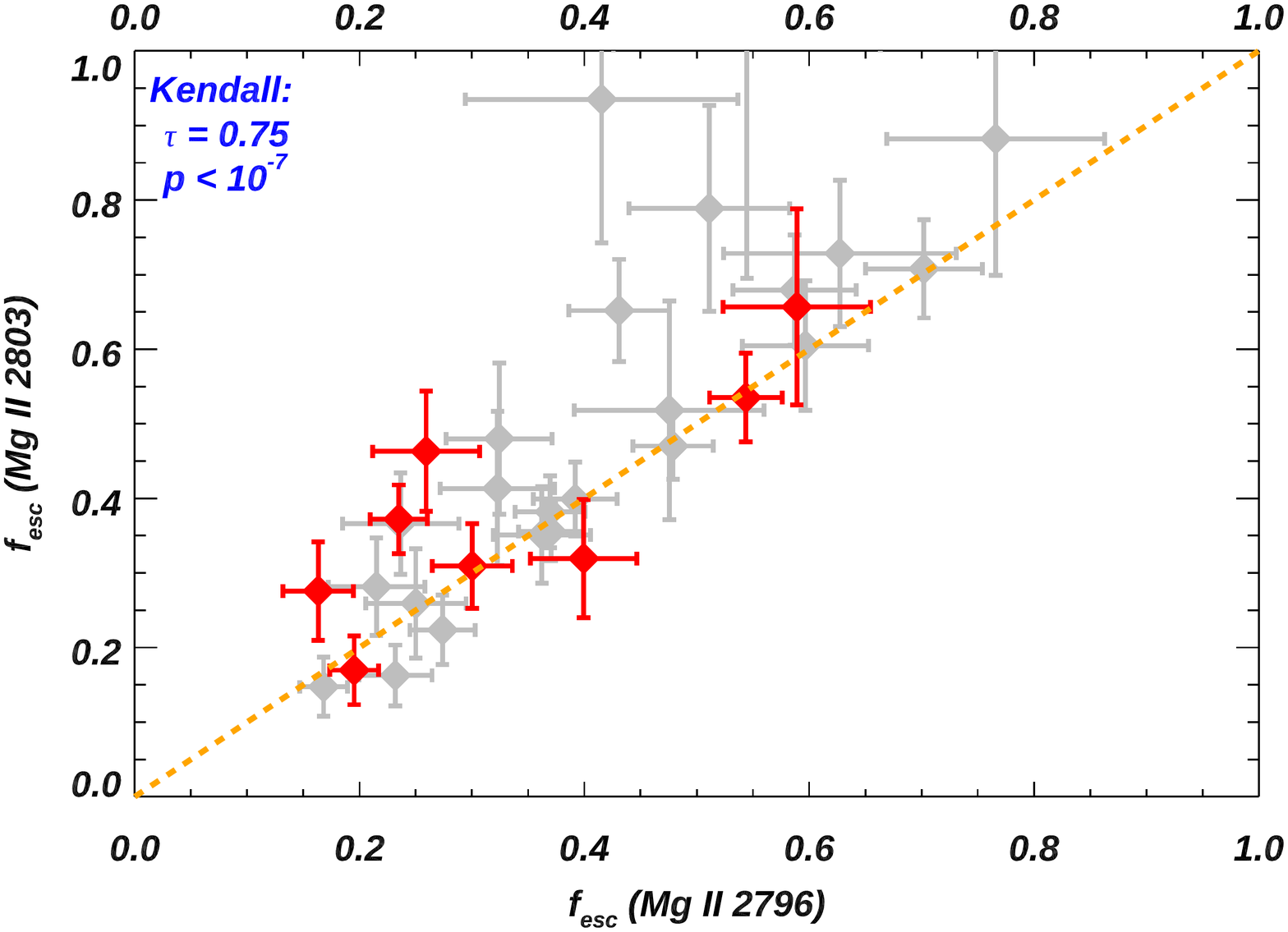}% trim: left lower right upper
	
\caption{\normalfont{Comparison of the \fescMgII\ values from the two limiting cases of DB and IB, and two doublet lines. The 8 galaxies in our sample are shown in red, while other possible LyC leakers from the literature with SNR of \mgii\ $>$ 3 are shown in gray (see Section \ref{sec:comp}). \textbf{Left:} The ratio of \fescMgII\ from DB to IB versus \fescMgII\ from DB. We find general consistencies between the derived \fescMgII\ values. But \fescMgII\ from DB are always larger than the ones from IB as expected from the CLOUDY models (see discussion in Section \ref{sec:CorrMgII}). The black line represents the best linear fit, while the green cross represents the average error bar. \textbf{Right:} Strong correlations between \fescMgII\ from \mgii\ \ly 2796 and \ly 2803, assuming DB. The scatter is mainly due to the low signal-to-noise ratio for \mgii\ \ly 2803 emission line from SDSS spectra. For each figure, Kendall's $\tau$ coefficient and the probability of the null hypothesis ($p$) are shown at the top-left corner. The dashed lines represent the 1:1 relationship.} }
\label{fig:MgIITwoCases}
\end{figure*}

Finally, given the ratios of R$_{2796}$ and R$_{2803}$, we derive the intrinsic flux of \mgii\ [\FMgIIINT], which then leads to the escape fraction of \mgii\ as \fescMgII =  \FMgIIOBS/\FMgIIINT. For each galaxy, we choose the model that has the closest gas-phase metallicity to the value we measured in Section \ref{sec:MeaSDSS}.  In this calculation, the observed \mgii\ flux [\FMgIIOBS] is corrected for Milky Way (MW) extinction \citep[see Section \ref{sec:CorrMgII} and][]{Henry18, Chisholm20}. The derived \fescMgII\ values for \mgii\ \ly 2796 are listed in Tables \ref{tab:SDSS} and \ref{tab:SDSS2}. Since we only have SDSS optical spectra, we do not report \fescMgII\ for \mgii\ \ly 2803 due to the relatively lower SNR. But in Figure \ref{fig:MgIITwoCases}, we compare \fescMgII\ from both lines, and we discuss possible implications in Section \ref{sec:CorrMgII}.

We have also tested CLOUDY models given the BPASS SED with burst SFR instead of the constant SFR discussed above. The difference in the resulting relationship is smaller than 5\%.

%See Section \ref{sec:CorrMgII} for more discussion about \mgii\ resonant scattering. 
%\textbf{6.1: Reproduce and extend the MgII-O32 relationship in Henry+18 in DB and IB cases.}

%\textbf{6.2: Estimate the intrinsic MgII flux from the best-fits and then f$_{esc,MgII}$ for our galaxies, and add a table to show the results.}

%\textbf{6.3: Relationships of f$_{esc,MgII}$ to other key parameters, such as f$_{esc,LyC}$, f$_{esc,LyA}$, O32, LyA-peak-sep, EBV, f900/f1100, EW2796.  Add multiple figures here.}

%\textbf{6.4: Draw conclusion that MgII could be good to predict  f$_{esc,LyA}$ (strong) and f$_{esc,LyC}$ (weak?). Emphysize the detection rates of LyC flux is high in our sample 6 out of 8.}

\subsection{Determining \lya\ Escape Fraction}
\label{Mea:LyA}
We calculate \fescLyA\ from the ratio of \lya\ and \hb\ flux. Similar to \mgii, \lya\ flux is only corrected for MW extinction due to the unknown resonant scattering, while \hb\ flux is corrected by both MW and internal extinction of the observed galaxy. We assume the case B with an intrinsic \lya/\hb\ ratio of $\sim$ 23.3 under \Te\ = 10,000 K and \ne\ = 100 cm$^{-3}$ \citep[see][]{Storey95}. We do not use the stronger \ha\ emission lines because of the clipping issues observed for some of the \ha\ emission lines (see Section \ref{sec:MeaSDSS} and \citealt{Flury22a}). The resulting values are shown in Table \ref{tab:LyA}.

%\textbf{5.2 Show LyA profiles and Estimate f$_{esc,LyA}$ using Hb}

%\textbf{5.3 Show correlations between LyA and LyC}
\iffalse
\begin{figure}
\center
	\includegraphics[angle=0,trim={0.5cm 0.5cm 0.1cm 0.0cm},clip=true,width=1\linewidth,keepaspectratio]{./Figures/EW_MgIItotal_vs_O32.pdf}% trim: left lower right upper

\caption{\normalfont{Comparisons between EW(\mgii\ 2796+2803) and O32 values for galaxies in our sample and other samples from the literature that focused on \lya\ and/or LyC escape. The eight galaxies from our sample are shown in red. Galaxies from Izotov et al. that have \mgii\ coverage \citep[][]{Izotov16a, Izotov16b, Izotov18a, Izotov18b, Izotov21} are shown in gray. The Green Pea galaxies published in \cite{Henry18} are shown in black, and the galaxies from the Low Redshift LyC Survey (LzLCS) that have high S/N \mgii\ detections are shown in light blue. } }
\label{fig:EWMgII-O32}
\end{figure}
%J1011_Guseva with O32~30 is not shown here.
%Our sample (in red) uncovers LyC leaker candidates that have high EW(\mgii) and not high O32 values that are mostly missed by other surveys.
\fi

\section{Results}
\label{sec:results}
The observations that we have presented thus-far show that at least 50\% (4 out of 8) of Mg II selected galaxies present LyC emission at 2$\sigma$ significance, and derived \fescLyC\ values range from $\sim$ 1.5\% to 14\%. We now explore the relationship between Mg II, \lya, and LyC, as we aim to build diagnostics that can be applied in EoR galaxies. In Section \ref{sec:comp}, we introduce several comparison samples to extend the dynamic range for the correlations. In Section \ref{sec:CorrMgII}, we compare \fescMgII\ with various galaxy and line properties. In Section \ref{sec:CorrLyA}, we contrast \fescLyA\ with \mgii\ and \lya\ line properties. Finally, we discuss various indirect indicators of \fescLyC, while comparing our \mgii\ selected galaxies with the whole LzLCS sample in Section \ref{sec:CorrLyC}.

\begin{figure*}
\center
	\includegraphics[angle=0,trim={0.2cm 0.5cm 0.1cm 2.2cm},clip=true,width=0.5\linewidth,keepaspectratio]{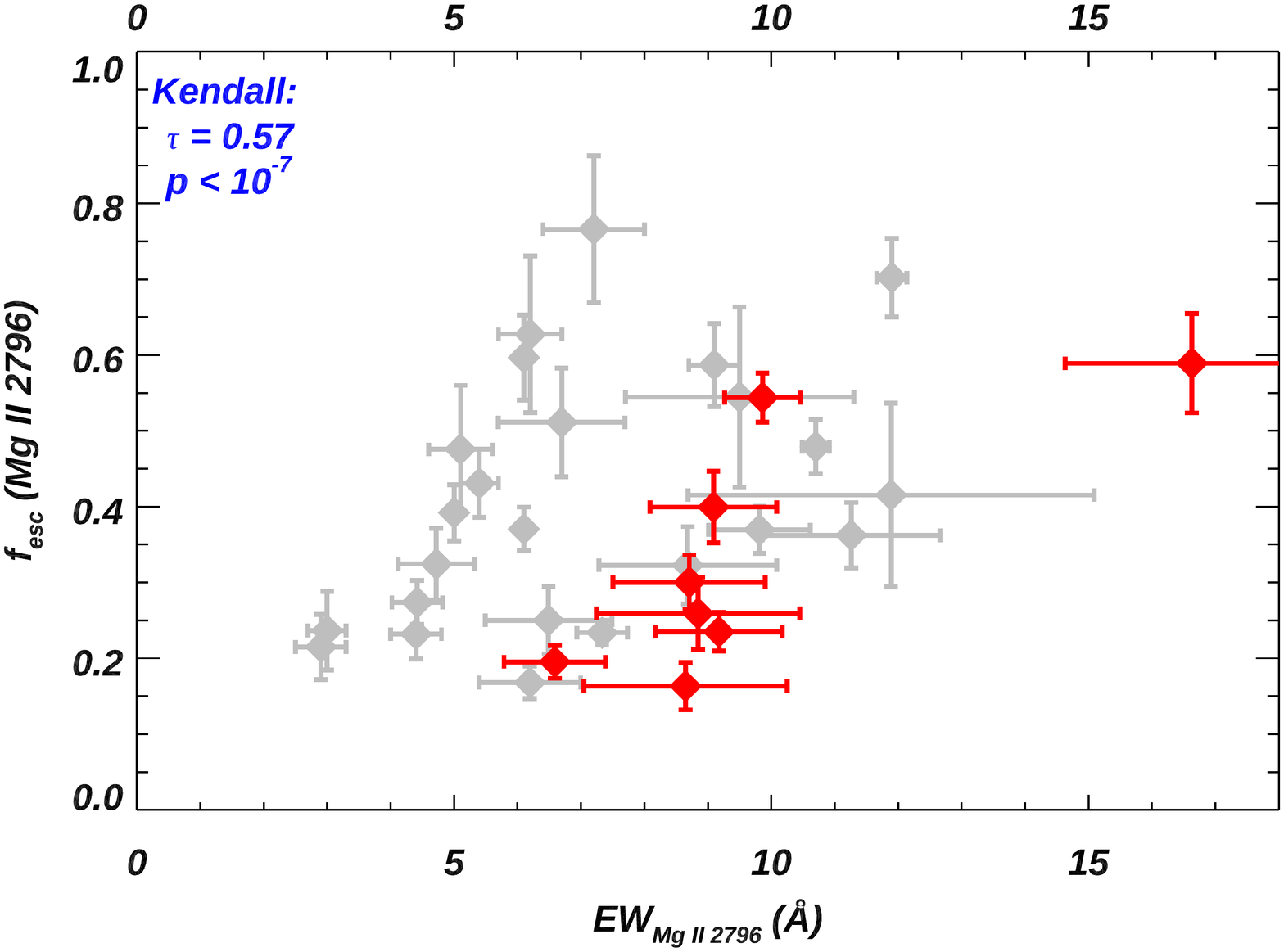}% trim: left lower right upper
	\includegraphics[angle=0,trim={0.2cm 0.5cm 0.1cm 2.2cm},clip=true,width=0.5\linewidth,keepaspectratio]{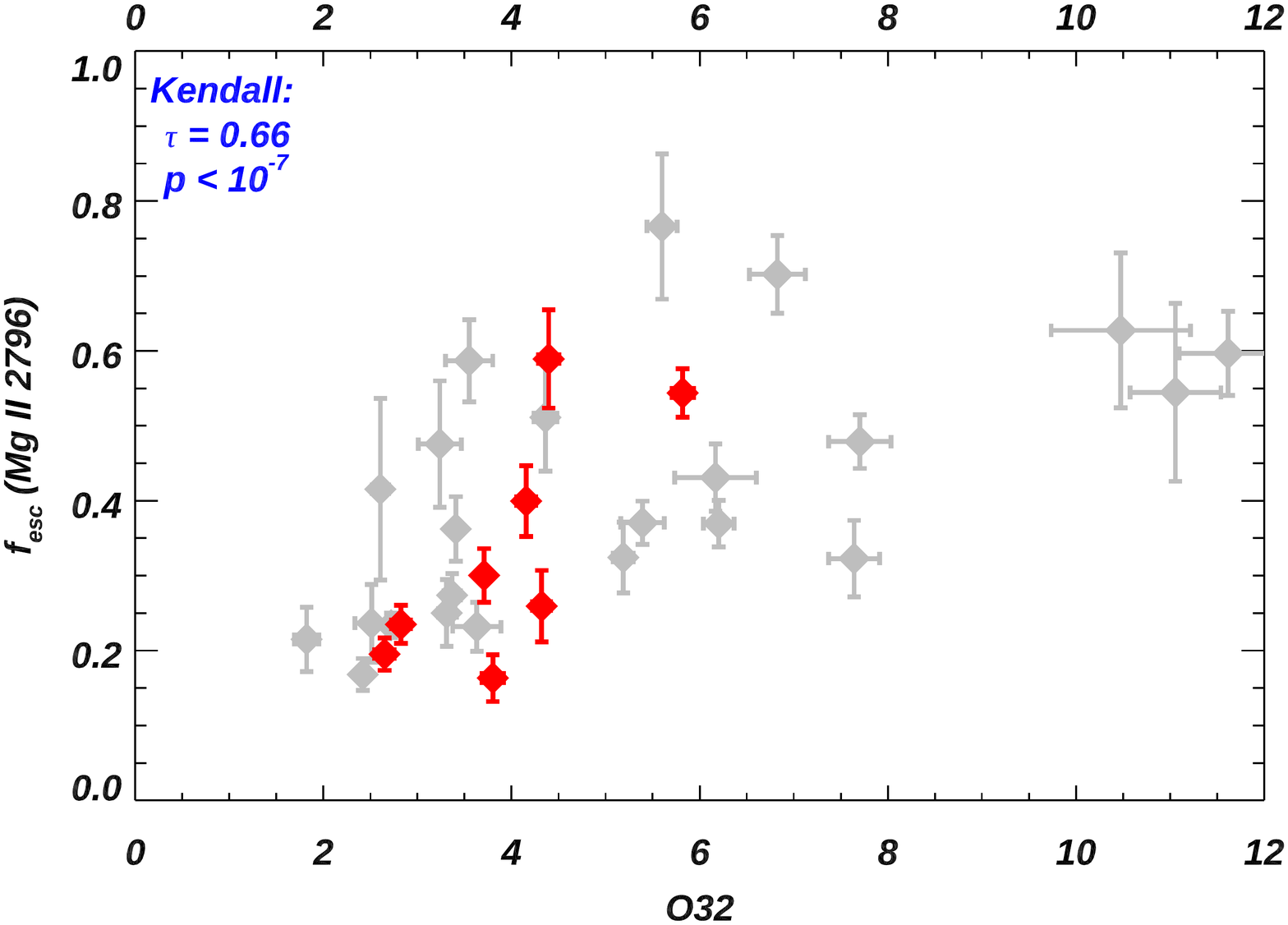}% trim: left lower right upper
	
	%,left if you want to put the figure on the left
	\includegraphics[angle=0,trim={0.2cm 0.5cm 0.1cm 2.2cm},clip=true,width=0.5\linewidth,keepaspectratio]{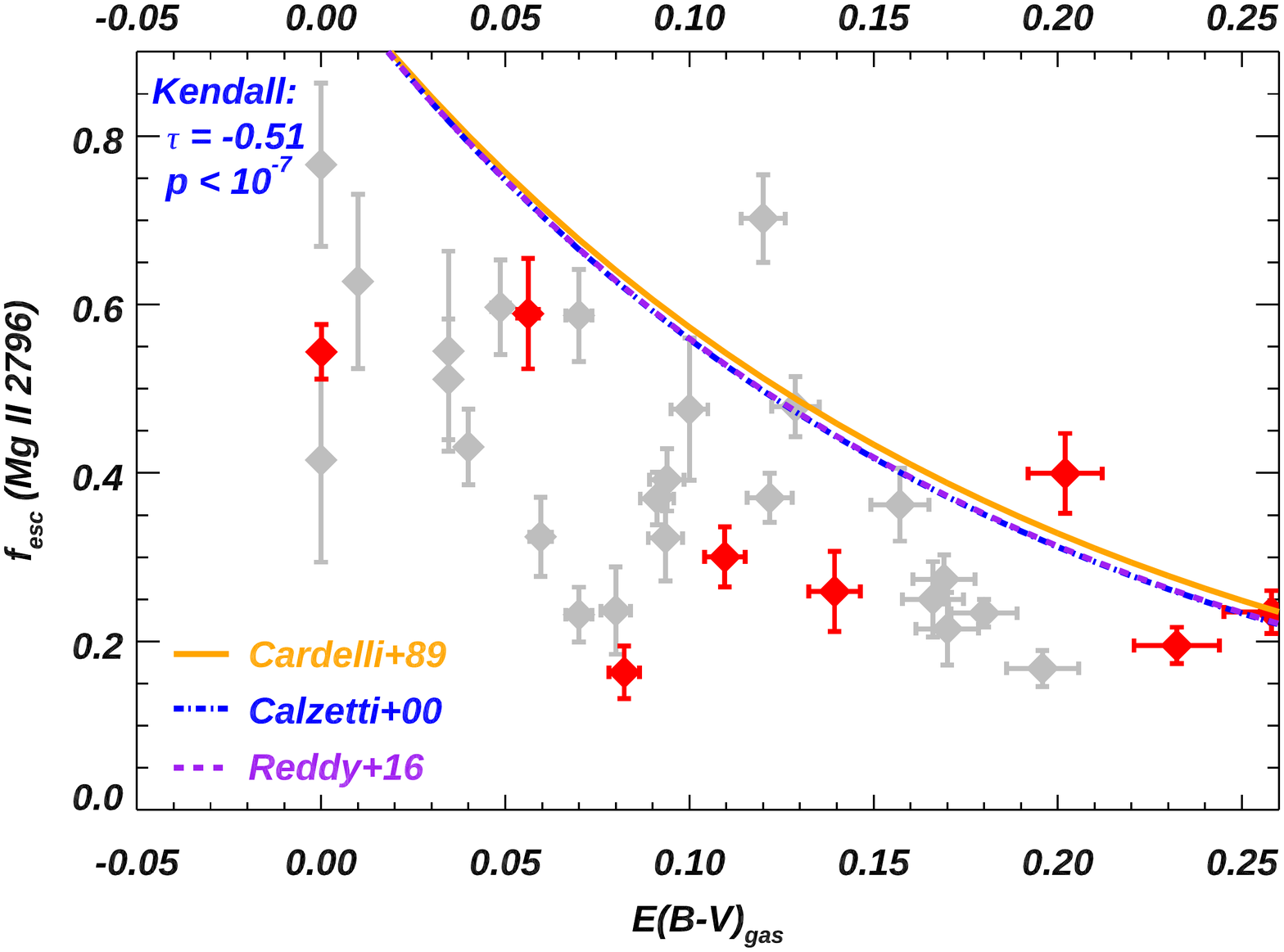}% trim: left lower right upper
	
\caption{\normalfont{Comparisons of \fescMgII\ (in y-axes) with equivalent width of \mgii\ \ly 2796 (\textbf{Top-Left}), O32 (\textbf{Top-Right}), and the internal extinction of the observed galaxy (\textbf{Bottom}). In the bottom panel, the dashed lines represent expectations from dust extinction laws without resonant scattering (note that \cite{Calzetti00} and \cite{Reddy16a} curves mostly overlap). Colors and labels are similar to Figure \ref{fig:MgIITwoCases}. See discussion in Section \ref{sec:CorrMgII}. } }
\label{fig:MgII-3panel}
\end{figure*}

\subsection{Comparison Samples}
\label{sec:comp}
In the remainder of this paper, we consider several comparison samples from the literature which focused on selecting LCE candidates. A subset of these galaxies have \mgii\ emission detected in their optical spectra, including from (1) LzLCS \citep{Flury22a}, (2) Izotov et al. \cite[][]{Izotov16a, Izotov16b, Izotov18a, Izotov18b, Izotov21}, (3) \cite{Henry18}, (4) \cite{Guseva20}, and (5) \cite{Malkan21}. For these samples, since they are not selected by strong \mgii\ emission, we only choose galaxies that have \mgii\ \ly\ly 2796, 2803 both detected with SNR $>$ 3. There are overlaps between the different samples, where the same object was observed in more than one of the studies listed above. In this case, we always adopt the one with higher SNR. Overall, 24 extra LCE candidates from the literature are selected. Basic information and \mgii\ measurements for these galaxies and the eight from the present sample (a total of 32) are 
listed in Tables \ref{tab:SDSS} and \ref{tab:SDSS2}.

%It is worth noting that the LyC leakers from Izotov et al. were selected to include some of the highest O32 objects in the SDSS, so our sample tends to fill out the lower O32, but high Mg II EW range.

%In Figure \ref{fig:EWMgII-O32}, we compare the O32 values with EW(\mgii\ 2796+2803) from our sample and three other samples from the literature that focused on studying \lya\ and/or LyC escape. For other samples, since they are not chosen by \mgii\ features, we only choose their galaxies that have clear \mgii\ spectra. Galaxies from our sample are shown in red. The selected galaxies from \cite[][]{Izotov16a, Izotov16b, Izotov18a, Izotov18b, Izotov21} are shown in gray, which emphasize high O32 ratios. But this selection could miss LyC leakers with low O32 ratios but strong \mgii\ emission (e.g., those galaxies with O32 $\la$ 6). Galaxies from the Low Redshift Lyman Continuum Survey (LzLCS)  and \cite{Henry18} are shown in light blue and black, respectively. These two surveys have EW(\mgii) $\lesssim$ than that of our sample. Overall, our sample highlights the LyC leaker candidates with high EW(\mgii) and relatively low O32 values, which are different from other samples.
%Discuss the overlapped object 1154 and 1442

For galaxies from Izotov et al. and LzLCS samples, optical spectra are from SDSS. To reduce systematic errors, we remeasure the optical lines adopting the same methodology in Sections \ref{sec:obs} and \ref{sec:analysis} to get \fescMgII. For objects in \cite{Henry18} and \cite{Guseva20}, since they have higher SNR optical data, we adopt their line measurements directly. For all comparison samples, we take the reported flux of \lya\ and recalculate \fescLyA\ as discussed in Section \ref{Mea:LyA}. For \fescLyC, we take the reported values from the literature directly, which are commonly based on similar SED fittings as we discussed in Section \ref{sec:SED}.
%(MMT and VLT spectra, respectively)

In all correlation figures later in this section, we calculate the Kendall's $\tau$ coefficients and the probability of a spurious correlation (p values). We show these values at the left-top corner for each figure. In the Kendall test, we have accounted for the upper limits (if any) following \cite{Akritas96}. For a significance level of $\alpha$ = 0.025, the sample size of 8 and 32 require $\tau$ $\gtrsim$ 0.64 and 0.25, respectively, to reject the null hypothesis (i.e., there is no correlation between the x and y values).

%For LzLCS and Izotov et al. samples, since they are not chosen by \mgii\ features, we only compare with their galaxies that have high S/N \mgii\ spectra. This could have prioritized ones with strong \mgii\ lines, while more galaxies in their sample with weaker and low S/N \mgii\ features.

\begin{figure*}
\center
	\includegraphics[angle=0,trim={0.0cm 0.8cm 0.1cm 0.0cm},clip=true,width=0.5\linewidth,keepaspectratio]{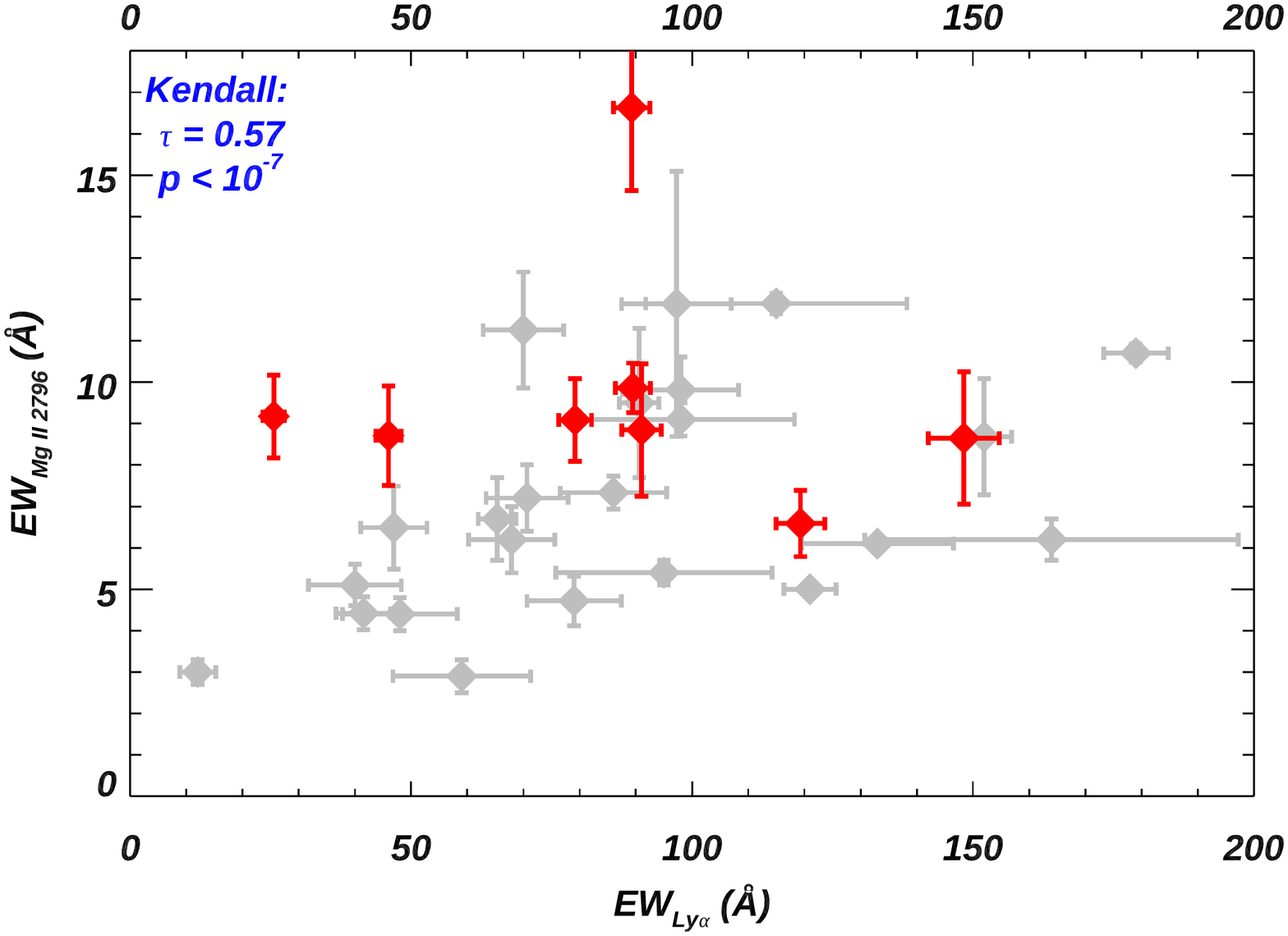}% trim: left lower right upper
	\includegraphics[angle=0,trim={0.0cm 0.8cm 0.1cm 0.0cm},clip=true,width=0.5\linewidth,keepaspectratio]{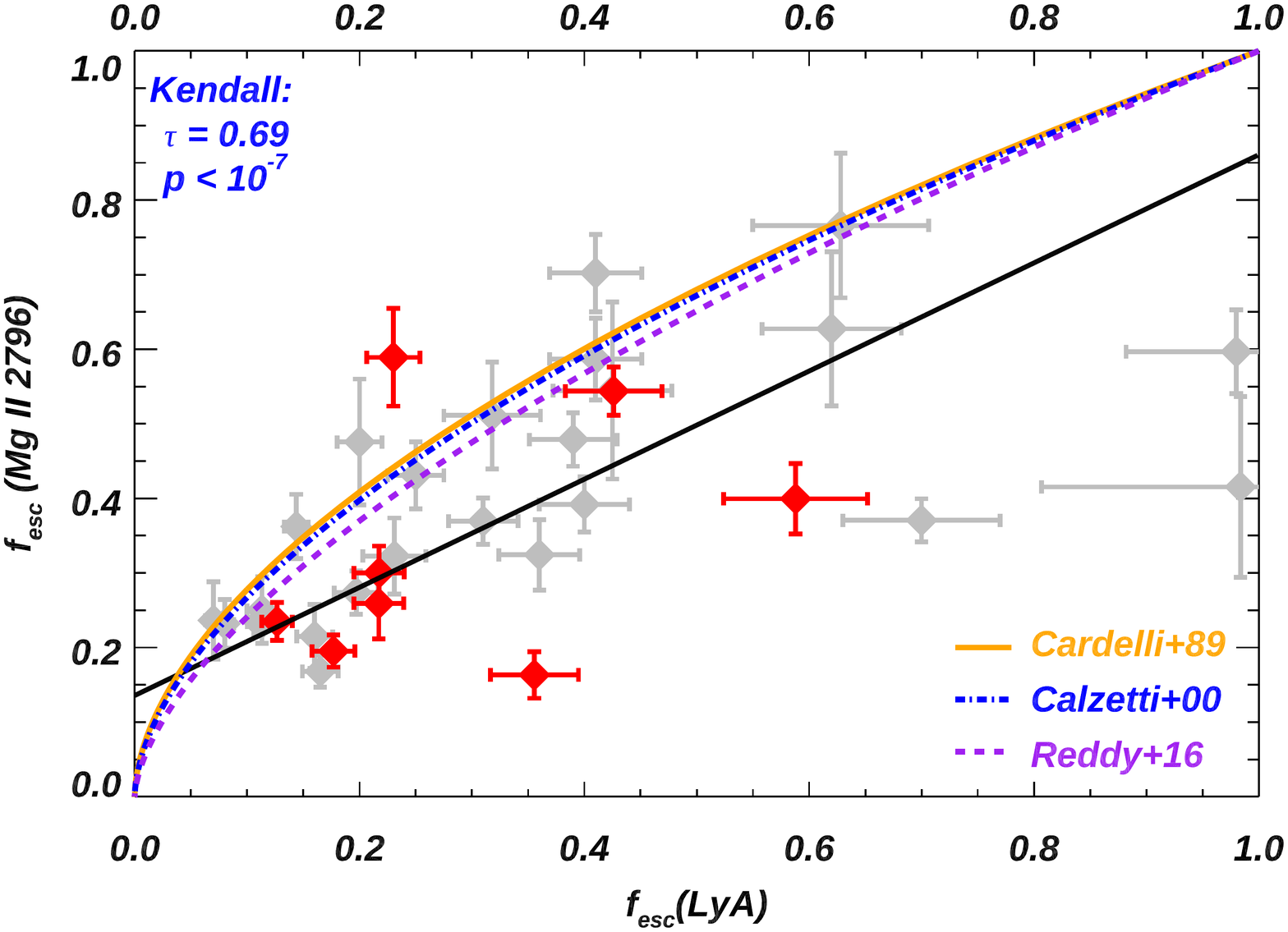}% trim: left lower right upper
	
\caption{\normalfont{Correlations between \mgii\ and \lya\ properties. Colors and labels are similar to Figure \ref{fig:MgIITwoCases}. \textbf{Left:} EW(\mgii\ \ly 2796) and EW(\lya) are positively correlated. \textbf{Right:} The escape fraction of \mgii\ \ly 2796 and \lya\ are tightly correlated. We show the expectations from different dust extinction laws \textit{without} resonant scattering, and their references are at the bottom-right corner. The black line represents best linear fit (\fescMgII\ = 0.725 + 0.136\fescLyA). See discussion in Section \ref{sec:CorrLyA}.} }
\label{fig:MgII-LyA1}
\end{figure*}
%For J0925_Guseva, it is originally from Izotov16c, but I didn't find its EW(LyA) there, so it is skipped in the plot
%, where we have excluded the two possible outliers that have \fescLyA\ $\sim$ 1.0.}

\subsection{Correlations for \fescMgII}
\label{sec:CorrMgII}

We first compare the derived \fescMgII\ from the two limiting cases of DB and IB, and from the two lines of the \mgii\ doublet in Figure \ref{fig:MgIITwoCases}. Our sample is shown in red, while galaxies from literature are shown in grey. In the left panel, we show the ratio of \fescMgII\ from DB to IB in the y-axis, and \fescMgII\ from IB in the x-axis.  We find that the \fescMgII\ values from DB are always slightly higher ($\sim$ 10 -- 20\%) than the ones from IB. This is as expected from Figure \ref{fig:MgII-O32} as follows. For each galaxy, the O32 value is fixed by the observation, and the predicted intrinsic flux of \mgii\ is lower in DB. Therefore, under the same observed flux of \mgii, we expect higher \fescMgII\ for the DB case. 
%The characteristic 1$\sigma$ uncertainties in $\tau$ estimated by bootstrapping are $\sim$ 0.1.

In the right panel of Figure \ref{fig:MgIITwoCases}, we show the \fescMgII\ values from the two lines of the doublet, i.e., \mgii\ \ly 2796 and \ly 2803, assuming DB. The oscillator strength ratio, and therefore the optical depth ratio between these two lines is 2:1.  However, since \mgii\ is a resonant line, this theoretical optical depth ratio between \mgii\ \ly 2796 and \ly 2803 does not directly translate into different escape fractions \citep[e.g.,][]{Henry18, Chisholm20}. Instead, we find that most of the galaxies have \fescMgII(2796) $\sim$ \fescMgII(2803). This suggests that, in these cases, the resonant scattering of \mgii\ is not significant enough to cause \ly 2796 to have a lower escape fraction. In other words, these galaxies' observed \mgii\ photons most likely leak through optically thin paths/channels \citep[possibly holes in ISM, see][]{Chisholm20, Saldana-Lopez22}. For \fescMgII\ $>$ 0.4, the trend in Figure \ref{fig:MgIITwoCases} has more dispersion. We also take caution here since most of the \mgii\ observations for these galaxies are from SDSS spectra, and the SNR of \mgii\ \ly 2803 is always smaller than \mgii\ \ly 2796. Future deeper observations  would shrink the error bars and clarify this trend. For all correlations later in this section, we focus on \mgii\ \ly 2796 from the DB case since 1) the higher SNR line yields more robust relationships, and 2) as discussed above, \mgii\ photons mostly likely escape from DB paths, especially when \fescMgII\ is not high.

%Two possible outliers (with \fescMgII(2796) $>$ 0.8) are J1233+4959 and J1127+4610, but they are still consistent with the trend within error bars.
%cases are more prominent in current observed LyC leakers (see Section \ref{Sec:MgII})

We next compare \fescMgII\ with different observed properties in Figure \ref{fig:MgII-3panel}. These correlations are consistent with the ones reported in \cite{Henry18} (see their Figures 5 and 9), but we show them in a larger sample here. In the top-left panel, we find that the galaxies with higher observed EW(\mgii) may have higher \fescMgII, although the relation shows large scatter. For the same \fescMgII, our galaxies (in red) tend to show stronger EW(\mgii) than the control sample (in grey). One explanation is that the intrinsic flux of \mgii\ is stronger for our galaxies since they were selected as strong \mgii\ emitters. In the top-right panel, we show the positive correlations between \fescMgII\ with O32 values. This may be as expected since \fescMgII\ and O32 are both possible tracers for the escape of LyC. However, caution needs to be taken since \fescMgII\ is also derived from O32. 

In the bottom panel, we present a negative correlation that galaxies with higher internal extinction (derived in Section \ref{sec:MeaSDSS}) tend to have lower \fescMgII. This is consistent with our expectations qualitatively, since the more dust, the fewer photons escape. Furthermore, we show the expectations from various extinction curves assuming no resonant scattering \citep[e.g.,][]{Cardelli89, Calzetti00, Reddy16a}. Since most of the galaxies in our combined sample fall below the extinction curves, we infer that \mgii\ is indeed more affected by dust than non-resonant lines. A similar relationship has also been found between \lya\ and dust \citep[see Figure 2 in][]{Hayes11}. 

The combination of the latter two panels also suggest that O32 may be correlated with the internal extinction. We indeed find a weak negative correlation between them, but there exist large scatters.

%Furthermore, it's also striking that most of the galaxies sit close to and follow a similar trend as the extinction curves. This suggests that, for these galaxies, \mgii\ suffer from non-zero but only small amount of dust. We reach similar conclusions in the next few subsections from different perspectives. 

%Applying extinction corrections to resonant lines such as \mgii\ and \lya\ is not straightforward. Instead of traveling straight out of the galaxy, emitted photons scatter within the galaxy some number of times before emerging.  This increases the probability of resonant photons experiencing a dust grain.  
%For a homogeneous distribution of gas/dust, this can be quantified using the Neufeld 1990 solution (eq 4.33, see also e.g. Verhamme+2006), but indeed it's probably beyond the scope of this paper.

%The two galaxies that are off from the trend are J1248+4259 and J1256+4509 (with O32 $>$ 13, from left to right).
%Later in this section, we mainly present correlations with \fescMgII(2796) since \mgii\ \ly 2796 is always strong than \mgii\ \ly 2803 and the results are more certain.

\subsection{Correlations for \fescLyA}
\label{sec:CorrLyA}

Since both \lya\ and \mgii\ are scattered by low-ionization gas while they escape the galaxy, a direct correlation between their escape fractions has been suggested by previous publication \citep{Henry18}. We verify this correlation in our larger sample in Figure \ref{fig:MgII-LyA1}. The left panel shows the positive correlation between EW(\mgii\ \ly 2796) and EW(\lya) but with moderate scattering. The right panel presents that \fescMgII\ and \fescLyA\ are tightly correlated, given the possibility of spurious correlations, $p$ $\lesssim$ 10$^{-3}$. The best fitting linear correlation is:
\begin{equation}\label{eq:MgII-LyA}
\begin{aligned} 
    f_\text{esc}^\text{Mg II} & = a + b \times f_\text{esc}^{\text{Ly}\alpha}\\
    a & = 0.136\pm0.05\\
    b & = 0.725\pm0.10
\end{aligned}
\end{equation}
%\textbf{where we have excluded the two possible outliers that have \fescLyA\ $\sim$ 1.0 with \fescMgII\ only $\sim$ 0.5.}

%May consider move this part to section 4.2.1
Our correlation in Equation (\ref{eq:MgII-LyA}) is similar to Equation (5) in \cite{Henry18}.   The \fescMgII\ and \fescLyA\ values are of the same order. This supports a scenario where \mgii\ and \lya\  mainly escape from optically thin (or DB) holes in ISM \citep[e.g.,][]{Gazagnes18, Chisholm20, Saldana-Lopez22}. In this case, \mgii\ and \lya\ photons travel through similar path lengths (likely in a single flight) when they escape from the galaxy. Therefore, the excess dust extinction due to resonant scattering is similar for both lines. Note that this does not contradict the point that \mgii\ and \lya\ still suffer more from dust extinction than non-resonant lines (Section \ref{sec:CorrMgII}). One explanation is that \mgii\ and \lya\ photons still undergo some (few) scatterings, thus, extending their path lengths and the susceptibility to dust destruction.

%is that the optically thin channels still have small (but non-zero) amounts of dust. 

%Since \lya\ and \mgii\ transitions have optical depths to resonant scattering that differs by orders of magnitude, one would expect \fescLyA/\fescMgII\ $<$ 1 if \lya\ and \mgii\ are optically thick and is affected strongly by resonant scattering and dust (so \lya\ would be affected more). Our correlation in Equation \ref{eq:MgII-LyA} [see also Equation (5) in \cite{Henry18}] yield that the \fescMgII\ and \fescLyA\ values are of the same order.
%(see the clumpy geometry discussed in Section \ref{sec:MgIITwoCases} above)

The optically thin \mgii\ emission line was also presented in \cite{Chisholm20}. They obtained Keck Cosmic Web Imager (KCWI) spatially-resolved \mgii\ maps for one LyC leaker (J1503+3644). From the spatially resolved ratio of the high SNR \mgii\ doublet lines, they conclude that \mgii\ is indeed optically thin in most regions. Similarly for our galaxies, future higher resolution, higher SNR spectra would reveal whether our observed \mgii\ emission lines are optically thin, and determine if the \mgii\ lines are broadened by resonant scattering. Overall, the correlation between \fescLyA\ and \fescMgII\ can add key constraints to the mechanisms that allow \mgii, \lya, and LyC to escape. 

Since \fescLyA\ traces the \hi\ column density and \fescMgII\ traces the \mgii\ column density, one may expect that the residuals in the right panel of Figure \ref{fig:MgII-LyA1} are related to the gas-phase metallicity. However, we have tested this hypothesis and there exist large scatter. Given the current sample size and scatter, we could not confirm or rule out the hypothesised relationship.

%and find no clear trends between the residuals and the gas-phase metallicity.
%by color-coding the figure by the metallicity that we derive in Section \ref{sec:MeaSDSS}.

%If \mgii\ is mainly escaped by scenario 1), both \lya\ and \mgii\ would be resonant scattered on their way out of the galaxy, but by different extent. This is because \lya\ and \mgii\ transitions have optical depths to resonant scattering that differ by orders of magnitudes. Thus, one would expect \fescLyA/\fescMgII\ $<<$ 1 given scenario 1). For scenario 2),

\iffalse
\begin{figure*}
\center
	\includegraphics[angle=0,trim={0.0cm 0.8cm 0.1cm 0.0cm},clip=true,width=0.5\linewidth,keepaspectratio]{./Figures/Sep_LyA_vs_f_esc_Hb.pdf}% trim: left lower right upper
	\includegraphics[angle=0,trim={0.0cm 0.8cm 0.1cm 0.0cm},clip=true,width=0.5\linewidth,keepaspectratio]{./Figures/Sep_LyA_vs_f_esc_BPASS.pdf}% trim: left lower right upper
	
\caption{\normalfont{Correlations between the peak separation of \lya\ profiles and \fescLyA\ or \fescMgII. Colors and labels are similar to Figure \ref{fig:MgIITwoCases}. There are anti-correlations in both panels (see discussion in Section \ref{sec:CorrLyA}).} }
\label{fig:MgII-LyA2}
\end{figure*}
\fi

\begin{figure*}
\center
	\includegraphics[angle=0,trim={0.0cm 0.8cm 0.1cm 0.0cm},clip=true,width=0.5\linewidth,keepaspectratio]{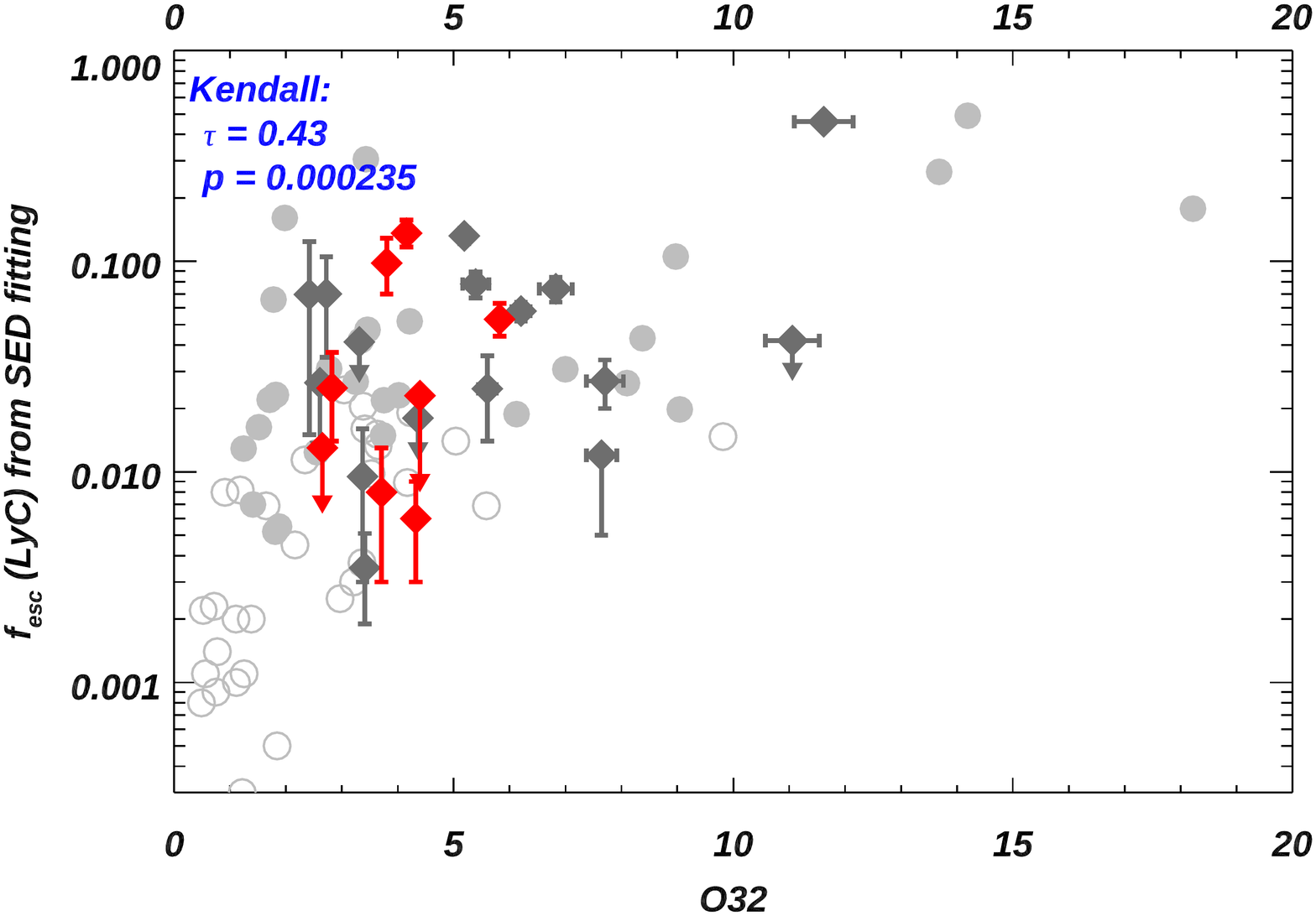}% trim: left lower right upper
	\includegraphics[angle=0,trim={0.0cm 0.8cm 0.1cm 0.0cm},clip=true,width=0.5\linewidth,keepaspectratio]{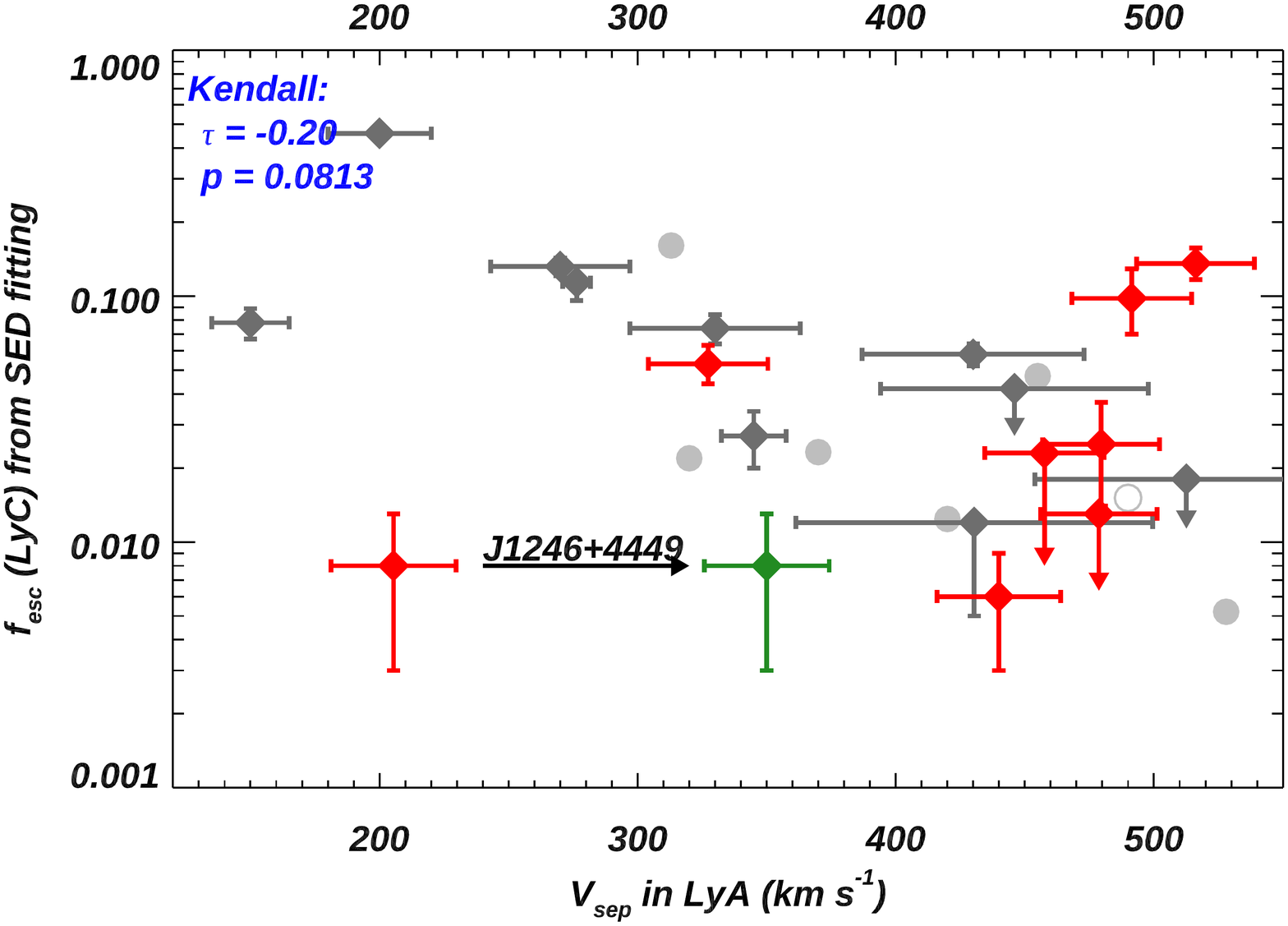}% trim: left lower right upper

    %,left	
	\includegraphics[angle=0,trim={0.0cm 0.8cm 0.1cm 0.0cm},clip=true,width=0.5\linewidth,keepaspectratio]{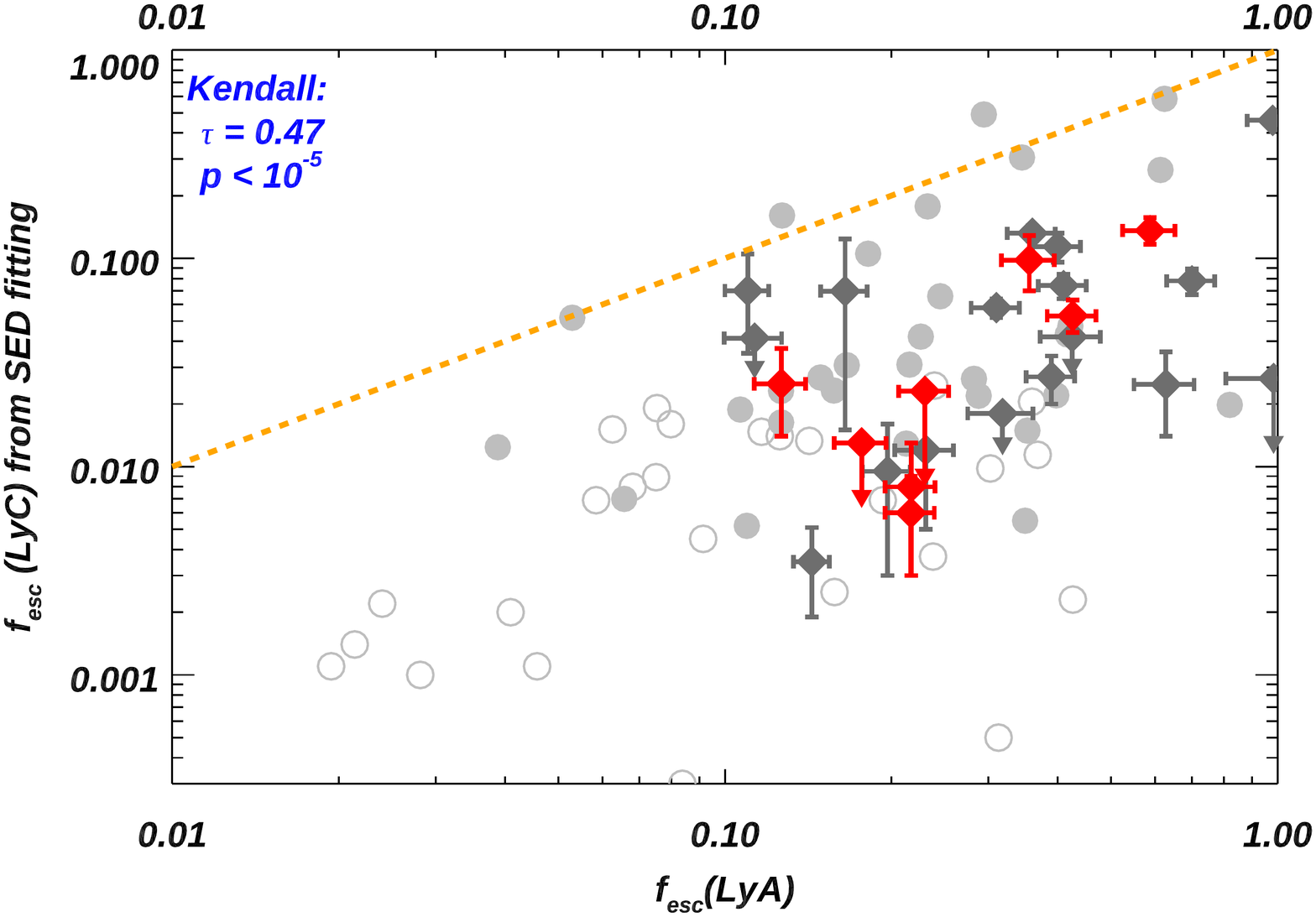}% trim: left lower right upper

\caption{\normalfont{Relationships between the escape fraction of Lyman continuum (\fescLyC, in y-axes) and three possible indirect indicators (x-axes): 1) \fescLyC\ vs O32, 2) \fescLyC\ vs the peak separation of \lya\ emission line (\vsep), and 3) \fescLyC\ vs the escape fraction of \lya. Galaxies from our \mgii\ selected sample are shown in red diamonds. Galaxies from comparison samples with high SNR \mgii\ detections \citep[][and LzLCS]{Izotov16a, Izotov18a,Izotov18b, Izotov21, Guseva20} are shown in dark gray diamonds. We also overlay other galaxies \textit{without} high SNR \mgii\ detections from the LzLCS sample. Their confirmed LyC leakers and non-leakers are shown in filled and hollow circles, respectively. In the right panel, galaxy J1246+4449 have two different \vsep\ (shown in red and green) due to its triple-peak \lya\ feature (see Section \ref{sec:CorrLyC}). In the bottom panel, we show where y = x as the dashed orange lines. } }
\label{fig:LyC}
\end{figure*}

\begin{figure}
\center

	\includegraphics[angle=0,trim={0.0cm 0.8cm 0.0cm 0.0cm},clip=true,width=1.0\linewidth,keepaspectratio]{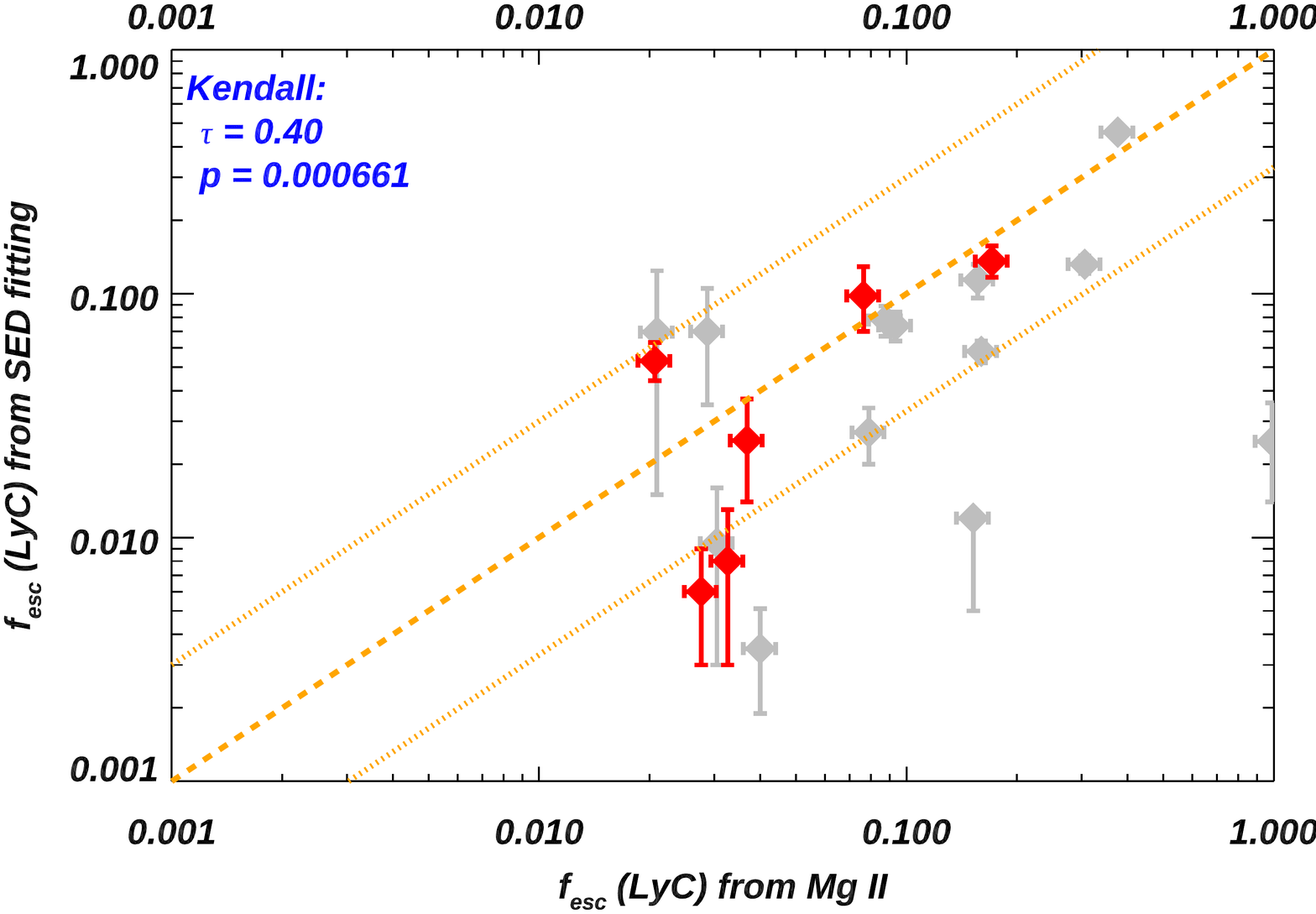}% trim: left lower right upper
	
\caption{\normalfont{Comparisons of measured \fescLyC\ with the predicted one from \mgii\ \ly 2796 emission lines and dust extinction (Section \ref{sec:pred}). Galaxies from our \mgii\ selected sample are shown in red diamonds, while galaxies from comparison samples with high SNR \mgii\ detections \citep[][and LzLCS]{Izotov16a, Izotov18a,Izotov18b, Izotov21, Guseva20} are shown in gray diamonds. We show where y = x as the dashed orange lines and y = 1/3x and 3x as the dotted orange lines. There is a positive 1:1 correlation, and the root-mean-square-error is $\sim$ 0.05. See discussion in Section \ref{sec:pred}. } }
\label{fig:LyCFromMgII}
\end{figure}

\subsection{Correlations for \fescLyC}
\label{sec:CorrLyC}
Direct detections of LyC flux for high redshift galaxies are difficult due to neutral IGM attenuation. Therefore, exploring and verifying indirect indicators from local galaxies is essential to interpret the growing samples of EoR galaxies that will be detected with JWST and the upcoming Extremely Large Telescope (ELT). A detailed discussion of all known indirect indicators in published LCE candidates has been presented in the Low-redshift Lyman Continuum Survey \citep[LzLCS, see][]{Flury22b}. In this subsection, we aim to contrast galaxies with high SNR \mgii\ emission lines to all of the other LCEs (i.e., LyC leakers) and non-LCEs in the LzLCS. This survey contains 35 LCEs, but the majority have low SNR ($<$ 3) detections of \mgii\ emission or completely lack spectroscopic coverage at the necessary blue wavelengths ($\sim$ 2800\angstrom\ in the rest-frame).

%We also present how to predict \fescLyC\ from the observed \mgii\ emission lines and argue \mgii\ is another good indirect indicator.

%In Figure \ref{fig:LyC}, we compare the measured \fescLyC\ to four possible indirect indicators in our sample, and we discuss the implications below. When possible, we contrast galaxies selected by high SNR \mgii\ emission to all other galaxies in LzLCS. The latter is not selected by \mgii\ emission, and have \mgii\ detected with SNR $<$ 3. Since most of the objects from \cite{Henry18} do not have direct LyC observations, we omit those galaxies in these figures.

In Figure \ref{fig:LyC}, we compare the measured \fescLyC\ to three possible indirect indicators, and we discuss the implications below. In all panels in Figure \ref{fig:LyC}, our 8 galaxies selected by strong \mgii\ emission are shown in red diamonds, while the other galaxies that have high SNR \mgii\ emission are shown in dark gray diamonds (see Section \ref{sec:comp}). In all panels, we also show galaxies from LzLCS that have low SNR detections or non-detections of \mgii\ emission in gray circles. The open and filled circles indicate the LyC non-leakers and leakers, respectively. We have omitted the error bars for LzLCS samples to avoid crowding, but their error bars are comparable to the \mgii\ selected galaxies.

%In the left two panels, we contrast the galaxies that have high SNR \mgii\ emission to galaxies in LzLCS that have .

The correlation between \fescLyC\ and O32 is shown in the first panel of Figure \ref{fig:LyC}. A high O32 value has been proposed to be an indirect indicator of escaping LyC photons \citep[e.g.,][]{Jaskot13, Nakajima14, Izotov21,Flury22b}. It is evident that \fescLyC\ increases with rising O32 values, and our \mgii\ selected galaxies (in red) fall into the same trend of all other galaxies. Consistent with previous publications \citep[e.g.,][]{Izotov21,Flury22b}, there still exists substantial scatter, which can be due to the dependence of O32 on other galaxy properties, e.g., metallicity, ionization, and geometry. Note our 8 galaxies show relatively low O32 values because they were not selected to have high O32. \cite{Izotov21} also found in their sample very low \fescLyC\ ($<$ 0.01) for galaxies with O32 $<$ 4. However, in our combined sample, there are $\sim$ 17 galaxies with O32 $<$ 4 and \fescLyC\ $>$ 0.01. This subset is comprised of our eight \mgii\ selected galaxies (red), along with nine from the LzLCS  (gray circles). It is clear that for galaxies with 2 $<$ O32 $<$ 4, a moderate amount of ionizing photons (1 -- 10\%) can still escape from the galaxies. 
%Overall, O32 correlates strong with \fescLyC\ and \mgii\ selected galaxies fall into the same trend as others in LzLCS.

Another indirect indicator proposed in the literature is the profile of \lya\ emission line. The peak separation (\vsep) of the \lya\ profile has been suggested to be one of the better proxies since it is sensitive to HI along the line of sight and can be directly measured from moderate to high-resolution spectra \citep[e.g., ][]{Dijkstra16, Verhamme17, Izotov18b, Gazagnes20, Kakiichi21}. In the second panel of Figure \ref{fig:LyC}, we show the correlation between \fescLyC\ and \vsep. From the LzLCS sample, only seven galaxies have the HST/COS G160M observations needed for accurate measurements of \vsep\ (gray circles). While most of the galaxies follow the negative trend that has been suggested by previous publications, e.g., \cite{Verhamme17,Izotov18b,Izotov21,Flury22b}, a few of our objects are off. The largest deviation is for J1246+4449 at \vsep\ $\sim$ 205 km s$^{-1}$. This could be because there are at least 3 peaks observed in J1246+4449's \lya\ profiles (see Figure \ref{fig:SEDFitsLyA1}), instead of the commonly observed blue+red peaks patterns. The two red peaks (at v $>$ 0) in J1246+4449 corresponds to \vsep\ $\sim$ 205 and 350 km s$^{-1}$, separately. We therefore show the separation between the blue peak and each of these red peaks as red and green diamonds at \fescLyC\ = 0.009. We find the wider \vsep\ (green) match better the negative trend between \fescLyC\ and \vsep. Similarly, other galaxies that may have $>$ 2 peaks in their \lya\ profiles are J0105+2349 and J1219+4814 (see Figure \ref{fig:SEDFitsLyA1}).

This multiple-peak \lya\ feature may be due to the two star-forming clumps in J1246+4449 (see its NUV image in Figure \ref{fig:acq}). In this case, its blue peak should also have two sub-peaks. However, it is unclear due to the low SNR on the blue peak in the COS spectrum. Overall, we caution that, in these cases, \vsep\ can be ill-defined and may affect the resulting trend with \fescLyC\ [see similar cases in J1243+4646 in \cite{Izotov18b}; and \cite{Rivera-Thorsen19, Naidu21}].

For all galaxies, we also check their kinematics in optical emission lines. We find they all show Gaussian-like profiles and only have one velocity center peaked at $v$ = 0. Thus, there is no evidence that the observed optical emission line flux is from only one of the clumps. But we also note that the spectral resolution in SDSS may be insufficient to check the detailed kinematics.

%For example, J1246+4449's red peak ($v$ $>$ 0) is split into two sub-peaks given a gap at +200 km s$^{-1}$, while the blue peak may also have two sub-peaks but is unclear due to the lower SNR. This could be because there are two star-forming clumps in this galaxy (see Figure \ref{fig:acq}). Similar multiple-peak \lya\ features have been discussed in \cite{Naidu21}, where they also conclude that \vsep\ is difficult to define and may affect the resulting trend with \fescLyC.

%These include (from left to right) J1246+4449 J0152--0431, J1105+5947 with \vsep\ $\sim$ --205, --490, --520 km s$^{-1}$, respectively.

%are omitted since they only have low resolution COS/G140L observations so that their \vsep\ are difficult to constrain (\Fluryab). 
%Future higher SNR data may reveal if there are indeed four peaks in these kind of \lya\ profile.

%We may need to take caution since some times the LyA show 3 peaks, need to discuss with ALaina.

In the third panel of Figure \ref{fig:LyC}, we compare \fescLyC\ with \fescLyA.  Our galaxies selected by \mgii\ emission follow the same trend as other galaxies from LzLCS. The positive trend with moderate scatter is consistent with previous publications \citep[e.g.,][]{Flury22b}.

%The orange line indicates where \fescLyC\ equals \fescLyA, and almost all galaxies are below this line. This is as expected since LyC photons are destroyed by neutral hydrogen while \lya\ photons are only be scattered and may still escape the galaxies (if not destroyed by dust). 

\subsection{Predicting \fescLyC\ using \mgii\ Emission Lines}
\label{sec:pred}
As shown in Section \ref{sec:MgIITwoCases} and Figure \ref{fig:zoneplot}, given similar ionization potentials, N(\mgii) can be used to trace N(\hi) in a large range from density-bound to nearly ionization-bound regions. Then in Section \ref{sec:MgIIFit}, we showed how photoionization models can be used to infer \fescMgII\ in both DB- and IB-regions. In this subsection, we aim to connect the observed \mgii\ emission and measured \fescMgII\ to the escape of LyC. We adopt the methodology that was previously discussed in \cite{Chisholm20}, and we briefly summarize it as follows.

%In the last panel of Figure \ref{fig:LyC}, we compare the measured \fescLyC\ with predicted ones adopting the same methodology described in \cite{Chisholm20} as follows. 

\mgii\ (or \lya\ and LyC) escape in the two scenarios discussed in Section \ref{Sec:MgII} can be generalized in to one partial-covering geometry:

\begin{equation}\label{eq:clumpy}
\begin{aligned} 
    f_\text{esc}(\text{Mg II}) = \frac{F_\text{obs}}{F_\text{int}} = (1-C_{f})e^{-\tau_\text{thin}} + C_{f}e^{-\tau_\text{thick}}
\end{aligned} 
\end{equation}
where F$_\text{obs}$ and F$_\text{int}$ are the observed and intrinsic flux of \mgii, respectively; \CF\ and 1 -- \CF\ are the covering fractions for the IB and DB paths, respectively; and \Tthick\ and \Tthin\ are the optical depths for \mgii\ at IB (optically thick) and DB (optically thin) paths, respectively. This equation is valid for both \mgii\ \ly 2796 and \ly 2803, where their optical depths ratios are related as $\tau_{2796}$ = 2$\tau_{2803}$, given the ratio of their oscillator strengths.

For IB paths, we have \Tthick\ $\gg$ 1, so photons cannot escape. Therefore, we can simplify Equation (\ref{eq:clumpy}), and for \mgii\ \ly2796:

\begin{equation}\label{eq:clumpy2}
\begin{aligned} 
    f_\text{esc}(\text{Mg II 2796}) = (1-C_{f})e^{-\tau_\text{2796,thin}}
%\fescMgII\ = (1 --C)$_{f}$e$^{\tau_\text{thin}}$, 
\end{aligned} 
\end{equation}
where we assume there is no dust in the DB paths (we discuss the effects of dust in Section \ref{sec:dust}). \cite{Chisholm20} shows that one can estimate \Tthin\ from doublet line ratios in this geometry [see their Equations (19) and (20)] as:

\begin{equation}\label{eq:MgIItau}
\begin{aligned} 
    \frac{F_\text{2796,obs}}{F_\text{2803,obs}} = 2e^{-\tau_{2803,thin}} = 2e^{-0.5 \times \tau_{2796,thin}}
\end{aligned} 
\end{equation}
where $\tau_{2796,thin}$ and $\tau_{2803,thin}$ are the optical depths for \mgii\ \ly 2796 and \ly 2803 at the optically thin (DB) paths, respectively. For our 8 galaxies, since their observed F$_\text{2796,obs}$/F$_\text{2803,obs}$ $\geq$ 1.0 (see Figure \ref{fig:SEDFitsLyA1}), we get $\tau_{2803,thin}$ $<$ --ln(0.5) = 0.7 or $\tau_{2796,thin}$ $< $ 1.4. This means that all of our galaxies have (nearly) optically thin \mgii\ lines.

From the photoionization solutions discussed in Section \ref{sec:MgIIFit}, we have presented how to estimate \fesc(\mgii\ 2796) given measured optical line flux. After that, we can solve the remaining two unknowns (\CF\ and $\tau_{2796,thin}$) from two equations (6) and (7) given the observed flux of \mgii\ \ly2796 and \ly2803 lines.

In Section \ref{sec:MgIITwoCases}, we have presented that N(\mgii) can be used to trace N(\hi) in wide range of conditions from DB to IB. Therefore, we get:

\begin{equation}\label{eq:NHI}
\begin{aligned} 
    N(\text{Mg II}) &= \frac{3.8\times10^{14}}{f_{2796}\times 2796.35} \times \tau_\text{2796,thin}\ (\text{cm}^{-2})\\
    N(\text{H I})   &=  \alpha \times N(\text{Mg II}) %\frac{\text{H}}{\text{Mg}}\frac{1}{\delta_\text{Mg}} \times N(\text{Mg II})
\end{aligned} 
\end{equation}
where f$_{2796}$ = 0.608 is the oscillator strength of \mgii\ \ly 2796, and $\alpha$ = N(\mgii)/N(\hi) is the column density ratios predicted from CLOUDY models in Section \ref{sec:MgIITwoCases}. For each galaxy, we have adopted their observed gas-phase metallicity and O32 to find the best-fit CLOUDY model. Finally, given the derived N(\hi) from Equation (\ref{eq:NHI}), we can predict the absolute escape fraction of LyC (i.e., \fescLyCPred) as:
%$\text{H}$/$\text{Mg}$ is the abundance ratio of hydrogen to magnisium, and $\delta_\text{Mg}$ is the amount of Mg that depleted into dust \citep[$\sim$ 27\% from][]{Jenkins09}. We have also ignored N(\mgi), given the fact that we do not detect obvious \mgi\ \ly 2853 emission line in any of our galaxies.

\begin{equation}\label{eq:LyC}
\begin{aligned} 
    f^\text{LyC}_\text{esc,pd} = e^{-N(\text{H I})\sigma_{ph}} \times 10^{-0.4E(B-V)k(912)}
\end{aligned} 
\end{equation}
where the first term represents the absorption of LyC photons by neutral hydrogen, and the second term stands for the stellar extinction by dust. $\sigma_{ph}$ is the photoionization cross section of \hi\ at 912\angstrom, and $E(B-V)$k(912) = A(912), which is the total magnitude of extinction at 912\angstrom. From extinction laws in \cite{Cardelli89, Calzetti00, Reddy16b}, k(912) = 21.32, 16.62, 12.87, respectively. We adopt k(912) from \cite{Calzetti00} extinction law to be consistent with our assumptions in the SED fittings in Section \ref{sec:SED}.
%Note that the reason equation 9 has not CF term is because we assume optically thick paths don't have LyC escape.

\begin{figure*}

	\includegraphics[angle=0,trim={0.0cm 0.8cm 0.1cm 0.0cm},clip=true,width=0.5\linewidth,keepaspectratio]{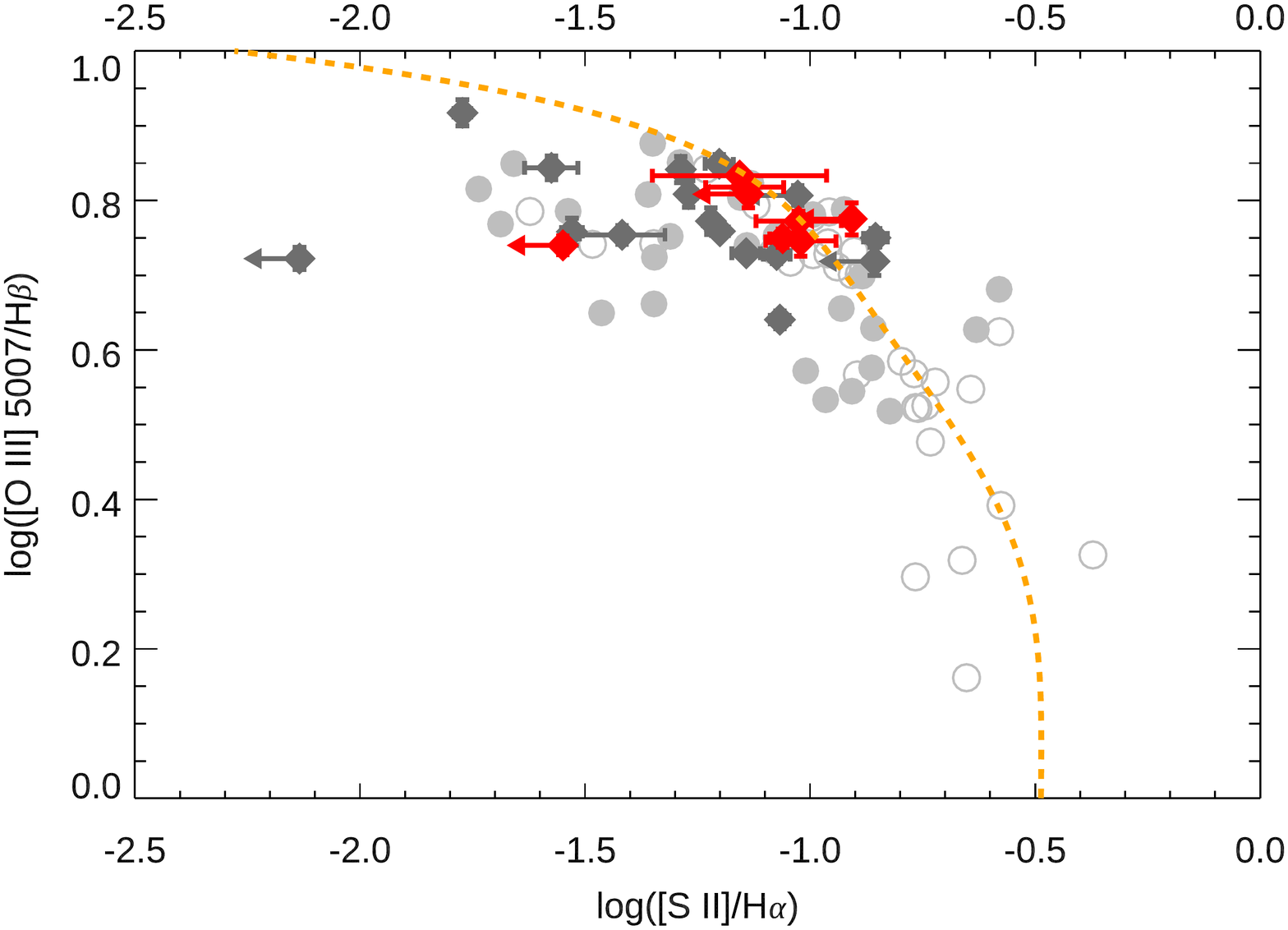}% trim: left lower right upper
	\includegraphics[angle=0,trim={0.0cm 0.8cm 0.1cm 0.0cm},clip=true,width=0.5\linewidth,keepaspectratio]{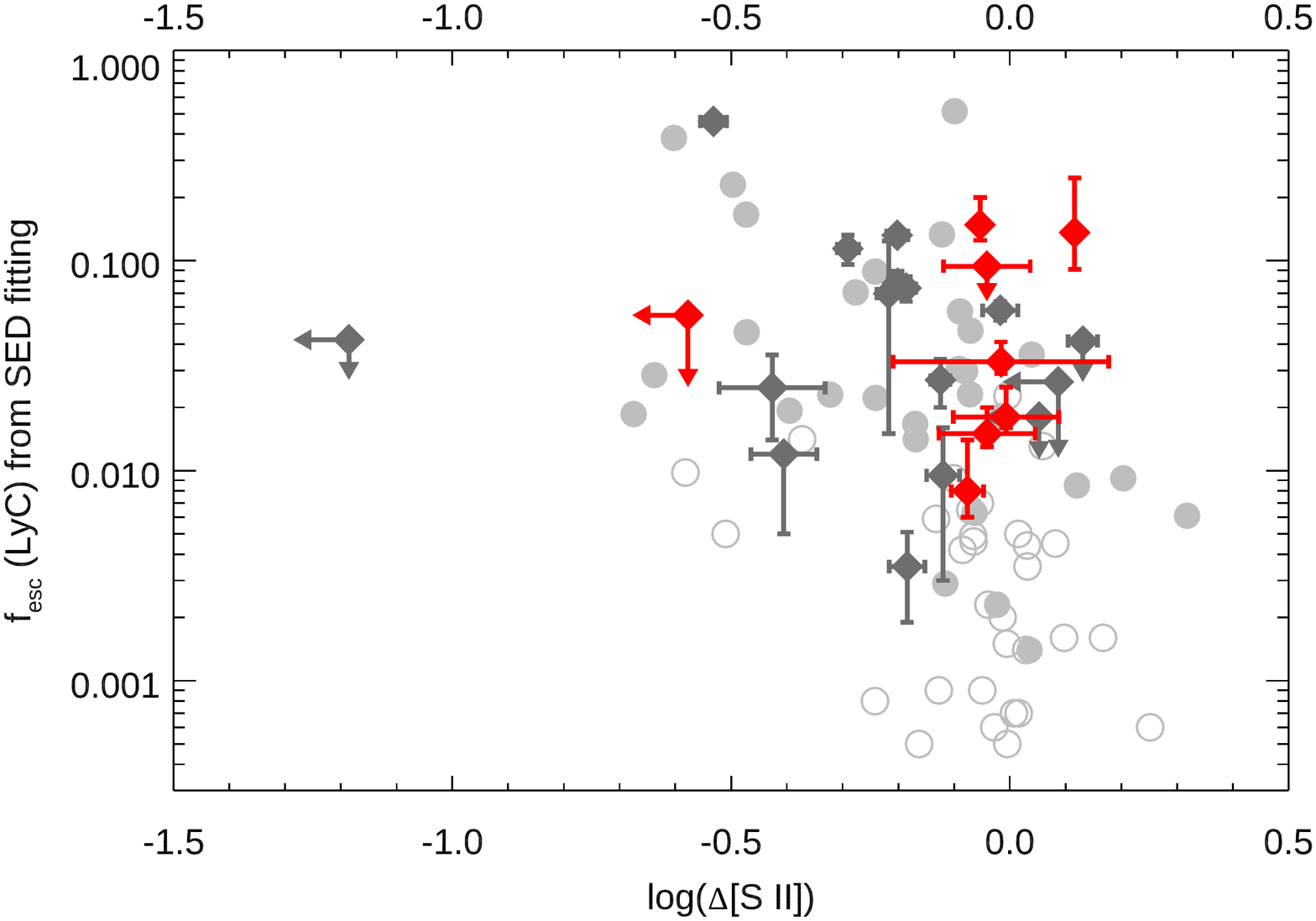}% trim: left lower right upper
	
\caption{\normalfont{Correlations involving the deficiency of [\sii] $\lambda \lambda 6716, 6731$. The colors and labels are the same as Figure \ref{fig:LyC}. \textbf{Left:} We compare the flux ratios between [\oiii] \ly 5007/\hb\ and [\sii] \ly\ly 6716, 6731/\ha. We show the fitted SDSS locus from \cite{Wang19} as the orange line. \textbf{Right:} We compare the \fescLyC\ with [\sii] deficiency. The latter is defined as the displacement of the measured log([\sii]/\ha) from the orange line in the left panel \citep{Wang19, Wang21}. See discussion in Section \ref{sec:SII}. } }
\label{fig:SII}
\end{figure*}

Given high enough SNR data on the required optical lines (i.e., \mgii\ \ly\ly 2796, 2803, [\oii] \ly\ 3727, and [\oiii] \ly 5007), one can solve Equations (\ref{eq:clumpy2}) -- (\ref{eq:LyC}) to get \fescLyCPred. However, it is difficult to do so in our 8 galaxies and most of the other comparison galaxies discussed in Section \ref{sec:comp}. This is because, for most of these galaxies, their only optical spectra are from SDSS, where the observed \mgii\ has relatively low SNR, especially insufficient to solve Equation (\ref{eq:MgIItau}). Therefore, we assume \CF\ = 0.0, i.e., the \mgii\ photons all escapes from DB paths. This was also adopted in \cite{Chisholm20}, where they confirmed it in one LCE, J1503+3644. Given this assumption, we can directly solve ${\tau_{2796,thin}}$ from Equation (\ref{eq:clumpy2}), which then leads to N$_\text{HI}$ and \fescLyCPred\ in Equations (\ref{eq:NHI}) and (\ref{eq:LyC}), respectively.

In Figure \ref{fig:LyCFromMgII}, we show  the measured \fescLyC\ derived from SED fitting versus \fescLyCPred\ predicted from \mgii. We find a positive 1:1 correlation which is consistent with the results in \cite{Chisholm20}. The root-mean-square-error (RMSE) between the predicted and measured \fescLyC\ is $\sim$ 0.05. This suggests that the \mgii\ emission can be used to infer the LyC escape fraction when \fescLyC\ is large ($>$ 5\%). The scatter and modorate RMSE could be because our assumption of CF = 0.0 is not valid for some of these galaxies, especially when \fescLyC\ is small. Those galaxies with very small \fescLyC\ could have larger coverage of IB paths (\CF\ $\to$ 1) instead of DB paths (\CF\ $\to$ 0). 
 
We caution that the derived \fescLyCPred\ in this way should only be considered rough estimates. In the future, we can reach more robust estimates of \fescLyCPred\ once we have higher SNR optical spectra for these galaxies (Xu et al. 2022b, in prep.).

  %For the majority of galaxies, \fescLyCPred\ values are within a factor of 3 of the measured ones.
  
%\cite{Chisholm20} has proposed an approach (see their Equation 22) to estimate CF using the photoionization solutions (see Section \ref{Sec:MgII}) and \mgii\ doublet line ratios. But this approach requires high SNR data of both lines of \mgii\ doublet, where SDSS spectra are usually insufficient. This necessties the needs for higher SNR data for these objects in order to constrain a tigher correlation in predicting \fescLyC\ using \mgii\ emissions.

%Try to limit the S/N on MgII objects and replot.

%We do not compare correlations with MgII 2803 since most of our galaxies's optical spectra is from SDSS and MgII 2803 is weaker with lower S/N.

%We only compare with DB, and DB = 1.0 dex before front, IB = 0.1 dex after the front.

\section{Discussion}
\label{sec:discuss}
%\subsection{Limitations of Prospector fittings}
%1. The BPASS models we have made available within FSPS are the “-bin-imf135all 100” models, i.e., they assume a Salpeter IMF with an upper mass cutoff of 100M⊙. This IMF cannot be changed in Prospector fittings, which use FSPS package.
%may not be useful anymore.

\subsection{A High Detection Rate of LCEs in \mgii\ Emitters}
\label{sec:Rate}
As shown in \cite{Flury22b}, the detection rate of LCEs rises when the observed O32 increases. They consider galaxies with known LyC observations from various samples including LzLCS, \cite{Izotov16a,Izotov16b,Izotov18a,Izotov18b,Izotov21,Wang19}. These galaxies are mainly selected by one or more of the following properties: high O32 values, high EW(\hb), high star formation rate surface density, or a deficiency of [\sii] emission. These studies conclude that, for galaxies with O32 $>$ 10, the detection rates are $>$ 60\%, while galaxies with 3 $<$ O32 $<$ 6 only have detection rates $<$ 20\%. For comparison, our galaxies are selected by strong \mgii\ emission with similar redshifts to the above samples. Given 3 $<$ O32 $<$ 6 for our 8 galaxies (see Figure \ref{fig:LyC}), our detection rate of LCEs is $>$  50\% (4 out of 8, with two other tentative detections). This suggests that strong \mgii\ emitters might be more likely to leak LyC than similar galaxies without strong \mgii. This is as expected given N(\mgii) traces N(\hi) (see Section \ref{sec:MgIITwoCases}), and the observed \fescLyC\ is positively related to \fescMgII\ (see Section \ref{sec:pred}). Therefore, future large surveys of LCEs can consider \mgii\ as a constraint to gain higher efficiency in detections of LyC.

%This leads to a factor of $>$2.5 boost of the detection rate, which suggests that galaxies selected by strong \mgii\ emission can add additional constraints to secure the detections of LyC.

%\subsection{Comparisons of Indirect Indicators of LyC}

%\subsection{\Siii\ Absorption Lines}

%checked J0105 in Vcut, there are blue-shifted absorption lines from HI 1025, 949, 930, but SiII 1260 is not clear due to the SNR.

\subsection{[\sii] Deficiency in \mgii\ Emitters}
\label{sec:SII}
Given the ionization potentials for creating and destroying [\sii] of 10.4 and 23.3 eV, respectively, [\sii] $\lambda \lambda 6716,6731$ would be weak or absent in an \hii\ region that is optically thin to ionizing photons. Therefore, it has been proposed that [\sii] deficiency is a probe of the optically thin cloud to LyC \citep[e.g.,][]{Wang19,Wang21}.

In Figure \ref{fig:SII}, we check the [\sii] deficiency (defined below) for our combined sample of \mgii\ emitters and LzLCS galaxies. In the left panel, we compare intensity ratios of [\oiii]/\hb\ with [\sii]/\ha, while the orange line is the best-fitted correlation from SDSS DR12 star-forming galaxies \citep{Wang19}. The colors and labels are similar to Figure \ref{fig:LyC}, i.e., our 8 galaxies are shown in red diamonds, the comparison samples that have high SNR \mgii\ emissions are shown in dark gray diamonds, and LzLCS galaxies are shown as circles. We find that galaxies with strong \mgii\ emission follow the same trend as the other LzLCS galaxies. In the right panel, we compare \fescLyC\ to the [\sii] deficiency, which is defined as the displacement of the measured log([\sii]/\ha) from the orange line in the left panel \citep{Wang19}. The negative correlation has already been discussed in \cite{Wang21}, and the \mgii\ emitters are consistent with the same trend. This suggests that selecting objects with strong \mgii\ emission probe galaxies with similar [\sii] deficiency as in the LzLCS.

%while our detection rate of LCEs is $>$2.5 higher given similar O32 values observed (see Section \ref{sec:Rate}).

%, which are selected by high O32, high SFR surface density, and low blue UV slope \Flurya, submitted.

\begin{figure*}

	\includegraphics[angle=0,trim={0.0cm 0.8cm 0.0cm 0.0cm},clip=true,width=0.5\linewidth,keepaspectratio]{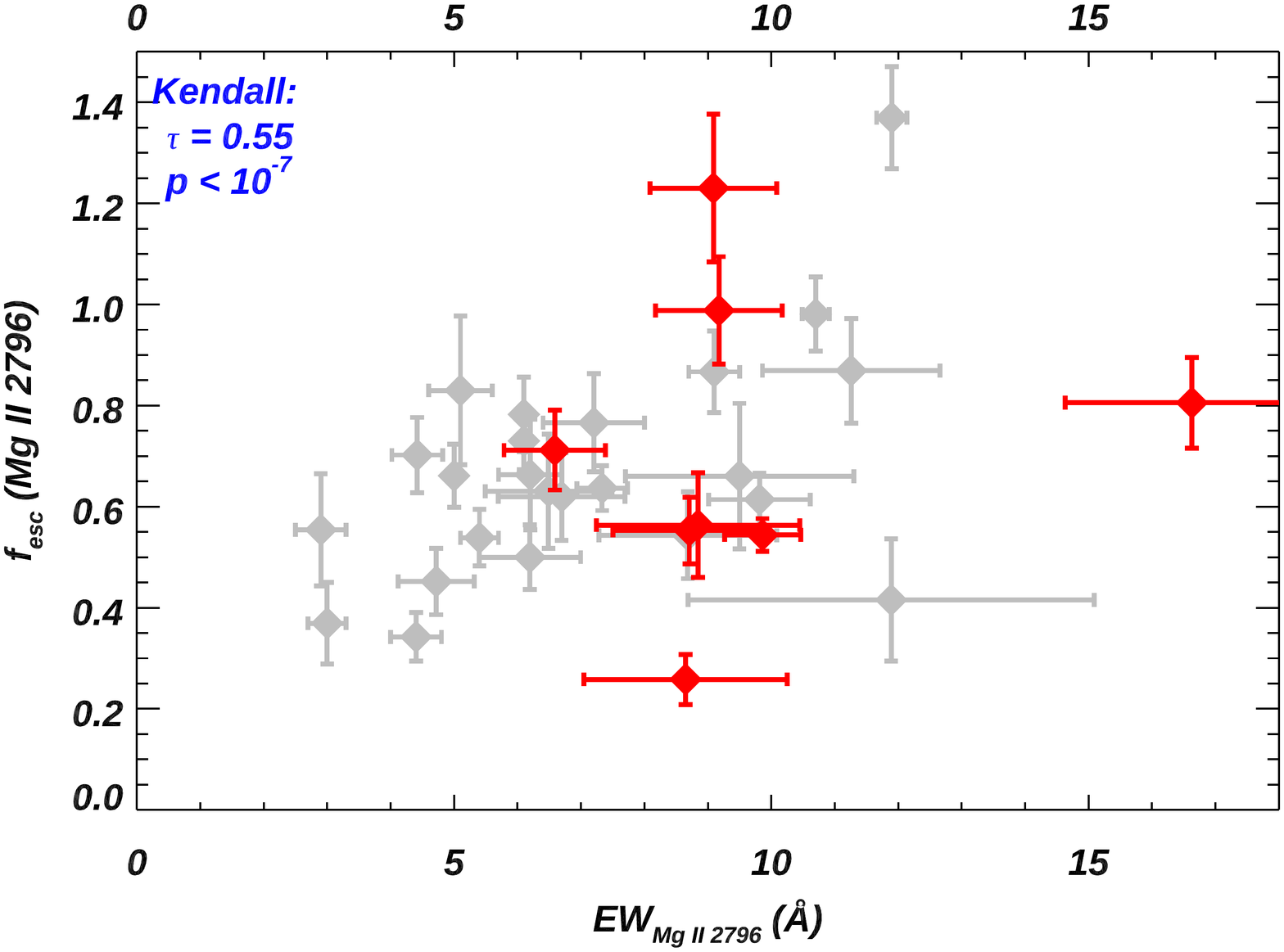}% trim: left lower right upper
	\includegraphics[angle=0,trim={0.0cm 0.8cm 0.0cm 0.0cm},clip=true,width=0.5\linewidth,keepaspectratio]{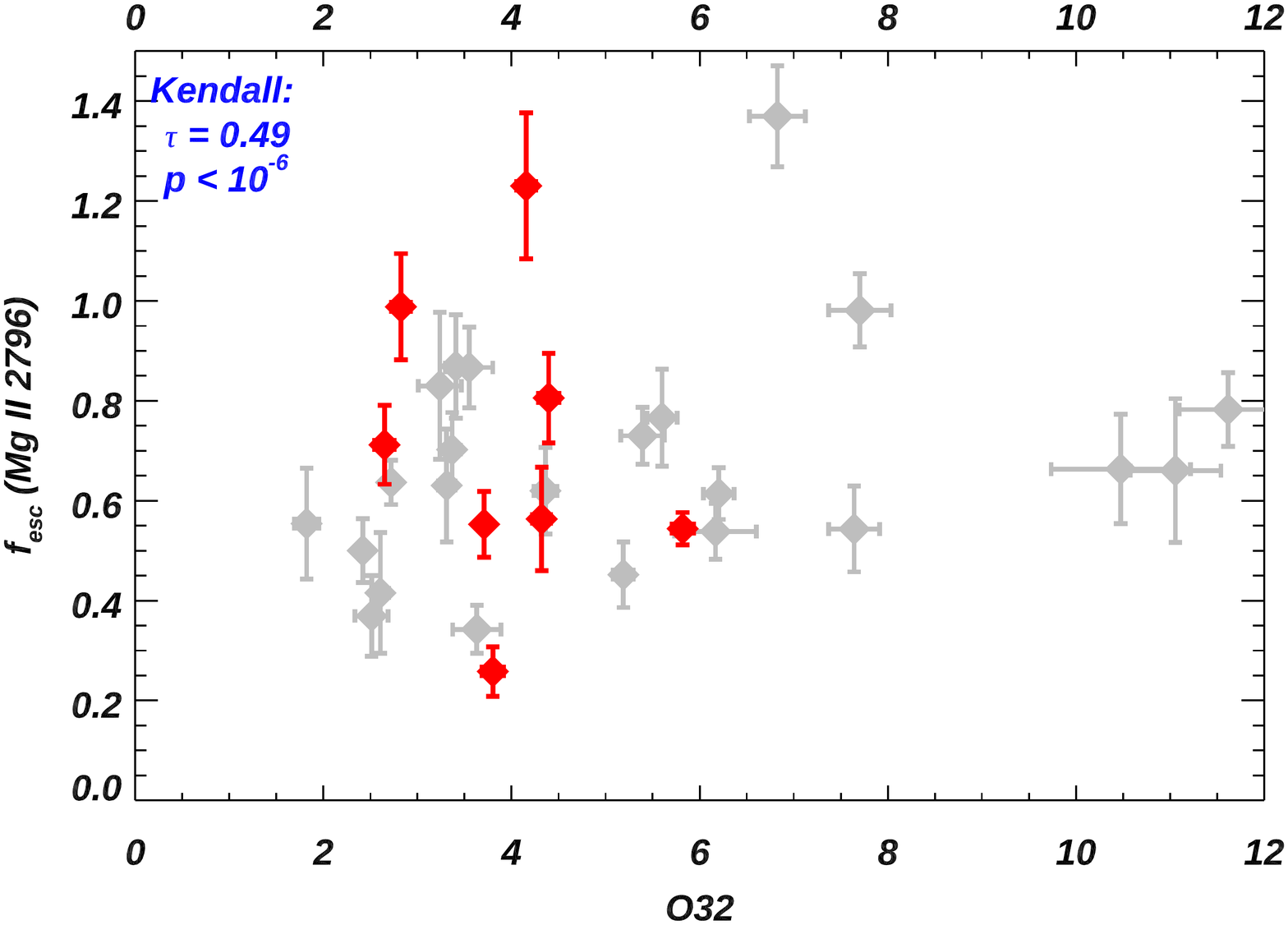}% trim: left lower right upper
	
	\includegraphics[angle=0,trim={0.0cm 0.8cm 0.00cm 0.0cm},clip=true,width=0.5\linewidth,keepaspectratio]{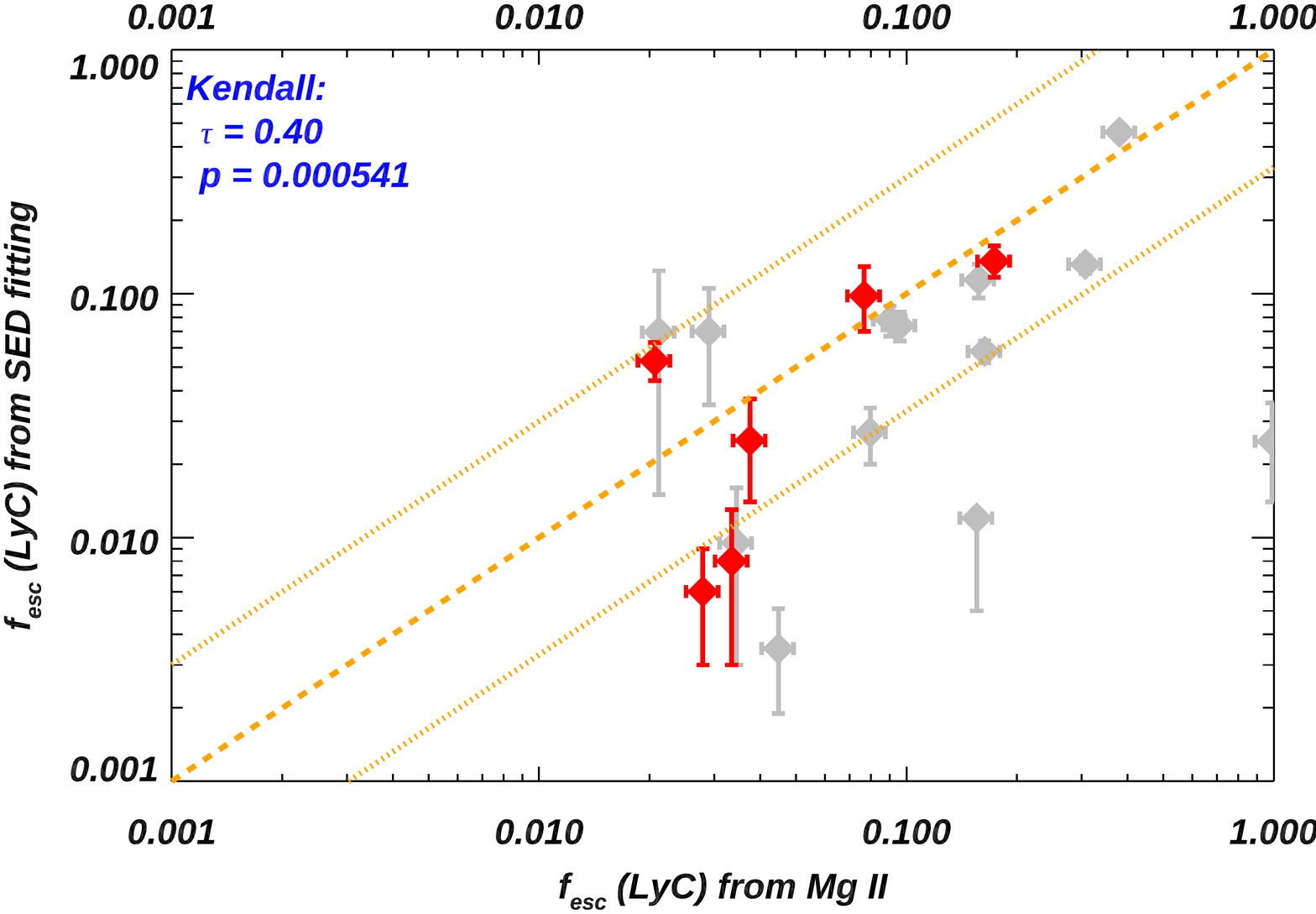}% trim: left lower right upper	
\caption{\normalfont{Same as Figures \ref{fig:MgII-3panel} and \ref{fig:LyCFromMgII}, but we now correct \mgii\ by non-resonant internal dust extinction of the observed galaxy. The correlations in these figures are similar to Figures \ref{fig:MgII-3panel} and \ref{fig:LyCFromMgII}. See discussion in Section \ref{sec:dust}.} }
\label{fig:dust}
\end{figure*}

\subsection{The Effects of Dust}
\label{sec:dust}

Dust extinction plays important roles in the escape of \mgii. In the last panel of Figure \ref{fig:MgII-3panel}, we have shown that \mgii\ is more affected by dust than non-resonant lines. In that figure, we do not correct the observed \mgii\ flux by internal dust extinction (of the observed galaxy) when calculating \fescMgII\ and adopting \mgii\ to predict \fescLyC\ above. This is because a robust correction is difficult to discern when \mgii\ photons could be resonantly scattered like \lya\ \citep[e.g.,][]{Henry18, Chisholm20}. A quantitative assessment of the extinction in these resonant lines depends on how much scattering there is, which is tightly related to geometry, kinematics, \hi\ column density, clumsiness, etc. \citep[e.g.,][]{Neufeld90, Verhamme06,Gronke17}, which is beyond the scope of this paper.

%For \mgii, it becomes more complex since, for most galaxies that we present, their only optical spectra are from SDSS. Therefore, their detailed kinematics due to resonant scattering are not resolved in the line profiles.

However, one interesting test is to treat \mgii\ as non-resonant lines and correct them by internal dust extinction similar to other optical lines. Since the observed \mgii\ lines in our sample should be close to optically thin (Section \ref{sec:pred}), this may be a fair first-order approximation. In Figure \ref{fig:dust}, we test this idea by remaking several correlations after correcting \mgii\ by internal dust extinction.

In the top panels of Figure \ref{fig:dust}, we compare \fescMgII\ to EW(\mgii) and O32, separately. As expected, the absolute value of \fescMgII\ for each galaxy increases comparing to that of Figure \ref{fig:MgII-3panel}. We also find that the resulting positive correlations in these two panels are similar (and slightly weaker) as in Figure \ref{fig:MgII-3panel}. For most galaxies, \fescMgII\ is still $<$ 1.0 even though we have corrected \mgii\ by dust (top panels of Figure \ref{fig:dust}). This suggests that, for the majority of galaxies, the derived \fescMgII\ values cannot be explained purely by non-resonant scattering. This is consistent with what we show in the bottom panel of Figure \ref{fig:MgII-3panel} (see Section  \ref{sec:CorrMgII}), where most galaxies fall below the prediction from dust extinction laws.  Curiously, however, some galaxies have dust-corrected \fescMgII\ values $>$ 1.0. This could be because our chosen extinction law is inaccurate for these galaxies, or the $E(B-V)$ calculated from Balmer lines (Section \ref{sec:MeaSDSS}) are too large for \mgii, or dust geometries are different \citep[e.g.,][]{Scarlata09}.

In the bottom panel, we compare the measured \fescLyC\ from SED fits with the predicted \fescLyC\ from \mgii\ (see Section \ref{sec:pred}, but \mgii\ here has been correct by internal dust). Notably, this figure is almost the same as Figure \ref{fig:LyCFromMgII}. This is as expected because we assume \CF\ = 0.0 in the calculations (see Section \ref{sec:pred}), i.e., the \mgii\ photons all escape from DB paths. In this case, the predicted \fescLyC\ from Equation (\ref{eq:LyC}) is insensitive to the change of optical depth of \hi\ [which is correlated with N(\mgii) in Equation (\ref{eq:NHI})].

%, and is instead dominated by dust extinction in the LyC.  

%Most of the galaxies presented have optically thin \mgii\ emission lines, and the derived N(\hi) is also optically thin given these galaxies' metallicities.

\section{Conclusion and Future Work}
\label{sec:conclude}
We have presented the analyses for 8 LCE candidates selected with strong \mgii\ emission lines, i.e., EW(\mgii) $\gtrsim$ 10\angstrom. These galaxies were observed with HST/COS G140L and G160M gratings (GO: 15865, PI: Henry) to cover their LyC and \lya\ regions, respectively. These galaxies' \mgii\ emission lines have been observed in SDSS.

In 50\% (4 out of 8) of the galaxies, we securely detected LyC flux at $>$ 2$\sigma$ level. We determined the intrinsic flux of LyC from both SED fittings and \hb\ emission lines, which are then adopted to predict the absolute LyC escape fraction (\fescLyC). We find that these two \fescLyC\ values are consistent with each other, and fall within the range of $\sim$ 1 -- 14\%.

We have discussed two geometries that allow \mgii\ (and \lya, LyC) photons to escape in galaxies. By truncating CLOUDY models at different radii, we have shown that N(\mgii) can be used to trace N(\hi) from density-bounded to ionization-bounded scenarios. To estimate the intrinsic flux of \mgii\ (which leads to \fescMgII), we have presented CLOUDY models under these two limiting scenarios as well as the best-fit correlations. We highlight the fact that our derived \fescMgII\ is insensitive to the prior knowledge of the limiting scenarios of the cloud around the galaxy.

We have built a larger comparison sample from published LCE candidates from the literature \citep{Izotov16a, Izotov16b, Izotov18a, Izotov18b, Izotov21,Henry18,Guseva20,Malkan21, Flury22a}. From them, we include 24 galaxies with high SNR ($>$ 3) detections of \mgii\ emission lines.   We find that our \mgii\ selected LCEs follow similar trends that have been established in the comparison sample.  We show that \fescMgII\ correlates positively with EW(\mgii) and O32, while moderate scatter exists for both correlations. Similar to \lya, we find that \fescMgII\ cannot be purely explained by dust extinction without resonant scattering. Furthermore, we study the correlation between \mgii\ and \lya. We find that the measured EW and escape fractions from both lines are correlated, although with significant scatter. For the latter, the fact that \fescMgII\ and \fescLyA\ are of the same order is consistent with both lines escaping from density-bounded optically-thin holes in the ISM.

%We further investigate a few indirect indicators of LyC, while also placing our \mgii\ selected LCEs in the context of the whole LzLCS sample. We find that our \mgii\ selected LCEs follow similar trends that have been established in the LzLCS.

%Overall, we find that the \mgii\ selected LCEs follow similar trends as the LzLCS galaxies, including a higher O32 or \fescLyA\ yield a higher \fescLyC, and smaller peak separations in \lya\ emission line leads to higher \fescLyC. However, 4 of our galaxies seem to show $>$ 2 peaks in their \lya\ emission lines, and their \lya\ peak separations are not well-defined.
%The latter consists of 35 LyC leakers and 31 non-leakers.

Finally, we have presented how to estimate \fescLyC\ from the information of \mgii\ emission lines and dust extinction. This method works because we can trace N(\hi) from N(\mgii), and then solve for the covering fraction of density-bounded clouds from the \mgii\ \ly\ly 2796, 2803 doublet. We find the root-mean-square-error between the measured and predicted \fescLyC\ is $\sim$ 0.05. This suggests that the \mgii\ emission can be used to infer the LyC escape fraction when \fescLyC\ is large ($>$ 5\%). Future deeper observations of \mgii\ for more LCEs would shrink the scattering.

%For the majority of galaxies, we find that the predicted \fescLyC\ values are within a factor of 3 of the measured ones.  
%Nonetheless, given the low SNR at the blue end of the SDSS spectra, \mgii\ \ly 2803 usually has lower SNR.

We have also noted that the detection rate of LCEs from our \mgii\ selected sample may be higher than from all other published LCEs {\it for galaxies with 3 $<$ O32 $<$ 6}. Our high detection rate suggests that strong \mgii\ emitters might be more likely to leak LyC than similar galaxies without strong \mgii. Therefore, future large surveys can consider \mgii\ as a constraint to gain higher efficiency in detections of LyC. 

%We have also tested the effect of non-resonant dust extinctions on \mgii. We find that the observed \fescMgII\ cannot be purely explained by non-resonant scattering.
%We have also briefly discussed the [\sii] deficiency in our sample. We find that the \mgii\ selected sample follows the same trend as the whole LzLCS sample, i.e.,  more [\sii] deficiency yields smaller \fescLyC.

There are various follow-ups to conduct in the future: 1) Consistent radiative transfer modeling of \lya\ and \mgii\ (and LyC) could help to explain the current correlation that we find between \fescLyA\ and \fescMgII\ (and LyC). These correlations compared to dust extinction should also be related to the geometry of the ISM, which currently remains an open question. 2) Deeper and higher resolution spectroscopic observations of the \mgii\ emitters than SDSS (e.g., from larger telescopes) would provide necessary information to predict \fescLyC\ from \mgii. This would further test our proposed correlation between the predicted and measured \fescLyC. 3) Large infrared and optical telescopes, e.g., JWST and future ELT, can detect \mgii\ at higher redshifts, thereby testing the correlations between \mgii\ and \lya\ (and maybe LyC) closer to the Epoch of Reionization.

\iffalse
In this paper, we present the analyses for possible Lyman continuum emitter (LCEs) selected with strong \mgii\ emission lines. The results are summarized as follows:

1. A total of 8 galaxies were selected and observed by HST/COS G140L and G160M observations (Section \ref{sec:obs}).

2. In 4 out of 8 galaxies, LyC flux is detected at $>$ 3$\sigma$ level (Section \ref{sec:meaLyC}).

3. We have determined the escape fraction of LyC (\fescLyC) from two different methods (i.e., SED fittings and \hb\ emission), and found they are consistent with each other and with UV continuum flux ratio (Section \ref{Mea:LyC}).

4. We have discussed the two scenarios for \mgii\ (and \lya, and LyC) photons to escape, and presented that N(\mgii) can trace N(\hi) before the cloud becomes neutral (Section \ref{sec:MgIITwoCases}). Then we update the photoionization models by CLOUDY to calculate the escape fraction of \mgii\ (\fescMgII, Section \ref{Sec:MgII}).

5. 
when updating the photoionization models to estiamte the escape fraction of \mgii.
\fi

%\pagebreak

\acknowledgments
X.X. and A.H. acknowledge support from NASA STScI grants GO 15865.

A.S.L. and D.S. acknowledge support from Swiss National Science Foundation.

M.T. acknowledges support from the NWO grant 0.16.VIDI.189.162 (``ODIN").

%For the dust extinction of MW:https://irsa.ipac.caltech.edu/applications/DUST/
This research has made use of the NASA/IPAC Infrared Science Archive, which is funded by the National Aeronautics and Space Administration and operated by the California Institute of Technology.\\


\vspace{5mm}
\facilities{HST(COS), SDSS, IRSA}

\software{
astropy (The Astropy Collaboration 2013, 2018),
BPASS (v2.2.1, Stanway \& Eldridge 2018),
CalCOS (STScI),
CLOUDY (v17.01; Ferland et al. 2017),
FaintCOS (Makan et al. 2021),
jupyter (Kluyver 2016),
MPFIT (Markwardt 2009),
python,
Prospector (Johnson et al. 2019; Leja et al. 2017),
PyNeb (Luridiana et al. 2015).
}



\newpage

\begin{table*}
	\centering
	\caption{Basic Measurements for Galaxies in Our Sample}
	\label{tab:SDSS}
	\begin{tabular}{lcccccccccc} % four columns, alignment for each
		\hline
		\hline
		Object 	    & O32   & O/H       &	$E(B-V)_{neb}$ &	$E(B-V)_{ste}$    &	F$_{2796}$    & F$_{2803}$  & EW$_{2796}$  & EW$_{2803}$    &   \fescMgII(DB)  &   \fescMgII(IB) \\
		\hline
		   (a)  &(b)  &(c)  &(d)  &(e)  &(f)  &(g)  &(h)  &(i)  &(j) &(k)         \\
		\hline
%\hline
%\multicolumn{10}{l}{Galaxies in this paper (HST-GO-15865):}\\
%\hline

J0105+2349 & 4.3 & 8.2  & 0.139     & 0.239 & 33.0$\pm{6.0}$ & 30.3$\pm{5.2}$ & 8.8$\pm{1.6}$ & 8.8$\pm{1.6}$ & 0.26$\pm{0.05}$ & 0.21$\pm{0.04}$\\
J0152--0431 & 3.8 & 8.3 & 0.082     & 0.151  & 12.7$\pm{2.4}$ & 11.0$\pm{2.6}$ & 8.6$\pm{1.6}$ & 7.5$\pm{1.8}$ & 0.16$\pm{0.03}$ & 0.13$\pm{0.03}$ \\
J0208--0401 & 5.8 & 7.2 & $<$1E-3   & 0.180  & 40.1$\pm{2.2}$ & 20.3$\pm{2.2}$ & 9.9$\pm{0.6}$ & 5.2$\pm{0.6}$ & 0.54$\pm{0.03}$ & 0.51$\pm{0.03}$\\
J1103+4834 & 2.8 & 8.0  & 0.258     & 0.149  & 33.6$\pm{3.4}$ & 27.1$\pm{3.2}$ & 9.2$\pm{1.0}$ & 7.4$\pm{0.8}$ & 0.23$\pm{0.03}$ & 0.19$\pm{0.02}$ \\
J1105+5947 & 4.2 & 8.1  & 0.202     & 0.100  & 13.8$\pm{1.6}$ & 5.7$\pm{1.4}$ & 9.1$\pm{1.0}$ & 4.1$\pm{1.0}$ & 0.40$\pm{0.05}$ & 0.32$\pm{0.04}$ \\
J1219+4814 & 2.6 & 8.1  & 0.232     & 0.130  & 20.3$\pm{2.0}$ & 9.0$\pm{2.4}$ & 6.6$\pm{0.8}$ & 3.0$\pm{0.8}$ & 0.20$\pm{0.02}$ & 0.15$\pm{0.02}$ \\
J1246+4449$^{*}$ & 3.7  & 7.7 & 0.110 & 0.124  & 70.6$\pm{8.2}$ & 37.6$\pm{6.8}$ & 8.7$\pm{1.2}$ & 4.9$\pm{0.8}$ & 0.30$\pm{0.04}$ & 0.26$\pm{0.03}$ \\
J1425+5249 & 4.4 & 7.8  & 0.056     & 0.115  & 28.0$\pm{3.0}$ & 16.2$\pm{3.2}$ & 16.6$\pm{2.0}$ & 8.9$\pm{1.6}$ & 0.59$\pm{0.07}$ & 0.51$\pm{0.06}$ \\

		\hline
		\hline
	\multicolumn{11}{l}{%
  	\begin{minipage}{18cm}%
	\textbf{Note.} --Measurements from the optical spectra galaxies from our sample. (b) [\oiii] 5007/[\oii] 3727 flux ratios; (c) Gas-phase metallicity in the form of 12+log(O/H); (d) the fitted dust extinction from nebular lines; (e) the stellar dust extinction from SED fits of HST/COS data; (f) and (g) Measured flux of \mgii\ \ly\ly 2796, 2803 in units of 10$^{-17}$ ergs s$^{-1}$ cm$^{-2}$, separately. These fluxes have been corrected for Milky Way dust extinction, but not for the dust internal extinction of the galaxy. (h) and (i) Rest-frame Equivalent widths in units of \angstrom\ for \mgii\ \ly\ly 2796, 2803, separately. (j) and (k) Escape fractions for \mgii\ \ly 2796 lines for density-bounded (DB) and ionization-bounded (IB) cases, separately, which are derived from the photoionization models in Section \ref{sec:MgIIFit}.
	
	$^{*}$ J1246+4449 is also included in LzLCS project \citep{Flury22a}.
    %Note that J1246 also in (4) and J1154 also in (2)
    
  	\end{minipage}%
	}\\
	\end{tabular}
	\\ [0mm]
	
\end{table*}

\begin{table*}
	\centering
	\caption{Measurements from Optical Spectra for the Comparison Sample}
	\label{tab:SDSS2}
	\begin{tabular}{lcccccccccc} % four columns, alignment for each
		\hline
		\hline
		Object 	    & O32   & O/H       &	$E(B-V)_{neb}$    &	F$_{2796}$    & F$_{2803}$  & EW$_{2796}$  & EW$_{2803}$    &   \fescMgII(DB)  &   \fescMgII(IB) & Ref.\\
		\hline
		   (a)  &(b)  &(c)  &(d)  &(e)  &(f)  &(g)  &(h)  &(i)  &(j) &(k)         \\
		\hline

J1152+3400 & 5.2 & 8.0 & 0.060 & 34.9$\pm{5.0}$ & 26.6$\pm{5.6}$ & 4.7$\pm{0.6}$ & 3.9$\pm{0.8}$ & 0.32$\pm{0.05}$ & 0.27$\pm{0.04}$ & (1)\\
J1503+3644 & 6.2 & 8.0 & 0.091 & 48.8$\pm{3.8}$ & 26.3$\pm{3.2}$ & 9.8$\pm{0.8}$ & 5.5$\pm{0.8}$ & 0.37$\pm{0.03}$ & 0.33$\pm{0.03}$ & (1)\\
J0232--0426 & 11.1 & 7.9 & 0.035 & 4.7$\pm{1.0}$ & 4.6$\pm{1.4}$ & 9.5$\pm{1.8}$ & 8.4$\pm{2.4}$ & 0.54$\pm{0.12}$ & 0.51$\pm{0.11}$ & (1)\\
J1046+5827 & 4.4 & 8.0 & 0.035 & 19.1$\pm{2.6}$ & 15.1$\pm{2.6}$ & 6.7$\pm{1.0}$ & 5.7$\pm{1.0}$ & 0.51$\pm{0.07}$ & 0.42$\pm{0.06}$ & (1)\\
J1355+1457 & 7.6 & 7.8 & 0.094 & 15.8$\pm{2.4}$ & 10.6$\pm{2.6}$ & 8.7$\pm{1.4}$ & 5.6$\pm{1.2}$ & 0.32$\pm{0.05}$ & 0.29$\pm{0.05}$ & (1)\\
J0911+1831 & 1.8 & 7.9 & 0.170 & 55.0$\pm{10.0}$ & 37.0$\pm{8.0}$ & 2.9$\pm{0.4}$ & 1.9$\pm{0.3}$ & 0.22$\pm{0.04}$ & 0.18$\pm{0.04}$ & (2)\\
J0926+4427 & 3.2 & 8.0 & 0.100 & 217.0$\pm{34.0}$ & 122.0$\pm{33.0}$ & 5.1$\pm{0.5}$ & 2.9$\pm{0.5}$ & 0.48$\pm{0.08}$ & 0.41$\pm{0.07}$ & (2)\\
J1054+5238 & 2.5 & 8.2 & 0.080 & 84.0$\pm{17.0}$ & 66.0$\pm{11.0}$ & 3.0$\pm{0.3}$ & 1.6$\pm{0.2}$ & 0.24$\pm{0.05}$ & 0.19$\pm{0.04}$ & (2)\\
J1219+1526 & 10.5 & 7.9 & 0.010 & 141.0$\pm{20.0}$ & 86.0$\pm{9.0}$ & 6.2$\pm{0.5}$ & 3.7$\pm{0.3}$ & 0.63$\pm{0.10}$ & 0.58$\pm{0.10}$ & (2)\\
J1244+0216 & 3.6 & 8.1 & 0.070 & 106.0$\pm{12.0}$ & 38.0$\pm{9.0}$ & 4.4$\pm{0.4}$ & 1.6$\pm{0.3}$ & 0.23$\pm{0.03}$ & 0.19$\pm{0.03}$ & (2)\\
J1249+1234 & 3.5 & 8.1 & 0.070 & 120.0$\pm{5.0}$ & 71.0$\pm{5.0}$ & 9.1$\pm{0.4}$ & 5.3$\pm{0.4}$ & 0.59$\pm{0.05}$ & 0.47$\pm{0.04}$ & (2)\\
J1424+4217 & 6.2 & 8.0 & 0.040 & 180.0$\pm{11.0}$ & 141.0$\pm{9.0}$ & 5.4$\pm{0.3}$ & 4.3$\pm{0.3}$ & 0.43$\pm{0.04}$ & 0.36$\pm{0.04}$ & (2)\\
J0122+0520 & 5.6 & 7.9 & $<$1E-3 & 34.2$\pm{4.2}$ & 20.4$\pm{4.2}$ & 7.2$\pm{0.8}$ & 4.9$\pm{0.8}$ & 0.77$\pm{0.10}$ & 0.68$\pm{0.09}$ & (3)\\
J1326+4218 & 3.3 & 8.2 & 0.166 & 52.4$\pm{9.2}$ & 27.7$\pm{7.8}$ & 6.5$\pm{1.0}$ & 3.2$\pm{1.0}$ & 0.25$\pm{0.04}$ & 0.20$\pm{0.04}$ & (3)\\
J0047+0154 & 3.4 & 8.1 & 0.169 & 39.6$\pm{4.0}$ & 16.5$\pm{3.4}$ & 4.4$\pm{0.4}$ & 1.9$\pm{0.4}$ & 0.27$\pm{0.03}$ & 0.22$\pm{0.02}$ & (3)\\
J1246+4449 & 3.4 & 8.0 & 0.157 & 90.8$\pm{10.6}$ & 44.9$\pm{8.2}$ & 11.3$\pm{1.4}$ & 6.0$\pm{1.0}$ & 0.36$\pm{0.04}$ & 0.29$\pm{0.03}$ & (3)\\
J1517+3705 & 2.4 & 8.3 & 0.196 & 37.0$\pm{4.6}$ & 16.4$\pm{4.4}$ & 6.2$\pm{0.8}$ & 2.3$\pm{0.6}$ & 0.17$\pm{0.02}$ & 0.13$\pm{0.02}$ & (3)\\
J1648+4957 & 2.6 & 8.2 & $<$1E-3 & 10.4$\pm{3.0}$ & 11.9$\pm{2.4}$ & 11.9$\pm{3.2}$ & 11.2$\pm{2.4}$ & 0.42$\pm{0.12}$ & 0.33$\pm{0.10}$ & (3)\\
J1154+2443 & 11.6 & 7.6 & 0.049 & 12.2$\pm{0.9}$ & 6.5$\pm{0.9}$ & 6.1$\pm{0.1}$ & 3.3$\pm{0.1}$ & 0.60$\pm{0.06}$ & 0.56$\pm{0.05}$ & (1,4)\\
J1442--0209 & 6.8 & 7.9 & 0.120 & 47.8$\pm{2.6}$ & 25.1$\pm{2.0}$ & 11.9$\pm{0.2}$ & 6.4$\pm{0.1}$ & 0.70$\pm{0.05}$ & 0.63$\pm{0.05}$ & (4)\\
J0901+2119 & 7.7 & 8.2 & 0.129 & 17.5$\pm{1.0}$ & 9.0$\pm{0.7}$ & 10.7$\pm{0.2}$ & 5.7$\pm{0.1}$ & 0.48$\pm{0.04}$ & 0.41$\pm{0.03}$ & (4)\\
J0925+1403 & 5.4 & 7.9 & 0.122 & 31.0$\pm{1.9}$ & 15.5$\pm{1.5}$ & 6.1$\pm{0.1}$ & 3.1$\pm{0.1}$ & 0.37$\pm{0.03}$ & 0.33$\pm{0.03}$ & (4)\\
J1011+1947 & 28.8 & 8.0 & 0.094 & 8.0$\pm{0.6}$ & 4.4$\pm{0.5}$ & 5.0$\pm{0.1}$ & 2.8$\pm{0.1}$ & 0.39$\pm{0.04}$ & 0.40$\pm{0.04}$ & (4)\\
J0207+0047 & 2.7 & 8.3 & 0.180 & 65.5$\pm{4.4}$ & 40.9$\pm{4.8}$ & 7.3$\pm{0.4}$ & 4.5$\pm{0.6}$ & 0.23$\pm{0.02}$ & 0.18$\pm{0.01}$ & (5)\\

		\hline
		\hline
	\multicolumn{11}{l}{%
  	\begin{minipage}{18cm}%
	\textbf{Note.} --Measurements from the optical spectra from published LCE candidates in the literature with high SNR \mgii\ detections. The labels are the same as Figure \ref{tab:SDSS}. Column (k): Reference for each object: (1) \cite{Izotov16a,Izotov16b,Izotov18a,Izotov18b,Izotov21}; (2) \cite{Henry18}, (3) \cite{Flury22a}; (4) \cite{Guseva20}; (5) \cite{Malkan21}.
    %Note that J1246 also in (4) and J1154 also in (2)
    
  	\end{minipage}%
	}\\
	\end{tabular}
	\\ [0mm]
	
\end{table*}


\bibliography{MgII_LyC_LyA}{}

\begin{thebibliography}{}
\expandafter\ifx\csname natexlab\endcsname\relax\def\natexlab#1{#1}\fi
\providecommand{\url}[1]{\href{#1}{#1}}
\providecommand{\dodoi}[1]{doi:~\href{http://doi.org/#1}{\nolinkurl{#1}}}
\providecommand{\doeprint}[1]{\href{http://ascl.net/#1}{\nolinkurl{http://ascl.net/#1}}}
\providecommand{\doarXiv}[1]{\href{https://arxiv.org/abs/#1}{\nolinkurl{https://arxiv.org/abs/#1}}}

\bibitem[{{Akritas} \& {Siebert}(1996)}]{Akritas96}
{Akritas}, M.~G., \& {Siebert}, J. 1996, \mnras, 278, 919,
  \dodoi{10.1093/mnras/278.4.919}

\bibitem[{{Alexandroff} {et~al.}(2015){Alexandroff}, {Heckman}, {Borthakur},
  {Overzier}, \& {Leitherer}}]{Alex15}
{Alexandroff}, R.~M., {Heckman}, T.~M., {Borthakur}, S., {Overzier}, R., \&
  {Leitherer}, C. 2015, \apj, 810, 104, \dodoi{10.1088/0004-637X/810/2/104}

\bibitem[{{Andrews} \& {Martini}(2013)}]{Andrews13}
{Andrews}, B.~H., \& {Martini}, P. 2013, \apj, 765, 140,
  \dodoi{10.1088/0004-637X/765/2/140}

\bibitem[{{Ba{\~n}ados} {et~al.}(2018){Ba{\~n}ados}, {Venemans},
  {Mazzucchelli}, {Farina}, {Walter}, {Wang}, {Decarli}, {Stern}, {Fan},
  {Davies}, {Hennawi}, {Simcoe}, {Turner}, {Rix}, {Yang}, {Kelson}, {Rudie}, \&
  {Winters}}]{Banados18}
{Ba{\~n}ados}, E., {Venemans}, B.~P., {Mazzucchelli}, C., {et~al.} 2018, \nat,
  553, 473, \dodoi{10.1038/nature25180}

\bibitem[{{Baldwin} {et~al.}(1981){Baldwin}, {Phillips}, \&
  {Terlevich}}]{Baldwin81}
{Baldwin}, J.~A., {Phillips}, M.~M., \& {Terlevich}, R. 1981, \pasp, 93, 5,
  \dodoi{10.1086/130766}

\bibitem[{{Becker} {et~al.}(2021){Becker}, {D'Aloisio}, {Christenson}, {Zhu},
  {Worseck}, \& {Bolton}}]{Becker21}
{Becker}, G.~D., {D'Aloisio}, A., {Christenson}, H.~M., {et~al.} 2021, \mnras,
  508, 1853, \dodoi{10.1093/mnras/stab2696}

\bibitem[{{Becker} {et~al.}(2001){Becker}, {Fan}, {White}, {Strauss},
  {Narayanan}, {Lupton}, {Gunn}, {Annis}, {Bahcall}, {Brinkmann}, {Connolly},
  {Csabai}, {Czarapata}, {Doi}, {Heckman}, {Hennessy}, {Ivezi{\'c}}, {Knapp},
  {Lamb}, {McKay}, {Munn}, {Nash}, {Nichol}, {Pier}, {Richards}, {Schneider},
  {Stoughton}, {Szalay}, {Thakar}, \& {York}}]{Becker01}
{Becker}, R.~H., {Fan}, X., {White}, R.~L., {et~al.} 2001, \aj, 122, 2850,
  \dodoi{10.1086/324231}

\bibitem[{{Berg} {et~al.}(2019){Berg}, {Chisholm}, {Erb}, {Pogge}, {Henry}, \&
  {Olivier}}]{Berg19}
{Berg}, D.~A., {Chisholm}, J., {Erb}, D.~K., {et~al.} 2019, \apjl, 878, L3,
  \dodoi{10.3847/2041-8213/ab21dc}

\bibitem[{{Bergvall} {et~al.}(2006){Bergvall}, {Zackrisson}, {Andersson},
  {Arnberg}, {Masegosa}, \& {{\"O}stlin}}]{Bergvall06}
{Bergvall}, N., {Zackrisson}, E., {Andersson}, B.~G., {et~al.} 2006, \aap, 448,
  513, \dodoi{10.1051/0004-6361:20053788}

\bibitem[{{Borthakur} {et~al.}(2014){Borthakur}, {Heckman}, {Leitherer}, \&
  {Overzier}}]{Borthakur14}
{Borthakur}, S., {Heckman}, T.~M., {Leitherer}, C., \& {Overzier}, R.~A. 2014,
  Science, 346, 216, \dodoi{10.1126/science.1254214}

\bibitem[{{Bouwens} {et~al.}(2021){Bouwens}, {Oesch}, {Stefanon},
  {Illingworth}, {Labb{\'e}}, {Reddy}, {Atek}, {Montes}, {Naidu},
  {Nanayakkara}, {Nelson}, \& {Wilkins}}]{Bouwens21}
{Bouwens}, R.~J., {Oesch}, P.~A., {Stefanon}, M., {et~al.} 2021, \aj, 162, 47,
  \dodoi{10.3847/1538-3881/abf83e}

\bibitem[{{Calzetti} {et~al.}(2000){Calzetti}, {Armus}, {Bohlin}, {Kinney},
  {Koornneef}, \& {Storchi-Bergmann}}]{Calzetti00}
{Calzetti}, D., {Armus}, L., {Bohlin}, R.~C., {et~al.} 2000, \apj, 533, 682,
  \dodoi{10.1086/308692}

\bibitem[{{Cardelli} {et~al.}(1989){Cardelli}, {Clayton}, \&
  {Mathis}}]{Cardelli89}
{Cardelli}, J.~A., {Clayton}, G.~C., \& {Mathis}, J.~S. 1989, \apj, 345, 245,
  \dodoi{10.1086/167900}

\bibitem[{{Chabrier}(2003)}]{Chabrier03}
{Chabrier}, G. 2003, \pasp, 115, 763, \dodoi{10.1086/376392}

\bibitem[{{Chevallard} {et~al.}(2018){Chevallard}, {Charlot}, {Senchyna},
  {Stark}, {Vidal-Garc{\'\i}a}, {Feltre}, {Gutkin}, {Jones}, {Mainali}, \&
  {Wofford}}]{Chevallard18}
{Chevallard}, J., {Charlot}, S., {Senchyna}, P., {et~al.} 2018, \mnras, 479,
  3264, \dodoi{10.1093/mnras/sty1461}

\bibitem[{{Chisholm} {et~al.}(2020){Chisholm}, {Prochaska}, {Schaerer},
  {Gazagnes}, \& {Henry}}]{Chisholm20}
{Chisholm}, J., {Prochaska}, J.~X., {Schaerer}, D., {Gazagnes}, S., \& {Henry},
  A. 2020, \mnras, 498, 2554, \dodoi{10.1093/mnras/staa2470}

\bibitem[{{Conroy} {et~al.}(2009){Conroy}, {Gunn}, \& {White}}]{Conroy09}
{Conroy}, C., {Gunn}, J.~E., \& {White}, M. 2009, \apj, 699, 486,
  \dodoi{10.1088/0004-637X/699/1/486}

\bibitem[{{de Kool} {et~al.}(2002){de Kool}, {Becker}, {Gregg}, {White}, \&
  {Arav}}]{deKool02}
{de Kool}, M., {Becker}, R.~H., {Gregg}, M.~D., {White}, R.~L., \& {Arav}, N.
  2002, \apj, 567, 58, \dodoi{10.1086/338490}

\bibitem[{{Dijkstra} {et~al.}(2016){Dijkstra}, {Gronke}, \&
  {Venkatesan}}]{Dijkstra16}
{Dijkstra}, M., {Gronke}, M., \& {Venkatesan}, A. 2016, \apj, 828, 71,
  \dodoi{10.3847/0004-637X/828/2/71}

\bibitem[{{Duncan} \& {Conselice}(2015)}]{Duncan15}
{Duncan}, K., \& {Conselice}, C.~J. 2015, \mnras, 451, 2030,
  \dodoi{10.1093/mnras/stv1049}

\bibitem[{{Eldridge} {et~al.}(2017){Eldridge}, {Stanway}, {Xiao}, {McClelland},
  {Taylor}, {Ng}, {Greis}, \& {Bray}}]{Eldridge17}
{Eldridge}, J.~J., {Stanway}, E.~R., {Xiao}, L., {et~al.} 2017, \pasa, 34,
  e058, \dodoi{10.1017/pasa.2017.51}

\bibitem[{{Fan} {et~al.}(2006){Fan}, {Strauss}, {Becker}, {White}, {Gunn},
  {Knapp}, {Richards}, {Schneider}, {Brinkmann}, \& {Fukugita}}]{Fan06}
{Fan}, X., {Strauss}, M.~A., {Becker}, R.~H., {et~al.} 2006, \aj, 132, 117,
  \dodoi{10.1086/504836}

\bibitem[{{Feldman} \& {Cousins}(1998)}]{Feldman98}
{Feldman}, G.~J., \& {Cousins}, R.~D. 1998, \prd, 57, 3873,
  \dodoi{10.1103/PhysRevD.57.3873}

\bibitem[{{Ferland} {et~al.}(2017){Ferland}, {Chatzikos}, {Guzm{\'a}n},
  {Lykins}, {van Hoof}, {Williams}, {Abel}, {Badnell}, {Keenan}, {Porter}, \&
  {Stancil}}]{Ferland17}
{Ferland}, G.~J., {Chatzikos}, M., {Guzm{\'a}n}, F., {et~al.} 2017, \rmxaa, 53,
  385.
\newblock \doarXiv{1705.10877}

\bibitem[{{Finkelstein} {et~al.}(2019){Finkelstein}, {D'Aloisio},
  {Paardekooper}, {Ryan}, {Behroozi}, {Finlator}, {Livermore}, {Upton
  Sanderbeck}, {Dalla Vecchia}, \& {Khochfar}}]{Finkelstein19}
{Finkelstein}, S.~L., {D'Aloisio}, A., {Paardekooper}, J.-P., {et~al.} 2019,
  \apj, 879, 36, \dodoi{10.3847/1538-4357/ab1ea8}

\bibitem[{{Flury} {et~al.}(2022{\natexlab{a}}){Flury}, {Jaskot}, {Ferguson},
  {Worseck}, {Makan}, {Chisholm}, {Saldana-Lopez}, {Schaerer}, {McCandless},
  {Wang}, {Ford}, {Heckman}, {Ji}, {Giavalisco}, {Amorin}, {Atek}, {Blaizot},
  {Borthakur}, {Carr}, {Castellano}, {Cristiani}, {de Barros}, {Dickinson},
  {Finkelstein}, {Fleming}, {Fontanot}, {Garel}, {Grazian}, {Hayes}, {Henry},
  {Mauerhofer}, {Micheva}, {Oey}, {Ostlin}, {Papovich}, {Pentericci},
  {Ravindranath}, {Rosdahl}, {Rutkowski}, {Santini}, {Scarlata}, {Teplitz},
  {Thuan}, {Trebitsch}, {Vanzella}, {Verhamme}, \& {Xu}}]{Flury22a}
{Flury}, S.~R., {Jaskot}, A.~E., {Ferguson}, H.~C., {et~al.}
  2022{\natexlab{a}}, arXiv e-prints, arXiv:2201.11716.
\newblock \doarXiv{2201.11716}

\bibitem[{{Flury} {et~al.}(2022{\natexlab{b}}){Flury}, {Jaskot}, {Ferguson},
  {Worseck}, {Makan}, {Chisholm}, {Saldana-Lopez}, {Schaerer}, {McCandliss},
  {Wang}, {Ford}, {Oey}, {Heckman}, {Ji}, {Giavalisco}, {Amorin}, {Atek},
  {Blaizot}, {Borthakur}, {Carr}, {Castellano}, {Cristiani}, {de Barros},
  {Dickinson}, {Finkelstein}, {Fleming}, {Fontanot}, {Garel}, {Grazian},
  {Hayes}, {Henry}, {Mauerhofer}, {Micheva}, {Ostlin}, {Papovich},
  {Pentericci}, {Ravindranath}, {Rosdahl}, {Rutkowski}, {Santini}, {Scarlata},
  {Teplitz}, {Thuan}, {Trebitsch}, {Vanzella}, {Verhamme}, \& {Xu}}]{Flury22b}
---. 2022{\natexlab{b}}, arXiv e-prints, arXiv:2203.15649.
\newblock \doarXiv{2203.15649}

\bibitem[{{Gazagnes} {et~al.}(2020){Gazagnes}, {Chisholm}, {Schaerer},
  {Verhamme}, \& {Izotov}}]{Gazagnes20}
{Gazagnes}, S., {Chisholm}, J., {Schaerer}, D., {Verhamme}, A., \& {Izotov}, Y.
  2020, \aap, 639, A85, \dodoi{10.1051/0004-6361/202038096}

\bibitem[{{Gazagnes} {et~al.}(2018){Gazagnes}, {Chisholm}, {Schaerer},
  {Verhamme}, {Rigby}, \& {Bayliss}}]{Gazagnes18}
{Gazagnes}, S., {Chisholm}, J., {Schaerer}, D., {et~al.} 2018, \aap, 616, A29,
  \dodoi{10.1051/0004-6361/201832759}

\bibitem[{{Giallongo} {et~al.}(2015){Giallongo}, {Grazian}, {Fiore}, {Fontana},
  {Pentericci}, {Vanzella}, {Dickinson}, {Kocevski}, {Castellano}, {Cristiani},
  {Ferguson}, {Finkelstein}, {Grogin}, {Hathi}, {Koekemoer}, {Newman}, \&
  {Salvato}}]{Giallongo15}
{Giallongo}, E., {Grazian}, A., {Fiore}, F., {et~al.} 2015, \aap, 578, A83,
  \dodoi{10.1051/0004-6361/201425334}

\bibitem[{{Grazian} {et~al.}(2020){Grazian}, {Giallongo}, {Fiore}, {Boutsia},
  {Civano}, {Cristiani}, {Cupani}, {Dickinson}, {Fontanot}, {Menci}, \&
  {Romano}}]{Grazian20}
{Grazian}, A., {Giallongo}, E., {Fiore}, F., {et~al.} 2020, \apj, 897, 94,
  \dodoi{10.3847/1538-4357/ab99a3}

\bibitem[{{Grevesse} {et~al.}(2010){Grevesse}, {Asplund}, {Sauval}, \&
  {Scott}}]{Grevesse10}
{Grevesse}, N., {Asplund}, M., {Sauval}, A.~J., \& {Scott}, P. 2010, \apss,
  328, 179, \dodoi{10.1007/s10509-010-0288-z}

\bibitem[{{Gronke} {et~al.}(2016){Gronke}, {Dijkstra}, {McCourt}, \&
  {Oh}}]{Gronke16}
{Gronke}, M., {Dijkstra}, M., {McCourt}, M., \& {Oh}, S.~P. 2016, \apjl, 833,
  L26, \dodoi{10.3847/2041-8213/833/2/L26}

\bibitem[{{Gronke} {et~al.}(2017){Gronke}, {Dijkstra}, {McCourt}, \&
  {Oh}}]{Gronke17}
---. 2017, \aap, 607, A71, \dodoi{10.1051/0004-6361/201731013}

\bibitem[{{Gronke} {et~al.}(2021){Gronke}, {Ocvirk}, {Mason}, {Matthee},
  {Bosman}, {Sorce}, {Lewis}, {Ahn}, {Aubert}, {Dawoodbhoy}, {Iliev},
  {Shapiro}, \& {Yepes}}]{Gronke21}
{Gronke}, M., {Ocvirk}, P., {Mason}, C., {et~al.} 2021, \mnras, 508, 3697,
  \dodoi{10.1093/mnras/stab2762}

\bibitem[{{Guseva} {et~al.}(2013){Guseva}, {Izotov}, {Fricke}, \&
  {Henkel}}]{Guseva13}
{Guseva}, N.~G., {Izotov}, Y.~I., {Fricke}, K.~J., \& {Henkel}, C. 2013, \aap,
  555, A90, \dodoi{10.1051/0004-6361/201221010}

\bibitem[{{Guseva} {et~al.}(2020){Guseva}, {Izotov}, {Schaerer},
  {V{\'\i}lchez}, {Amor{\'\i}n}, {P{\'e}rez-Montero}, {Iglesias-P{\'a}ramo},
  {Verhamme}, {Kehrig}, \& {Ramambason}}]{Guseva20}
{Guseva}, N.~G., {Izotov}, Y.~I., {Schaerer}, D., {et~al.} 2020, \mnras, 497,
  4293, \dodoi{10.1093/mnras/staa2197}

\bibitem[{{Hayes} {et~al.}(2011){Hayes}, {Schaerer}, {{\"O}stlin}, {Mas-Hesse},
  {Atek}, \& {Kunth}}]{Hayes11}
{Hayes}, M., {Schaerer}, D., {{\"O}stlin}, G., {et~al.} 2011, \apj, 730, 8,
  \dodoi{10.1088/0004-637X/730/1/8}

\bibitem[{{Hayes} {et~al.}(2021){Hayes}, {Runnholm}, {Gronke}, \&
  {Scarlata}}]{Hayes21}
{Hayes}, M.~J., {Runnholm}, A., {Gronke}, M., \& {Scarlata}, C. 2021, \apj,
  908, 36, \dodoi{10.3847/1538-4357/abd246}

\bibitem[{{Heckman} {et~al.}(2001){Heckman}, {Sembach}, {Meurer}, {Leitherer},
  {Calzetti}, \& {Martin}}]{Heckman01}
{Heckman}, T.~M., {Sembach}, K.~R., {Meurer}, G.~R., {et~al.} 2001, \apj, 558,
  56, \dodoi{10.1086/322475}

\bibitem[{{Heckman} {et~al.}(2011){Heckman}, {Borthakur}, {Overzier},
  {Kauffmann}, {Basu-Zych}, {Leitherer}, {Sembach}, {Martin}, {Rich},
  {Schiminovich}, \& {Seibert}}]{Heckman11}
{Heckman}, T.~M., {Borthakur}, S., {Overzier}, R., {et~al.} 2011, \apj, 730, 5,
  \dodoi{10.1088/0004-637X/730/1/5}

\bibitem[{{Henry} {et~al.}(2018){Henry}, {Berg}, {Scarlata}, {Verhamme}, \&
  {Erb}}]{Henry18}
{Henry}, A., {Berg}, D.~A., {Scarlata}, C., {Verhamme}, A., \& {Erb}, D. 2018,
  \apj, 855, 96, \dodoi{10.3847/1538-4357/aab099}

\bibitem[{{Henry} {et~al.}(2015){Henry}, {Scarlata}, {Martin}, \&
  {Erb}}]{Henry15}
{Henry}, A., {Scarlata}, C., {Martin}, C.~L., \& {Erb}, D. 2015, \apj, 809, 19,
  \dodoi{10.1088/0004-637X/809/1/19}

\bibitem[{{Henry} {et~al.}(2021){Henry}, {Rafelski}, {Sunnquist}, {Pirzkal},
  {Pacifici}, {Atek}, {Bagley}, {Baronchelli}, {Barro}, {Bunker}, {Colbert},
  {Dai}, {Elmegreen}, {Elmegreen}, {Finkelstein}, {Kocevski}, {Koekemoer},
  {Malkan}, {Martin}, {Mehta}, {Pahl}, {Papovich}, {Rutkowski}, {Sanchez
  Almeida}, {Scarlata}, {Snyder}, \& {Teplitz}}]{Henry21}
{Henry}, A., {Rafelski}, M., {Sunnquist}, B., {et~al.} 2021, arXiv e-prints,
  arXiv:2107.00672.
\newblock \doarXiv{2107.00672}

\bibitem[{{Hopkins} {et~al.}(2008){Hopkins}, {Hernquist}, {Cox}, \&
  {Kere{\v{s}}}}]{Hopkins08}
{Hopkins}, P.~F., {Hernquist}, L., {Cox}, T.~J., \& {Kere{\v{s}}}, D. 2008,
  \apjs, 175, 356, \dodoi{10.1086/524362}

\bibitem[{{Inoue} {et~al.}(2014){Inoue}, {Shimizu}, {Iwata}, \&
  {Tanaka}}]{Inoue14}
{Inoue}, A.~K., {Shimizu}, I., {Iwata}, I., \& {Tanaka}, M. 2014, \mnras, 442,
  1805, \dodoi{10.1093/mnras/stu936}

\bibitem[{{Izotov} {et~al.}(2016{\natexlab{a}}){Izotov}, {Orlitov{\'a}},
  {Schaerer}, {Thuan}, {Verhamme}, {Guseva}, \& {Worseck}}]{Izotov16a}
{Izotov}, Y.~I., {Orlitov{\'a}}, I., {Schaerer}, D., {et~al.}
  2016{\natexlab{a}}, \nat, 529, 178, \dodoi{10.1038/nature16456}

\bibitem[{{Izotov} {et~al.}(2016{\natexlab{b}}){Izotov}, {Schaerer}, {Thuan},
  {Worseck}, {Guseva}, {Orlitov{\'a}}, \& {Verhamme}}]{Izotov16b}
{Izotov}, Y.~I., {Schaerer}, D., {Thuan}, T.~X., {et~al.} 2016{\natexlab{b}},
  \mnras, 461, 3683, \dodoi{10.1093/mnras/stw1205}

\bibitem[{{Izotov} {et~al.}(2018{\natexlab{a}}){Izotov}, {Schaerer}, {Worseck},
  {Guseva}, {Thuan}, {Verhamme}, {Orlitov{\'a}}, \& {Fricke}}]{Izotov18a}
{Izotov}, Y.~I., {Schaerer}, D., {Worseck}, G., {et~al.} 2018{\natexlab{a}},
  \mnras, 474, 4514, \dodoi{10.1093/mnras/stx3115}

\bibitem[{{Izotov} {et~al.}(2021){Izotov}, {Worseck}, {Schaerer}, {Guseva},
  {Chisholm}, {Thuan}, {Fricke}, \& {Verhamme}}]{Izotov21}
{Izotov}, Y.~I., {Worseck}, G., {Schaerer}, D., {et~al.} 2021, \mnras, 503,
  1734, \dodoi{10.1093/mnras/stab612}

\bibitem[{{Izotov} {et~al.}(2018{\natexlab{b}}){Izotov}, {Worseck}, {Schaerer},
  {Guseva}, {Thuan}, {Fricke}, \& {Orlitov{\'a}}}]{Izotov18b}
---. 2018{\natexlab{b}}, \mnras, 478, 4851, \dodoi{10.1093/mnras/sty1378}

\bibitem[{{Jaskot} {et~al.}(2019){Jaskot}, {Dowd}, {Oey}, {Scarlata}, \&
  {McKinney}}]{Jaskot19}
{Jaskot}, A.~E., {Dowd}, T., {Oey}, M.~S., {Scarlata}, C., \& {McKinney}, J.
  2019, \apj, 885, 96, \dodoi{10.3847/1538-4357/ab3d3b}

\bibitem[{{Jaskot} \& {Oey}(2013)}]{Jaskot13}
{Jaskot}, A.~E., \& {Oey}, M.~S. 2013, \apj, 766, 91,
  \dodoi{10.1088/0004-637X/766/2/91}

\bibitem[{{Johnson} {et~al.}(2021){Johnson}, {Leja}, {Conroy}, \&
  {Speagle}}]{Johnson21}
{Johnson}, B.~D., {Leja}, J., {Conroy}, C., \& {Speagle}, J.~S. 2021, \apjs,
  254, 22, \dodoi{10.3847/1538-4365/abef67}

\bibitem[{{Kakiichi} \& {Gronke}(2021)}]{Kakiichi21}
{Kakiichi}, K., \& {Gronke}, M. 2021, \apj, 908, 30,
  \dodoi{10.3847/1538-4357/abc2d9}

\bibitem[{{Kennicutt} \& {Evans}(2012)}]{Kennicutt12}
{Kennicutt}, R.~C., \& {Evans}, N.~J. 2012, \araa, 50, 531,
  \dodoi{10.1146/annurev-astro-081811-125610}

\bibitem[{{Kulkarni} {et~al.}(2019){Kulkarni}, {Worseck}, \&
  {Hennawi}}]{Kulkarni19}
{Kulkarni}, G., {Worseck}, G., \& {Hennawi}, J.~F. 2019, \mnras, 488, 1035,
  \dodoi{10.1093/mnras/stz1493}

\bibitem[{{Leitet} {et~al.}(2013){Leitet}, {Bergvall}, {Hayes}, {Linn{\'e}}, \&
  {Zackrisson}}]{Leitet13}
{Leitet}, E., {Bergvall}, N., {Hayes}, M., {Linn{\'e}}, S., \& {Zackrisson}, E.
  2013, \aap, 553, A106, \dodoi{10.1051/0004-6361/201118370}

\bibitem[{{Leitherer} {et~al.}(2016){Leitherer}, {Hernandez}, {Lee}, \&
  {Oey}}]{Leitherer16}
{Leitherer}, C., {Hernandez}, S., {Lee}, J.~C., \& {Oey}, M.~S. 2016, \apj,
  823, 64, \dodoi{10.3847/0004-637X/823/1/64}

\bibitem[{{Leitherer} {et~al.}(1999){Leitherer}, {Schaerer}, {Goldader},
  {Delgado}, {Robert}, {Kune}, {de Mello}, {Devost}, \&
  {Heckman}}]{Leitherer99}
{Leitherer}, C., {Schaerer}, D., {Goldader}, J.~D., {et~al.} 1999, \apjs, 123,
  3, \dodoi{10.1086/313233}

\bibitem[{{Luridiana} {et~al.}(2015){Luridiana}, {Morisset}, \&
  {Shaw}}]{Luridiana15}
{Luridiana}, V., {Morisset}, C., \& {Shaw}, R.~A. 2015, \aap, 573, A42,
  \dodoi{10.1051/0004-6361/201323152}

\bibitem[{{Madau} \& {Dickinson}(2014)}]{Madau14}
{Madau}, P., \& {Dickinson}, M. 2014, \araa, 52, 415,
  \dodoi{10.1146/annurev-astro-081811-125615}

\bibitem[{{Madau} \& {Haardt}(2015)}]{Madau15}
{Madau}, P., \& {Haardt}, F. 2015, \apjl, 813, L8,
  \dodoi{10.1088/2041-8205/813/1/L8}

\bibitem[{{Madau} {et~al.}(1999){Madau}, {Haardt}, \& {Rees}}]{Madau99}
{Madau}, P., {Haardt}, F., \& {Rees}, M.~J. 1999, \apj, 514, 648,
  \dodoi{10.1086/306975}

\bibitem[{{Makan} {et~al.}(2021){Makan}, {Worseck}, {Davies}, {Hennawi},
  {Prochaska}, \& {Richter}}]{Makan21}
{Makan}, K., {Worseck}, G., {Davies}, F.~B., {et~al.} 2021, \apj, 912, 38,
  \dodoi{10.3847/1538-4357/abee17}

\bibitem[{{Malkan} \& {Malkan}(2021)}]{Malkan21}
{Malkan}, M.~A., \& {Malkan}, B.~K. 2021, \apj, 909, 92,
  \dodoi{10.3847/1538-4357/abd84e}

\bibitem[{{Mason} {et~al.}(2018){Mason}, {Treu}, {Dijkstra}, {Mesinger},
  {Trenti}, {Pentericci}, {de Barros}, \& {Vanzella}}]{Mason18}
{Mason}, C.~A., {Treu}, T., {Dijkstra}, M., {et~al.} 2018, \apj, 856, 2,
  \dodoi{10.3847/1538-4357/aab0a7}

\bibitem[{{Matsuoka} {et~al.}(2018){Matsuoka}, {Strauss}, {Kashikawa}, {Onoue},
  {Iwasawa}, {Tang}, {Lee}, {Imanishi}, {Nagao}, {Akiyama}, {Asami}, {Bosch},
  {Furusawa}, {Goto}, {Gunn}, {Harikane}, {Ikeda}, {Izumi}, {Kawaguchi},
  {Kato}, {Kikuta}, {Kohno}, {Komiyama}, {Lupton}, {Minezaki}, {Miyazaki},
  {Murayama}, {Niida}, {Nishizawa}, {Noboriguchi}, {Oguri}, {Ono}, {Ouchi},
  {Price}, {Sameshima}, {Schulze}, {Shirakata}, {Silverman}, {Sugiyama},
  {Tait}, {Takada}, {Takata}, {Tanaka}, {Toba}, {Utsumi}, {Wang}, \&
  {Yamashita}}]{Matsuoka18}
{Matsuoka}, Y., {Strauss}, M.~A., {Kashikawa}, N., {et~al.} 2018, \apj, 869,
  150, \dodoi{10.3847/1538-4357/aaee7a}

\bibitem[{{Naidu} {et~al.}(2020){Naidu}, {Tacchella}, {Mason}, {Bose}, {Oesch},
  \& {Conroy}}]{Naidu20}
{Naidu}, R.~P., {Tacchella}, S., {Mason}, C.~A., {et~al.} 2020, \apj, 892, 109,
  \dodoi{10.3847/1538-4357/ab7cc9}

\bibitem[{{Naidu} {et~al.}(2021){Naidu}, {Matthee}, {Oesch}, {Conroy},
  {Sobral}, {Pezzulli}, {Hayes}, {Erb}, {Amor{\'\i}n}, {Gronke}, {Schaerer},
  {Tacchella}, {Kerutt}, {Paulino-Afonso}, {Calhau}, {Llerena}, \&
  {R{\"o}ttgering}}]{Naidu21}
{Naidu}, R.~P., {Matthee}, J., {Oesch}, P.~A., {et~al.} 2021, arXiv e-prints,
  arXiv:2110.11961.
\newblock \doarXiv{2110.11961}

\bibitem[{{Nakajima} {et~al.}(2020){Nakajima}, {Ellis}, {Robertson}, {Tang}, \&
  {Stark}}]{Nakajima20}
{Nakajima}, K., {Ellis}, R.~S., {Robertson}, B.~E., {Tang}, M., \& {Stark},
  D.~P. 2020, \apj, 889, 161, \dodoi{10.3847/1538-4357/ab6604}

\bibitem[{{Nakajima} \& {Ouchi}(2014)}]{Nakajima14}
{Nakajima}, K., \& {Ouchi}, M. 2014, \mnras, 442, 900,
  \dodoi{10.1093/mnras/stu902}

\bibitem[{{Neufeld}(1990)}]{Neufeld90}
{Neufeld}, D.~A. 1990, \apj, 350, 216, \dodoi{10.1086/168375}

\bibitem[{{Oey} {et~al.}(2015){Oey}, {Pellegrini}, {Winkler}, {Points},
  {Smith}, {Jaskot}, \& {Zastrow}}]{Oey15}
{Oey}, M.~S., {Pellegrini}, E.~W., {Winkler}, P.~F., {et~al.} 2015, Highlights
  of Astronomy, 16, 587, \dodoi{10.1017/S1743921314012290}

\bibitem[{{Orlitov{\'a}} {et~al.}(2018){Orlitov{\'a}}, {Verhamme}, {Henry},
  {Scarlata}, {Jaskot}, {Oey}, \& {Schaerer}}]{Orlitova18}
{Orlitov{\'a}}, I., {Verhamme}, A., {Henry}, A., {et~al.} 2018, \aap, 616, A60,
  \dodoi{10.1051/0004-6361/201732478}

\bibitem[{{Parsa} {et~al.}(2018){Parsa}, {Dunlop}, \& {McLure}}]{Parsa18}
{Parsa}, S., {Dunlop}, J.~S., \& {McLure}, R.~J. 2018, \mnras, 474, 2904,
  \dodoi{10.1093/mnras/stx2887}

\bibitem[{{Planck Collaboration} {et~al.}(2020){Planck Collaboration},
  {Aghanim}, {Akrami}, {Ashdown}, {Aumont}, {Baccigalupi}, {Ballardini},
  {Banday}, {Barreiro}, {Bartolo}, {Basak}, {Battye}, {Benabed}, {Bernard},
  {Bersanelli}, {Bielewicz}, {Bock}, {Bond}, {Borrill}, {Bouchet}, {Boulanger},
  {Bucher}, {Burigana}, {Butler}, {Calabrese}, {Cardoso}, {Carron},
  {Challinor}, {Chiang}, {Chluba}, {Colombo}, {Combet}, {Contreras}, {Crill},
  {Cuttaia}, {de Bernardis}, {de Zotti}, {Delabrouille}, {Delouis}, {Di
  Valentino}, {Diego}, {Dor{\'e}}, {Douspis}, {Ducout}, {Dupac}, {Dusini},
  {Efstathiou}, {Elsner}, {En{\ss}lin}, {Eriksen}, {Fantaye}, {Farhang},
  {Fergusson}, {Fernandez-Cobos}, {Finelli}, {Forastieri}, {Frailis},
  {Fraisse}, {Franceschi}, {Frolov}, {Galeotta}, {Galli}, {Ganga},
  {G{\'e}nova-Santos}, {Gerbino}, {Ghosh}, {Gonz{\'a}lez-Nuevo}, {G{\'o}rski},
  {Gratton}, {Gruppuso}, {Gudmundsson}, {Hamann}, {Handley}, {Hansen},
  {Herranz}, {Hildebrandt}, {Hivon}, {Huang}, {Jaffe}, {Jones}, {Karakci},
  {Keih{\"a}nen}, {Keskitalo}, {Kiiveri}, {Kim}, {Kisner}, {Knox},
  {Krachmalnicoff}, {Kunz}, {Kurki-Suonio}, {Lagache}, {Lamarre}, {Lasenby},
  {Lattanzi}, {Lawrence}, {Le Jeune}, {Lemos}, {Lesgourgues}, {Levrier},
  {Lewis}, {Liguori}, {Lilje}, {Lilley}, {Lindholm}, {L{\'o}pez-Caniego},
  {Lubin}, {Ma}, {Mac{\'\i}as-P{\'e}rez}, {Maggio}, {Maino}, {Mandolesi},
  {Mangilli}, {Marcos-Caballero}, {Maris}, {Martin}, {Martinelli},
  {Mart{\'\i}nez-Gonz{\'a}lez}, {Matarrese}, {Mauri}, {McEwen}, {Meinhold},
  {Melchiorri}, {Mennella}, {Migliaccio}, {Millea}, {Mitra},
  {Miville-Desch{\^e}nes}, {Molinari}, {Montier}, {Morgante}, {Moss}, {Natoli},
  {N{\o}rgaard-Nielsen}, {Pagano}, {Paoletti}, {Partridge}, {Patanchon},
  {Peiris}, {Perrotta}, {Pettorino}, {Piacentini}, {Polastri}, {Polenta},
  {Puget}, {Rachen}, {Reinecke}, {Remazeilles}, {Renzi}, {Rocha}, {Rosset},
  {Roudier}, {Rubi{\~n}o-Mart{\'\i}n}, {Ruiz-Granados}, {Salvati}, {Sandri},
  {Savelainen}, {Scott}, {Shellard}, {Sirignano}, {Sirri}, {Spencer},
  {Sunyaev}, {Suur-Uski}, {Tauber}, {Tavagnacco}, {Tenti}, {Toffolatti},
  {Tomasi}, {Trombetti}, {Valenziano}, {Valiviita}, {Van Tent}, {Vibert},
  {Vielva}, {Villa}, {Vittorio}, {Wandelt}, {Wehus}, {White}, {White},
  {Zacchei}, \& {Zonca}}]{Planck20}
{Planck Collaboration}, {Aghanim}, N., {Akrami}, Y., {et~al.} 2020, \aap, 641,
  A6, \dodoi{10.1051/0004-6361/201833910}

\bibitem[{{Puschnig} {et~al.}(2017){Puschnig}, {Hayes}, {{\"O}stlin},
  {Rivera-Thorsen}, {Melinder}, {Cannon}, {Menacho}, {Zackrisson}, {Bergvall},
  \& {Leitet}}]{Puschnig17}
{Puschnig}, J., {Hayes}, M., {{\"O}stlin}, G., {et~al.} 2017, \mnras, 469,
  3252, \dodoi{10.1093/mnras/stx951}

\bibitem[{{Rafelski} {et~al.}(2014){Rafelski}, {Neeleman}, {Fumagalli},
  {Wolfe}, \& {Prochaska}}]{Rafelski14}
{Rafelski}, M., {Neeleman}, M., {Fumagalli}, M., {Wolfe}, A.~M., \&
  {Prochaska}, J.~X. 2014, \apjl, 782, L29, \dodoi{10.1088/2041-8205/782/2/L29}

\bibitem[{{Ramambason} {et~al.}(2020){Ramambason}, {Schaerer}, {Stasi{\'n}ska},
  {Izotov}, {Guseva}, {V{\'\i}lchez}, {Amor{\'\i}n}, \&
  {Morisset}}]{Ramambason20}
{Ramambason}, L., {Schaerer}, D., {Stasi{\'n}ska}, G., {et~al.} 2020, \aap,
  644, A21, \dodoi{10.1051/0004-6361/202038634}

\bibitem[{{Reddy} {et~al.}(2016{\natexlab{a}}){Reddy}, {Steidel}, {Pettini}, \&
  {Bogosavljevi{\'c}}}]{Reddy16a}
{Reddy}, N.~A., {Steidel}, C.~C., {Pettini}, M., \& {Bogosavljevi{\'c}}, M.
  2016{\natexlab{a}}, \apj, 828, 107, \dodoi{10.3847/0004-637X/828/2/107}

\bibitem[{{Reddy} {et~al.}(2016{\natexlab{b}}){Reddy}, {Steidel}, {Pettini},
  {Bogosavljevi{\'c}}, \& {Shapley}}]{Reddy16b}
{Reddy}, N.~A., {Steidel}, C.~C., {Pettini}, M., {Bogosavljevi{\'c}}, M., \&
  {Shapley}, A.~E. 2016{\natexlab{b}}, \apj, 828, 108,
  \dodoi{10.3847/0004-637X/828/2/108}

\bibitem[{{Rivera-Thorsen} {et~al.}(2019){Rivera-Thorsen}, {Dahle}, {Chisholm},
  {Florian}, {Gronke}, {Rigby}, {Gladders}, {Mahler}, {Sharon}, \&
  {Bayliss}}]{Rivera-Thorsen19}
{Rivera-Thorsen}, T.~E., {Dahle}, H., {Chisholm}, J., {et~al.} 2019, Science,
  366, 738, \dodoi{10.1126/science.aaw0978}

\bibitem[{{Robertson} {et~al.}(2015){Robertson}, {Ellis}, {Furlanetto}, \&
  {Dunlop}}]{Robertson15}
{Robertson}, B.~E., {Ellis}, R.~S., {Furlanetto}, S.~R., \& {Dunlop}, J.~S.
  2015, \apjl, 802, L19, \dodoi{10.1088/2041-8205/802/2/L19}

\bibitem[{{Robertson} {et~al.}(2013){Robertson}, {Furlanetto}, {Schneider},
  {Charlot}, {Ellis}, {Stark}, {McLure}, {Dunlop}, {Koekemoer}, {Schenker},
  {Ouchi}, {Ono}, {Curtis-Lake}, {Rogers}, {Bowler}, \&
  {Cirasuolo}}]{Robertson13}
{Robertson}, B.~E., {Furlanetto}, S.~R., {Schneider}, E., {et~al.} 2013, \apj,
  768, 71, \dodoi{10.1088/0004-637X/768/1/71}

\bibitem[{{Rosdahl} {et~al.}(2018){Rosdahl}, {Katz}, {Blaizot}, {Kimm},
  {Michel-Dansac}, {Garel}, {Haehnelt}, {Ocvirk}, \& {Teyssier}}]{Rosdahl18}
{Rosdahl}, J., {Katz}, H., {Blaizot}, J., {et~al.} 2018, \mnras, 479, 994,
  \dodoi{10.1093/mnras/sty1655}

\bibitem[{{Saldana-Lopez} {et~al.}(2022){Saldana-Lopez}, {Schaerer},
  {Chisholm}, {Flury}, {Jaskot}, {Worseck}, {Makan}, {Gazagnes}, {Mauerhofer},
  {Verhamme}, {Amor{\'\i}n}, {Ferguson}, {Giavalisco}, {Grazian}, {Hayes},
  {Heckman}, {Henry}, {Ji}, {Marques-Chaves}, {McCandliss}, {Oey},
  {{\"O}stlin}, {Pentericci}, {Thuan}, {Trebitsch}, {Vanzella}, \&
  {Xu}}]{Saldana-Lopez22}
{Saldana-Lopez}, A., {Schaerer}, D., {Chisholm}, J., {et~al.} 2022, arXiv
  e-prints, arXiv:2201.11800.
\newblock \doarXiv{2201.11800}

\bibitem[{{Salpeter}(1955)}]{Salpeter55}
{Salpeter}, E.~E. 1955, \apj, 121, 161, \dodoi{10.1086/145971}

\bibitem[{{Scarlata} {et~al.}(2009){Scarlata}, {Colbert}, {Teplitz}, {Panagia},
  {Hayes}, {Siana}, {Rau}, {Francis}, {Caon}, {Pizzella}, \&
  {Bridge}}]{Scarlata09}
{Scarlata}, C., {Colbert}, J., {Teplitz}, H.~I., {et~al.} 2009, \apjl, 704,
  L98, \dodoi{10.1088/0004-637X/704/2/L98}

\bibitem[{{Schenker} {et~al.}(2014){Schenker}, {Ellis}, {Konidaris}, \&
  {Stark}}]{Schenker14}
{Schenker}, M.~A., {Ellis}, R.~S., {Konidaris}, N.~P., \& {Stark}, D.~P. 2014,
  \apj, 795, 20, \dodoi{10.1088/0004-637X/795/1/20}

\bibitem[{{Schlafly} \& {Finkbeiner}(2011)}]{Schlafly11}
{Schlafly}, E.~F., \& {Finkbeiner}, D.~P. 2011, \apj, 737, 103,
  \dodoi{10.1088/0004-637X/737/2/103}

\bibitem[{{Shen} {et~al.}(2020){Shen}, {Hopkins}, {Faucher-Gigu{\`e}re},
  {Alexander}, {Richards}, {Ross}, \& {Hickox}}]{Shen20}
{Shen}, X., {Hopkins}, P.~F., {Faucher-Gigu{\`e}re}, C.-A., {et~al.} 2020,
  \mnras, 495, 3252, \dodoi{10.1093/mnras/staa1381}

\bibitem[{{Shivaei} {et~al.}(2020){Shivaei}, {Reddy}, {Rieke}, {Shapley},
  {Kriek}, {Battisti}, {Mobasher}, {Sanders}, {Fetherolf}, {Azadi}, {Coil},
  {Freeman}, {de Groot}, {Leung}, {Price}, {Siana}, \& {Zick}}]{Shivaei20}
{Shivaei}, I., {Reddy}, N., {Rieke}, G., {et~al.} 2020, \apj, 899, 117,
  \dodoi{10.3847/1538-4357/aba35e}

\bibitem[{{Stanway} \& {Eldridge}(2018)}]{Stanway18}
{Stanway}, E.~R., \& {Eldridge}, J.~J. 2018, \mnras, 479, 75,
  \dodoi{10.1093/mnras/sty1353}

\bibitem[{{Stark} {et~al.}(2011){Stark}, {Ellis}, \& {Ouchi}}]{Stark11}
{Stark}, D.~P., {Ellis}, R.~S., \& {Ouchi}, M. 2011, \apjl, 728, L2,
  \dodoi{10.1088/2041-8205/728/1/L2}

\bibitem[{{Steidel} {et~al.}(2016){Steidel}, {Strom}, {Pettini}, {Rudie},
  {Reddy}, \& {Trainor}}]{Steidel16}
{Steidel}, C.~C., {Strom}, A.~L., {Pettini}, M., {et~al.} 2016, \apj, 826, 159,
  \dodoi{10.3847/0004-637X/826/2/159}

\bibitem[{{Storey} \& {Hummer}(1995)}]{Storey95}
{Storey}, P.~J., \& {Hummer}, D.~G. 1995, \mnras, 272, 41,
  \dodoi{10.1093/mnras/272.1.41}

\bibitem[{{Tang} {et~al.}(2019){Tang}, {Stark}, {Chevallard}, \&
  {Charlot}}]{Tang19}
{Tang}, M., {Stark}, D.~P., {Chevallard}, J., \& {Charlot}, S. 2019, \mnras,
  489, 2572, \dodoi{10.1093/mnras/stz2236}

\bibitem[{{Trebitsch} {et~al.}(2017){Trebitsch}, {Blaizot}, {Rosdahl},
  {Devriendt}, \& {Slyz}}]{Trebitsch17}
{Trebitsch}, M., {Blaizot}, J., {Rosdahl}, J., {Devriendt}, J., \& {Slyz}, A.
  2017, \mnras, 470, 224, \dodoi{10.1093/mnras/stx1060}

\bibitem[{{Trebitsch} {et~al.}(2021){Trebitsch}, {Dubois}, {Volonteri},
  {Pfister}, {Cadiou}, {Katz}, {Rosdahl}, {Kimm}, {Pichon}, {Beckmann},
  {Devriendt}, \& {Slyz}}]{Trebitsch21}
{Trebitsch}, M., {Dubois}, Y., {Volonteri}, M., {et~al.} 2021, \aap, 653, A154,
  \dodoi{10.1051/0004-6361/202037698}

\bibitem[{{Verhamme} {et~al.}(2015){Verhamme}, {Orlitov{\'a}}, {Schaerer}, \&
  {Hayes}}]{Verhamme15}
{Verhamme}, A., {Orlitov{\'a}}, I., {Schaerer}, D., \& {Hayes}, M. 2015, \aap,
  578, A7, \dodoi{10.1051/0004-6361/201423978}

\bibitem[{{Verhamme} {et~al.}(2017){Verhamme}, {Orlitov{\'a}}, {Schaerer},
  {Izotov}, {Worseck}, {Thuan}, \& {Guseva}}]{Verhamme17}
{Verhamme}, A., {Orlitov{\'a}}, I., {Schaerer}, D., {et~al.} 2017, \aap, 597,
  A13, \dodoi{10.1051/0004-6361/201629264}

\bibitem[{{Verhamme} {et~al.}(2006){Verhamme}, {Schaerer}, \&
  {Maselli}}]{Verhamme06}
{Verhamme}, A., {Schaerer}, D., \& {Maselli}, A. 2006, \aap, 460, 397,
  \dodoi{10.1051/0004-6361:20065554}

\bibitem[{{Wang} {et~al.}(2019){Wang}, {Heckman}, {Leitherer}, {Alexandroff},
  {Borthakur}, \& {Overzier}}]{Wang19}
{Wang}, B., {Heckman}, T.~M., {Leitherer}, C., {et~al.} 2019, \apj, 885, 57,
  \dodoi{10.3847/1538-4357/ab418f}

\bibitem[{{Wang} {et~al.}(2021){Wang}, {Heckman}, {Amor{\'\i}n}, {Borthakur},
  {Chisholm}, {Ferguson}, {Flury}, {Giavalisco}, {Grazian}, {Hayes}, {Henry},
  {Jaskot}, {Ji}, {Makan}, {McCandliss}, {Oey}, {{\"O}stlin}, {Saldana-Lopez},
  {Schaerer}, {Thuan}, {Worseck}, \& {Xu}}]{Wang21}
{Wang}, B., {Heckman}, T.~M., {Amor{\'\i}n}, R., {et~al.} 2021, \apj, 916, 3,
  \dodoi{10.3847/1538-4357/ac0434}

\bibitem[{{Worseck} {et~al.}(2014){Worseck}, {Prochaska}, {O'Meara}, {Becker},
  {Ellison}, {Lopez}, {Meiksin}, {M{\'e}nard}, {Murphy}, \&
  {Fumagalli}}]{Worseck14}
{Worseck}, G., {Prochaska}, J.~X., {O'Meara}, J.~M., {et~al.} 2014, \mnras,
  445, 1745, \dodoi{10.1093/mnras/stu1827}

\bibitem[{{Zackrisson} {et~al.}(2013){Zackrisson}, {Inoue}, \&
  {Jensen}}]{Zackrisson13}
{Zackrisson}, E., {Inoue}, A.~K., \& {Jensen}, H. 2013, \apj, 777, 39,
  \dodoi{10.1088/0004-637X/777/1/39}

\end{thebibliography}
\bibliographystyle{aasjournal}

\end{document}